\documentclass[final,authoryear,3p,times]{elsarticle}

\usepackage{graphicx,color,colortbl}
\usepackage{enumerate}
\usepackage[colorlinks]{hyperref}
\usepackage{amsmath,amssymb,amsthm,bm}
\usepackage{multicol}
\usepackage[dvipsnames]{xcolor}
\usepackage{ulem}
\usepackage{tikz}
\usetikzlibrary{automata,positioning}
\usepackage{afterpage}
\usepackage{calc}
\usepackage{ifthen}
\usepackage{algorithm}
\usepackage{algpseudocode}
\usepackage{caption}
\usepackage{subcaption}
\usepackage{siunitx}
\usepackage{booktabs}
\usepackage{physics}
\usepackage{svg}
\usepackage[nameinlink,capitalise]{cleveref} 

\usepackage{tikz}
\usetikzlibrary{shapes.geometric, arrows}

\edef\svtheparindent{\the\parindent}
\usepackage{parskip}
\parindent=\svtheparindent\relax

\newcommand{\RR}[1]{#1}

\tikzstyle{process} = [rectangle, rounded corners, minimum width=13cm, minimum height=1cm, text centered, draw=black, fill=white!30]
\tikzstyle{process1} = [rectangle, rounded corners, minimum width=2cm, minimum height=1cm, text centered, draw=black, fill=white!30]
\tikzstyle{decision} = [diamond, minimum width=5cm, minimum height=0.5cm, text centered, draw=black, fill=white!30]
\tikzstyle{arrow} = [thick,->,>=stealth]

\newcommand{\boldface}[1]{\boldsymbol{#1}}  
\newcommand{\bfa}{\boldface{a}}
\newcommand{\bfb}{\boldface{b}}

\newcommand{\bfn}{\boldface{n}}

\newcommand{\bfx}{\boldface{x}}

\newcommand{\bfA}{\vb{A}}

\newcommand{\bfC}{\vb{C}}

\newcommand{\bfF}{\vb{F}}

\newcommand{\bfN}{\boldface{N}}

\newcommand{\bfP}{\vb{P}}
\newcommand{\bfQ}{\boldface{Q}}

\newcommand{\bfT}{\vb{T}}

\newcommand{\bftheta}{\boldsymbol{\theta}}

\newcommand{\Rset}{\mathbb{R}}

\newlength{\boxwidth}
\def\btheorem{\begin{theorem}}
\def\etheorem{\end{theorem}}
\def\blemma{\begin{lemma}}
\def\elemma{\end{lemma}}
\def\bproposition{\begin{proposition}}
\def\eproposition{\end{proposition}}
\def\bcorollary{\begin{corollary}}
\def\ecorollary{\end{corollary}}
\def\bdefinition{\begin{definition}}
\def\edefinition{\end{definition}}
\def\bexample{\begin{example}}
\def\eexample{\end{example}}
\def\bremark{\begin{remark}}
\def\eremark{\end{remark}}

\newcommand{\eps}{\varepsilon}

\newcommand{\be}{\begin{equation}}
\newcommand{\ee}{\end{equation}}
\newcommand{\beq}{\begin{eqnarray}}
\newcommand{\eeq}{\end{eqnarray}}
\newcommand{\bem}{\begin{multline}}
\newcommand{\eem}{\end{multline}}
\newcommand{\ba}{\begin{align}}
\newcommand{\ea}{\end{align}}

\newcommand{\argminWithArgs}{\operatornamewithlimits{arg\ min}}

\begin{document}

\begin{frontmatter}

\title{Automated discovery of interpretable hyperelastic material models for human brain tissue with EUCLID}

\author[eth]{Moritz Flaschel\corref{cor1}}
\author[eth]{Huitian Yu\corref{cor1}}
\author[fau]{Nina Reiter}
\author[fau]{Jan Hinrichsen}
\author[fau]{Silvia Budday}
\author[fau]{Paul Steinmann}
\author[tud]{Siddhant Kumar}
\author[eth]{Laura De Lorenzis\corref{cor2}}

\cortext[cor1]{These authors contributed equally.}
\cortext[cor2]{Correspondence: ldelorenzis@ethz.ch}

\address[eth]{Department of Mechanical and Process Engineering, ETH Z\"{u}rich, 8092 Z\"{u}rich, Switzerland}
\address[fau]{Department of Mechanical Engineering, Friedrich-Alexander-Universit\"at of Erlangen–N\"urnberg, 91058 Erlangen, Germany}
\address[tud]{Department of Materials Science and Engineering, Delft University of Technology, 2628 CD Delft, The Netherlands}

\begin{abstract}

We propose an automated computational algorithm for simultaneous model selection and parameter identification for the hyperelastic mechanical characterization of human brain tissue.
Following the motive of the recently proposed computational framework EUCLID (Efficient Unsupervised Constitutive Law Identitication and Discovery) and in contrast to conventional parameter calibration methods, we construct an extensive set of candidate hyperelastic models, i.e., a model library including popular models known from the literature, and develop a computational strategy for automatically selecting a model from the library that conforms to the available experimental data while being represented as an interpretable symbolic mathematical expression.
This computational strategy comprises sparse regression, i.e., a regression problem that is regularized by a sparsity promoting penalty term that filters out irrelevant models from the model library, and a clustering method for grouping together highly correlated and thus redundant features in the model library.
The model selection procedure is driven by labelled data pairs stemming from mechanical tests under different deformation modes, i.e., uniaxial compression/tension and simple torsion, and can thus be interpreted as a supervised counterpart to the originally proposed EUCLID that is informed by full-field displacement data and global reaction forces.
The proposed method is verified on synthetical data with artificial noise and validated on experimental data acquired through mechanical tests of human brain specimens, proving that the method is capable of discovering hyperelastic models that exhibit both high fitting accuracy to the data as well as concise and thus interpretable mathematical representations.
\end{abstract}


\end{frontmatter}


\section{Introduction}

Despite extensive research over the past decades, the characterization of the mechanical response of biological tissues like those of the human brain remains an active field with many open questions, see the review article by \cite{budday_fifty_2020}.
The amount of data is usually limited by ethical restrictions and by the availability of only few specimens with short durability \citep{faber_tissue-scale_2022}.
The general objective is therefore to leverage these limited data in an efficient manner to discover an adequate mathematical description of the tissue response.
The conventional strategy is to a priori assume a constitutive model and to calibrate the tunable model parameters by means of the experimental measurements.
However, the a priori choice of the constitutive model is driven by human experience and intuition; thus, this classical strategy is highly susceptible to introducing modeling errors that lead to poor fitting accuracy and/or the need for tedious and time-consuming iterative model correction procedures.
Data-driven methods and machine learning approaches promise a powerful remedy.
In this work, one of such methods, which falls under the umbrella of methods denoted as EUCLID (Efficient Unsupervised Constitutive Law Identification and Discovery) \citep{flaschel_unsupervised_2021}, is used for automatically discovering symbolic expressions of hyperelastic material models for human brain tissue based on experimental measurements.

The general idea behind machine-learning-based material modeling is to choose a model ansatz that exhibits a high expressiveness owing to a vast amount of tunable parameters, and to calibrate these parameters by leveraging the available data.
Popular examples of versatile machine-learning-based material models include neural networks \citep{ghaboussi_knowledgebased_1991}, splines \citep{sussman_model_2009}, Gaussian processes \citep{frankel_tensor_2020}, and neural ordinary differential equations \citep{tac_data-driven_2022}.
In the context of hyperelasticity, these methods can be used to learn the characteristic strain energy density function while special attention needs to be attributed to not violating physical constraints like objectivity or (poly-)convexity of the strain energy density function \citep{asad_mechanics-informed_2022,klein_polyconvex_2022,klein_finite_2022,kalina_automated_2022,thakolkaran_nn-euclid_2022,chen_polyconvex_2022,linka_new_2022,asad_mechanics-informed_2022-1,tac_data-driven_2022}.
Until now, the majority of the proposed machine learning models for hyperelasticy are informed by artificially generated data, e.g., data generated through simulations at the microscopic level.
An exception is for example the work by \cite{linka_unraveling_2021}, who train neural networks based on real experimental data for the characterization of human brain tissue.
In contrast to machine-learning-based approaches stand model-free data-driven methods \citep{kirchdoerfer_data-driven_2016,ibanez_manifold_2018}, which seek to avoid the formulation of a material model altogether by solving forward problems that are directly informed by the data.

Both the previously mentioned machine-learning-based approaches as well as model-free data-driven methods have in common that the constitutive behavior is encoded in a black box that does not allow for physical interpretation, e.g., it is difficult to interpret the meaning of the many functions and parameters that compose models encoded in machine learning tools like neural networks.
Furthermore, these methods typically rely on a vast amount of labeled data pairs that are not available under experimental conditions. 
The recognition of these issues motivated the development of the EUCLID method, which aims to discover interpretable symbolic expressions of material models using only experimentally available data like displacement fields and global reaction forces.
Initially applied to hyperelasticity \citep{flaschel_unsupervised_2021,joshi_bayesian-euclid_2022} (see also the related and independent work by \cite{wang_inference_2021}), the framework was later extended to viscoelasticity \citep{marino_automated_2023}, elastoplasticity \citep{flaschel_discovering_2022}, and generalized standard materials \citep{flaschel_automated_2023-1}, see \cite{flaschel_automated_2023} for an overview.
Recently, discovering symbolic expressions for material models is gaining more and more attention in the field, see, e.g., the works by \cite{bomarito_development_2021,kabliman_application_2021,park_multiscale_2021,abdusalamov_automatic_2023,meyer_thermodynamically_2023}.

In this work, EUCLID is used to automatically discover hyperelastic strain energy density functions for describing the mechanical behavior of human brain tissue.
The idea is to construct a set of candidate material models called the model library, which can be assembled for example from the vast available literature on hyperelastic material models (see the non-comprehensive list of reviewing articles by \cite{boyce_constitutive_2000,marckmann_comparison_2006,steinmann_hyperelastic_2012,dal_performance_2021,he_comparative_2021}), and to use a specifically designed algorithm, based on sparse regression and feature clustering, to select from the library a material model which is encoded by an interpretable, simple mathematical expression and which is able to well capture the mechanical behavior observed in the experimental data.
Due to the wet and shiny surface of brain tissues, displacement field data are difficult to acquire; thus, against the original philosophy of EUCLID \citep{flaschel_automated_2023}, the method in this paper is not informed by full-field displacement and global reaction force data. Instead, we rely here on labeled data pairs stemming from uniaxial compression/tension and simple torsion tests of human brain tissue specimens.

The paper is organized as follows: In \cref{sec:model_library}, we construct the model library in which EUCLID searches for a suitable material model.
In \cref{sec:model_library}, the proposed algorithm for model selection is discussed in detail.
Subsequently, the method is tested on artificial data and actual experimental data in \cref{sec:numerical_verification} and \cref{sec:experimental_validation}, respectively, and conclusions are drawn in \cref{sec:conclusion}.

\textbf{Notation}:
Tensors and matrices may appear in compact or index notation, e.g., $\bfF$ or $F_{ij}$, respectively. In compact notation, first-order tensors (vectors) and second-order tensors are denoted by bold letters, e.g., $\bfF$.
The transpose of a tensor is denoted by $\square^T$. When appearing in index notation, the order of the tensor equals the number of the indices.
If not stated otherwise, the Einstein convention for summation over repeated indices is used in equations appearing in index notation, e.g., $a_i b_i = \sum_i a_i b_i$.
Inner products are denoted by $\cdot$, e.g., $\bfa \cdot \bfb = a_i b_i$, and outer products by $\otimes$, e.g., $\{\bfa \otimes \bfb\}_{ij} = a_i b_j$.
If no operation is indicated between two tensors, the juxtaposition implies tensor contraction, e.g., $\{\bfF^T\bfF\}_{ij}=F_{ki}F_{kj}$.
The trace of a tensor is denoted by $\text{tr}(\square)$, e.g., $\text{tr}(\bfC) = C_{ii}$, the volumetric part by $\text{vol}(\square)$, the deviatoric part by $\text{dev}(\square)$, and the determinant by $\text{det}(\square)$.

\section{Model library}
\label{sec:model_library}
\subsection{Strain energy density}
The constitutive response of a hyperelastic material is completely characterized by the material strain energy density function.
In the spirit of EUCLID, we construct a material model library, i.e., a large set of potential candidate material models, by introducing a general parametric ansatz for the unknown strain energy density.
Under the assumption of incompressibility, the strain energy density $W$ of a hyperelastic material is postulated as \citep{holzapfel_nonlinear_2000}
\be
W = \tilde{W}(\bfF) - p \cdot (J-1),
\ee
where $\bfF$ is the deformation gradient, $p$ is a scalar Lagrange multiplier that can be physically interpreted as the hydrostatic (or volumetric) pressure, and $J=\det\bfF\overset{!}{=}1$ is the determinant of the deformation gradient. A sufficient condition to satisfy the objectivity requirement for the material model is that $W$ depends on $\bfF$ through the right Cauchy-Green tensor $\bfC=\bfF^T\bfF$, i.e. $\tilde{W}(\bfF)=\hat{W}(\bfC)$.
Assuming furthermore isotropic material behavior, the contribution $\hat{W}$ can be expressed as a function of the invariants of $\bfC$, defined as $I_1=\text{tr} (\bfC)$,  $I_2=\frac{1}{2}(\text{tr}^2 (\bfC) - \text{tr} (\bfC^2))$ and $I_3=\det(\bfC)$. An equally objective possibility for isotropic hyperelasticity is to express the strain energy density as a function of the principal stretches $\lambda_1$, $\lambda_2$, $\lambda_3$ defined as the eigenvalues of the right stretch tensor $\vb{U}$, which in turn is defined through the polar decomposition $\bfF=\vb{R}\vb{U}$ \citep{holzapfel_nonlinear_2000}.
As we seek to avoid a priori assumptions on the material response, we do not limit ourselves to a strain energy density that depends solely on the strain invariants or solely on the principal stretches, but instead consider the general case in which the strain energy density includes a contribution $\tilde{W}_I$ that depends on the strain invariants and a contribution $\tilde{W}_{\lambda}$ that depends on the principal stretches
\be
W = \tilde{W}_I(I_1,I_2,I_3) + \tilde{W}_{\lambda}(\lambda_1,\lambda_2,\lambda_3) - p \cdot (J-1).
\ee
By doing so, we obtain a highly general expression for the strain energy density that encompasses many well-known hyperelastic constitutive models e.g. of the Mooney-Rivlin or Ogden types.

To introduce a general parametric ansatz for $\tilde{W}_I$, we assume that it can be expressed as a linear combination of a priori chosen nonlinear feature functions
\be
\label{eq:energy_I}
\tilde{W}_I(I_1,I_2,I_3) = \bftheta_I \cdot \bfQ_I(I_1,I_2,I_3),
\ee
where the nonlinear feature functions have been collected in the vector $\bfQ_I$ and $\bftheta_I$ is a vector of unknown real-valued material parameters.
We construct the feature vector $\bfQ_I$ such that it contains features of the generalized Mooney-Rivlin model \citep{rivlin_torsion_1947} and a logarithmic feature as it appears in the Gent-Thomas model \citep{gent_forms_1958}
\be
\bfQ_I(I_1,I_2,I_3) = 
\underbrace{\left[ (I_1-3)^m(I_2-3)^{n-m} : n\in \{1,\dots,N_{\text{Mooney}}\}, m\in\{0,\dots,n\}\right]^T}_{\text{generalized Mooney-Rivlin features}}
\oplus 
\underbrace{\left[ \log \left(I_2 / 3\right)\right]}_{\text{logarithmic feature}},
\ee
where $\oplus$ denotes vector concatenation and the choice of $N_{\text{Mooney}}$ dictates the size of the feature vector.
This selection of feature functions was proven in \cite{flaschel_unsupervised_2021} to have a high approximation power.
We choose $N_{\text{Mooney}} = 3$ in this work, such that $\bftheta_I$ comprises 10 unknown parameters.

To obtain a general parametric ansatz for the principal stretch dependent contribution $\tilde{W}_{\lambda}$, we consider the generalized Ogden model \citep{ogden_large_1972}
\be
\tilde{W}_{\lambda}(\lambda_1,\lambda_2,\lambda_3)  = \sum_{i=1}^{N_{\text{Ogden}}} \frac{2\mu_{i}}{\alpha_i^2}
\left(\lambda_1^{\alpha_i} + \lambda_2^{\alpha_i} + \lambda_3^{\alpha_i} - 3\right),
\ee
where $\mu_{i}$ and $\alpha_{i}$ are unknown real-valued material parameters and $N_{\text{Ogden}}$ is the number of considered terms in the generalized Ogden model.
In contrast to the model library chosen in \cref{eq:energy_I}, the strain energy density of the generalized Ogden model depends nonlinearly on the unknown material parameters.
In general, this complicates the inference of the material parameters from experimental measurements.
Therefore, we assume in the following an a priori fixed set of $N_{\text{Ogden}}$ distinct values of $\alpha_i$.
By choosing $N_{\text{Ogden}}$ sufficiently large, it is expected that this assumption does not significantly restrict the versatility of the model library.
In this work, we choose $N_{\text{Ogden}}=2 \cdot 10^4$ values of $\alpha_i$ evenly distributed between $-100$ and $100$ excluding zero, i.e., $\alpha_i \in \{ -100, -99.99, \dots, -0.01, 0.01, \dots, 99.99, 100 \}$.
With this assumption, $\mu_{i}$ remain the only unknowns in the model ansatz for $\tilde{W}_{\lambda}$ which can now be written as a linear combination of nonlinear feature functions
\be
\label{eq:energy_lambda}
\tilde{W}_{\lambda}(\lambda_1,\lambda_2,\lambda_3) = \bftheta_{\lambda} \cdot \bfQ_{\lambda}(\lambda_1,\lambda_2,\lambda_3),
\ee
where the nonlinear feature functions have been collected in the vector $\bfQ_{\lambda}$ with components
\be
\left\{Q_{\lambda}(\lambda_1,\lambda_2,\lambda_3)\right\}_i = \lambda_1^{\alpha_i} + \lambda_2^{\alpha_i} + \lambda_3^{\alpha_i} - 3,
\ee
and $\bftheta_{\lambda}$ is a vector of unknown real-valued material parameters that are related to the parameters $\mu_i$ through
$\left\{\theta_{\lambda}\right\}_{i} = 2\mu_i / \alpha_i^2$.

By defining
\be
\bftheta=
\begin{bmatrix}
\bftheta_{I} \\
\bftheta_{\lambda} \\
\end{bmatrix}, 
\quad
\bfQ=
\begin{bmatrix}
\bfQ_{I}(I_1,I_2,I_3) \\
\bfQ_{\lambda}(\lambda_1,\lambda_2,\lambda_3) \\
\end{bmatrix},
\ee
the model library for the strain energy density can be written as
\be
W = \bftheta \cdot \bfQ - p \cdot (J-1).
\ee

To ensure that the strain energy density describes physically meaningful material behavior, we assume that $\theta_i \geq 0$ for all $i$.
The non-negativity of the material parameters is a sufficient (but not necessary) requirement for stability, see \cite{hartmann_parameter_2001}.

\subsection{Stress-strain relation}
After having defined the strain energy density, a bijective relation between the kinematic state and the stress state of the material is obtained through its differentiation.
Specifically, the Piola stress $\bfP$ is computed by differentiating $W$ with respect to the deformation gradient
\be
\label{eq:stress}
\bfP
= \frac{\partial W}{\partial \bfF}
= \frac{\partial \tilde{W}}{\partial \bfF} - p\bfF^{-T},
\ee
where we used $\frac{\partial J}{\partial \bfF} = J\bfF^{-T}$.
Noting that $\tilde{W} = \bftheta \cdot \bfQ = \bftheta_I \cdot \bfQ_I + \bftheta_{\lambda} \cdot \bfQ_{\lambda}$, we apply the chain rule to obtain
\be\label{eq:stress_chain_rule}
\frac{\partial \tilde{W}}{\partial F_{ij}}
= \bftheta_I \cdot \left(\frac{\partial \bfQ_I}{\partial I_a}\frac{\partial I_a}{\partial F_{ij}}\right)
+ \bftheta_{\lambda} \cdot \left(\frac{\partial \bfQ_{\lambda}}{\partial \lambda_b}\frac{\partial \lambda_b}{\partial F_{ij}}\right).
\ee
The derivatives of the strain invariants and of the principal stretches are provided in \ref{sec:derivatives_strain_invariants} and \ref{sec:derivatives_principal_stretches}, respectively.
The derivatives of the feature vectors follow in \ref{sec:derivatives_feature_vectors}.
The Cauchy stress $\bfT$ is calculated from the Piola stress as $\bfT = \bfP\bfF^T$.

\subsubsection{Uniaxial compression and tension}
Experimental measurements of specimens under uniaxial compression and tension deliver labeled data pairs in the form $(u_{\text{UT}},F)$, where $u_{\text{UT}}$ is the longitudinal displacement at the displacement controlled end of the specimen and $F$ is the resulting force.
Data pairs of the form $(\lambda_{\text{UT}},P_{11})$, where $\lambda_{\text{UT}}$ is the longitudinal stretch applied to the specimen and $P_{11}$ is the longitudinal normal component of the Piola stress, are obtained through
\be
\lambda_{\text{UT}} = \frac{u_{\text{UT}} + h}{h}, \quad P_{11} = \frac{F}{\pi r_{\text{out}}^2},
\ee
where $h$ and $r_{\text{out}}$ are the height and the outer radius of the undeformed specimen, respectively.
The previously introduced material model library can be used to derive a relation $P_{11}(\lambda_{\text{UT}};\bftheta)$ that links the experimental inputs $\lambda_{\text{UT}}$ and outputs $P_{11}$ depending on the material parameters $\bftheta$.

Applying a longitudinal stretch $\lambda_{\text{UT}}$ to the specimen implies that $F_{11}=\lambda_{\text{UT}}$ and, due to symmetry, $F_{22}=F_{33}$. As a result of the incompressibility assumption $\det\bfF\overset{!}{=}1$, the deformation gradient under uniaxial compression/tension must hence be
\be
\begin{aligned}
\bfF_{\text{UT}} =
	\begin{bmatrix}
		\lambda_{\text{UT}} & 0 & 0\\
		0 & \frac{1}{\sqrt{\lambda_{\text{UT}}}} & 0\\
		0 & 0 & \frac{1}{\sqrt{\lambda_{\text{UT}}}}\\
	\end{bmatrix}.
\end{aligned}
\ee
The kinematic state is thus completely determined by the applied longitudinal stretch.

To obtain the relationship $P_{11}(\lambda_{\text{UT}};\bftheta)$, the unknown hydrostatic pressure $p$ in \cref{eq:stress} needs to be computed.
As the material is allowed to freely deform in the $x_2$- and $x_3$-directions, it is $P_{22}=P_{33}\overset{!}{=}0$.
This condition is used along with \cref{eq:stress}
\be
P_{33} = \frac{\partial \tilde{W}}{\partial F_{33}}- pF^{-1}_{33} \overset{!}{=} 0,
\ee
to find the hydrostatic pressure
\be
p = \frac{\partial \tilde{W}}{\partial F_{33}} F_{33}.
\ee
We hence obtain the desired relationship by substituting the pressure in \cref{eq:stress}
\be
\label{eq:constitutive_map_uniaxial_tension}
P_{11}(\lambda_{\text{UT}};\bftheta)
= \frac{\partial \tilde{W}}{\partial F_{11}} - pF^{-1}_{11}
= \frac{\partial \tilde{W}}{\partial F_{11}} - \frac{F_{33}}{F_{11}}\frac{\partial \tilde{W}}{\partial F_{33}}\\
= \bftheta \cdot \underbrace{\left( \frac{\partial\bfQ}{\partial F_{11}} - \frac{F_{33}}{F_{11}} \frac{\partial\bfQ}{\partial F_{33}} \right)}_{\bfQ'_{\text{UT}}\left(\lambda_{\text{UT}}\right)},
\ee
where we denoted all the terms that depend on the longitudinal stretch as $\bfQ'_{\text{UT}}(\lambda_{\text{UT}})$. This can be interpreted as a feature function derivative vector whose scalar product with the material parameters computes the longitudinal normal component of the Piola stress. More information for computing $\bfQ'_{\text{UT}}(\lambda_{\text{UT}})$ is provided in \ref{sec:derivatives_uniaxial_tension}.

\subsubsection{Simple torsion}
Now we consider a cylindrical specimen in a cylindrical coordinate system $(r,\vartheta,z)$, where $r$ is the radial coordinate, $\vartheta$ is the polar angle, and $z$ is the longitudinal coordinate.
Assuming the specimen to undergo simple torsion around the $z$-axis,
experimental measurements on specimens under simple torsion deliver labeled data pairs in the form $(\phi,M)$,
where $\phi$ is the applied angle of deformation at the sheared end of the specimen and $M$ is the resulting torque.
The twist $\psi$ of the specimen can be deduced from the applied angle of deformation through $\psi = \phi/h$.
The data is transformed into data pairs of the form $(\tilde\psi,\tau)$, where $\tilde\psi=r_{\text{out}}\psi$ and $\tau = M/r_{\text{out}}^3$ are respectively a normalized twist and normalized torque.

After normalizing the radial coordinate through $\rho=r/r_{\text{out}}$, the deformation gradient in the cylindrical coordinate system then reads \citep{hartmann_parameter_2001}
\be
\label{FST}
\begin{aligned}
\bfF_{\text{ST}} =
\begin{bmatrix}
F_{rr} & F_{r\vartheta} & F_{rz}\\
F_{\vartheta r} & F_{\vartheta\vartheta} & F_{\vartheta z}\\
F_{zr} & F_{z\vartheta} & F_{zz}\\
\end{bmatrix}
=
	\begin{bmatrix}
		1 & 0 & 0\\
		0 & 1 & \rho\tilde{\psi}\\
		0 & 0 & 1\\
	\end{bmatrix},
\end{aligned}
\ee
The material model library can be used to derive a relation $\tau(\tilde{\psi};\bftheta)$ that links the experimental inputs $\tilde\psi$ and outputs $\tau$ depending on the material parameters $\bftheta$.

The normalized torque is given by
\be
\label{eq:torque}
\tau = \int_0^1 \ 2\pi\rho^2T_{\vartheta z} \ \mathrm{d}\rho,
\ee
where the component of the Cauchy stress $T_{\vartheta z}$ is, due to \cref{FST},
\be
T_{\vartheta z} = P_{\vartheta i} F_{zi} = P_{\vartheta z}.
\ee
With reference to \cref{eq:stress}, we notice that there is no hydrostatic pressure contribution to $P_{\vartheta z}$ under simple torsion, such that
\be
\label{Tthetaz}
T_{\vartheta z} = \frac{\partial {W}}{\partial F_{\vartheta z}} = \frac{\partial \tilde{W}}{\partial F_{\vartheta z}} = \bftheta \cdot \frac{\partial \bfQ}{\partial F_{\vartheta z}}.
\ee
We finally obtain the desired relationship $\tau(\tilde{\psi};\bftheta)$ by substituting \cref{Tthetaz} in \cref{eq:torque}
\be
\label{eq:constitutive_map_simple_torsion}
\tau(\tilde\psi;\bftheta) = \bftheta \cdot \underbrace{\int_0^1 \ 2\pi\rho^2 \frac{\partial \bfQ}{\partial F_{\vartheta z}} \ \mathrm{d}\rho}_{\bfQ'_{\text{ST}}\left(\tilde\psi\right)},
\ee
where we denoted all the terms that depend on the twist as $\bfQ'_{\text{ST}}(\tilde\psi)$. This can be interpreted as a feature function derivative vector whose scalar product with the material parameters computes the normalized torque. More information for computing $\bfQ'_{\text{ST}}(\tilde\psi)$ is provided in \ref{sec:derivatives_simple_torsion}.

\section{Model discovery}
\label{sec:model_discovery}
After having defined the material model library, i.e., a versatile ansatz for the hyperelastic strain energy density, we seek an optimal choice of the material model parameters $\bftheta$.
Specifically, in the spirit of EUCLID - see \cite{flaschel_unsupervised_2021} - we seek a parameter vector $\bftheta$ which leads at the same time to a small mismatch between the material model predictions and the experimental measurements, and to a simple discovered material model, i.e. one with a small number of terms.
Unlike in the original work on EUCLID \citep{flaschel_unsupervised_2021}, in which the model selection was driven by a physics-informed optimization problem that was fed by unlabelled data, in this paper, labelled data pairs (i.e., stress versus strain data pairs under uniaxial compression/tension, and twist versus torque data pairs under simple torsion) are available to drive the model selection.
Thus, the method proposed here can be interpreted as a supervised counterpart to the originally proposed EUCLID.
The proposed algorithm for inferring $\bftheta$ is described in the following and
a step-by-step overview is provided in \ref{algorithm}, see \cref{fig:flowchart_algorithm}.

\subsection{Model-data mismatch}
First, a quantitative measure of the mismatch between the material model predictions and the experimental measurements needs to be introduced.
In the case of uniaxial compression/tension, the measurements are provided in the form of pairs $(\lambda_{\text{UT}}^{(l)},P_{11}^{(l)})$ for a number of load steps $l=1,\dots,n_{\text{UT}}$.
\cref{eq:constitutive_map_uniaxial_tension} provides a mapping between the input and output data, thus leading to $n_{\text{UT}}$ equations
\be
P_{11}(\lambda_{\text{UT}}^{(l)};\bftheta) = P_{11}^{(l)}.
\ee
Noting that each of the equations above depends linearly on $\bftheta$, we write them as a linear system of equations
\be
\label{eq:linear_system_UT}
\bfA_{\text{UT}} \bftheta = \bfb_{\text{UT}},
\ee
with
\be
\bfA_{\text{UT}} = 
\begin{bmatrix}
    \left(\bfQ'_{\text{UT}}(\lambda_{\text{UT}}^{(1)})\right)^T\\
    \vdots\\
    \left(\bfQ'_{\text{UT}}(\lambda_{\text{UT}}^{(n_{\text{UT}})})\right)^T\\
\end{bmatrix}, \quad
\bfb_{\text{UT}} = 
\begin{bmatrix}
    P_{11}^{(1)}\\
    \vdots\\
    P_{11}^{(n_{\text{UT}})}\\
\end{bmatrix}.
\ee
Analogously, in the case of simple torsion, the measurements are provided in the form of pairs $(\tilde\psi^{(l)},\tau^{(l)})$ for a number of load steps $l=1,\dots,n_{\text{ST}}$.
\cref{eq:constitutive_map_simple_torsion} provides a mapping between the input and output data, thus leading to $n_{\text{ST}}$ equations
\be
\tau(\tilde\psi^{(l)};\bftheta) = \tau^{(l)}.
\ee
which can be expressed once again as a linear system of equations
\be
\label{eq:linear_system_ST}
\bfA_{\text{ST}} \bftheta = \bfb_{\text{ST}},
\ee
with
\be
\bfA_{\text{ST}} = 
\begin{bmatrix}
    \left(\bfQ'_{\text{ST}}(\tilde\psi^{(1)})\right)^T\\
    \vdots\\
    \left(\bfQ'_{\text{ST}}(\tilde\psi^{(n_{\text{ST}})})\right)^T\\
\end{bmatrix}, \quad
\bfb_{\text{ST}} = 
\begin{bmatrix}
    \tau^{(1)}\\
    \vdots\\
    \tau^{(n_{\text{ST}})}\\
\end{bmatrix}.
\ee
The linear systems in \cref{eq:linear_system_UT} and \cref{eq:linear_system_ST} are concatenated to the system $\bfA\bftheta=\bfb$ with
\be
\bfA = \begin{bmatrix}
		r_{\text{UT}}\bfA_{\text{UT}}\\
        r_{\text{ST}}\bfA_{\text{ST}}\\
	\end{bmatrix},\quad
\vb*{b} = \begin{bmatrix}
		r_{\text{UT}}\bfb_{\text{UT}}\\
        r_{\text{ST}}\bfb_{\text{ST}}\\
	\end{bmatrix},
\ee
where $r_{\text{UT}}>0$ and $r_{\text{ST}}>0$ scale the contributions of the uniaxial compression/tension data and the simple torsion data, respectively. 
During the experimental investigations conducted in the context of this work (see, e.g., \cref{sec:experimental_validation}), the chosen range of applied deformation in uniaxial compression/tension and simple torsion led to measurements $\bfb_{\text{UT}}$ and $\bfb_{\text{ST}}$, respectively, that were different in magnitude.
Choosing $r_{\text{UT}} = 0.3$ and $r_{\text{ST}} = 1$ was found to result in a proper scaling of the contributions from uniaxial compression/tension and simple torsion, leading to similar maximum absolute entries in the scaled measurements $r_{\text{UT}}\bfb_{\text{UT}}$ and $r_{\text{ST}}\bfb_{\text{ST}}$.
Finally, the total mismatch between the material model prediction and the experimental measurements is quantified by computing the mean squared error $\mathrm{MSE}$ of the residuals of the linear system of equations
\be
\mathrm{MSE}(\bftheta) =  \frac{1}{n_{\text{UT}} + n_{\text{ST}}} \|\bfA\bftheta - \bfb\|_2^2.
\ee

\subsection{Feature scaling}
To obtain a system of equations with dimensionless parameters and system coefficients, we standardize the system of equations. 
Defining the mean and standard deviation of a vector $\bfx\in\Rset^n$ as 
\be
\mathrm{Mean}(\bfx) = \frac{1}{n} \sum_i x_i, \quad
\mathrm{Std}(\bfx) = \sqrt{ \frac{1}{n} \sum_i \left( x_i - \mathrm{Mean}(\bfx) \right)^2 },
\ee
the standardized system of equations reads
\be
\bar\bfA\bar\bftheta=\bar\bfb
\ee
with
\be
\label{eq:standardization}
\bar{A}_{ij} = \frac{A_{ij}}{\mathrm{Std}(\bm{A}_{j})}, \quad
\bar{b}_i = \frac{b_i}{\mathrm{Std}(\bfb)}, \quad
\bar{\theta}_{j} = \frac{ \mathrm{Std}(\bm{A}_{j}) }{ \mathrm{Std}(\bfb) } {\theta}_{j},
\ee
where we defined $\bm{A}_{j}$ as the $j^{\text{th}}$ column of $\bfA$.
The mean squared error of the residuals of the standardized linear system of equations is defined as
\be
\overline{\mathrm{MSE}}(\bar\bftheta) =  \frac{1}{n_{\text{UT}} + n_{\text{ST}}} \|\bar\bfA\bar\bftheta - \bar\bfb\|_2^2.
\ee

\subsection{Optimization problem}

Our objectives are to calibrate the unknown material parameters such that the material model is in agreement with the experimental data, and to discard those terms in the model ansatz which have a minor influence on the fitting accuracy by setting their corresponding parameters to zero.  
To this end, we solve an $L_1$-regularized optimization problem (also denoted as sparse regression problem) of the form
\be
\label{eq:lasso_problem}
\bar\bftheta^{\text{opt}} = \argminWithArgs_{\bar\bftheta \geq \bm{0}}  \left( \frac{1}{2}\overline{\mathrm{MSE}}(\bar\bftheta) + \lambda_p \| \bar\bftheta \|_1 \right),
\ee
where $\overline{\mathrm{MSE}}(\bar\bftheta)$ is the previously introduced mean squared error function that quantifies the mismatch between the material model predictions and the experimental data, and $\| \bar\bftheta \|_1 = \sum_{i}|\bar\theta_i|$ is an $L_1$-regularization term, also denoted as \textit{Lasso} (least absolute shrinkage and selection operator).
This term, first used for model selection by \cite{frank_statistical_1993} and \cite{tibshirani_regression_1996} and later applied to problems in dynamics by \cite{brunton_discovering_2016}, can be interpreted as a convex approximation of an operator that counts the number of nonzero entries in the material parameter vector.
Thus, adding the $L_1$-regularization term to the minimization problem promotes sparsity of the material parameter vector.
In previous works \citep{flaschel_unsupervised_2021,flaschel_discovering_2022}, the more general $L_p$-regularization term $\| \bar\bftheta \|_p^p = \sum_{i}|\bar\theta_i|^p$ with $0 < p \leq 1$ was adopted to promote sparsity in the solution vector. 
The $L_p$ term has the advantage that it converges to the operator that counts the number of nonzero entries in the material parameter vector as $p$ approaches zero.
Thus, it can be considered as a better measure of sparsity than the $L_1$-regularization term.
However, the $L_p$ term is non-convex for $p<1$; this implies that the optimization problem becomes non-convex and thus multiple solutions are to be expected, corresponding to local minima of the objective function. In our previous papers \citep{flaschel_unsupervised_2021,flaschel_discovering_2022}, this issue was addressed by solving the optimization problem multiple times and selecting the minimum of the computed local minima, which obviously increases the computational complexity and cost of the minimization problem. In this work, we rely on the $L_1$ regularization term and thus solve a convex minimization problem which thus admits a unique solution.
The influence of the regularization is governed by the choice of the weighting factor $\lambda_p \geq 0$.
As proposed in \cite{flaschel_discovering_2022,flaschel_automated_2023-1,marino_automated_2023}, an appropriate choice of $\lambda_p$ for striking a balance between fitting accuracy and mathematical complexity of the material model is obtained through a Pareto analysis (see \cref{sec:Pareto}).

\subsection{Solver}
To solve the problem in \cref{eq:lasso_problem}, we leverage a coordinate descent algorithm as proposed by \cite{friedman_regularization_2010}, which is implemented in the \textit{Lasso} subroutine of the open-source Python library \textit{sklearn}.
The subroutine takes $\bar\bfA$ and $\bar\bfb$ as well as the value of $\lambda_p$ as input and returns the sparse solution vector $\bar\bftheta$ whose entries are constrained to be greater than or equal to zero.
The intercept in the \textit{Lasso} subroutine is turned off and the maximum number of iterations is set to $10^4$.

\begin{figure}[H]
	\centering
    \includegraphics[width=0.5\textwidth]{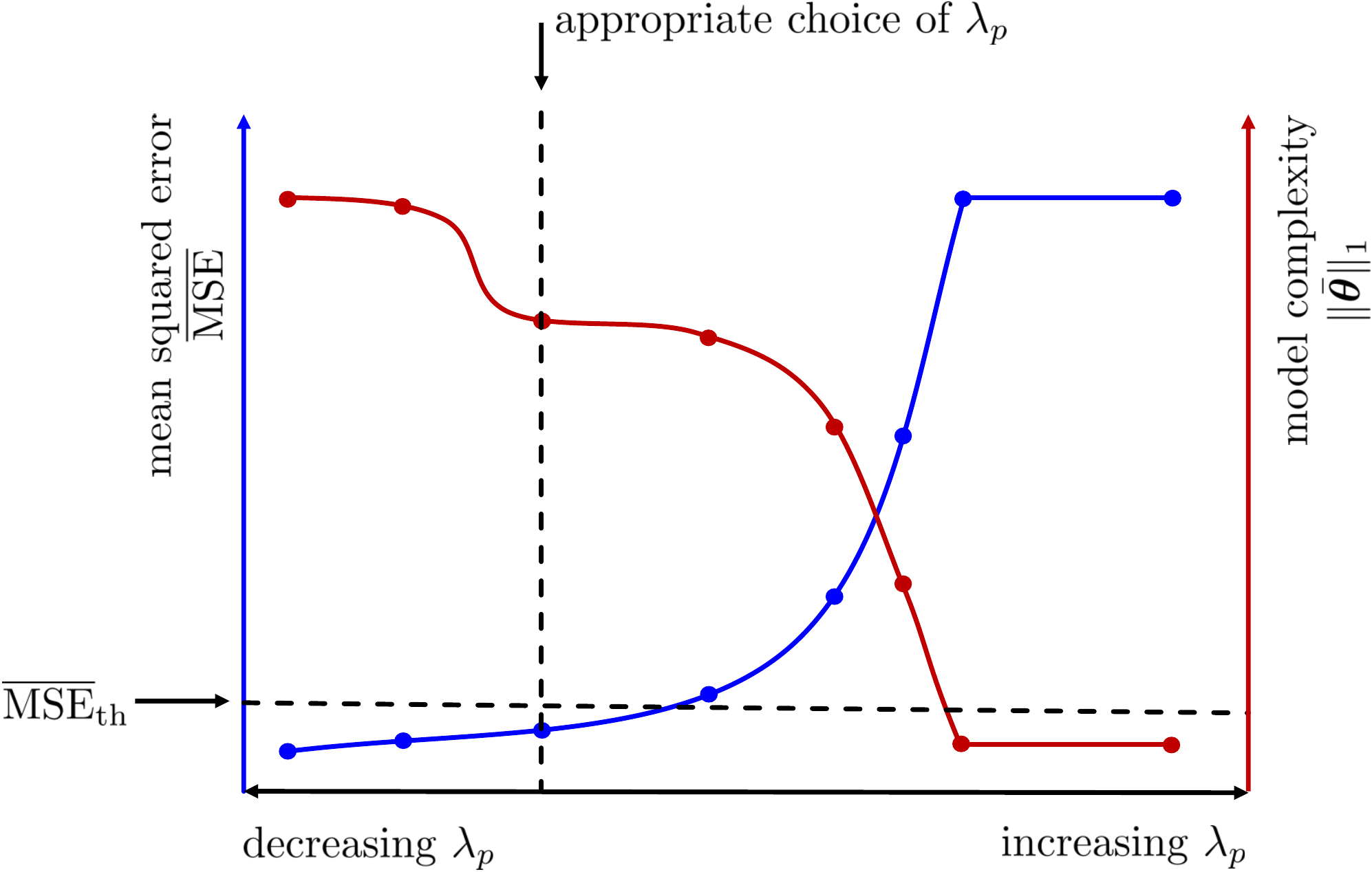}
    \caption{Qualitative illustration of the Pareto analysis for selecting an appropriate value of $\lambda_p$.}
	\label{fig:Pareto}
\end{figure}

\subsection{Pareto analysis}
\label{sec:Pareto}
The solution of the problem in \cref{eq:lasso_problem} is highly dependent on the choice of $\lambda_p$.
As qualitatively illustrated in \cref{fig:Pareto}, a small value of $\lambda_p$ yields a dense solution vector and therefore a complex expression for the material model that is in general associated with a high fitting accuracy, i.e., a small mean squared error. 
Increasing the value of $\lambda_p$ results in a sparser solution vector and thus a simpler expression for the material model at the cost of a reduced fitting accuracy with a larger mean squared error. In the limit of very large values of $\lambda_p$, the optimization problem returns a solution vector with only zero entries and the mean squared error saturates.

To strike a balance between the accuracy and the complexity of the material model, a Pareto analysis is applied to the problem.
To this end, the problem in \cref{eq:lasso_problem} is solved multiple times using a set of different $\lambda_p$ values, leading to a solution vector $\bar\bftheta$ for every choice of $\lambda_p$.
Here, we choose $41$ values of $\lambda_p$ that are evenly distributed on a logarithmic scale between $10^{-2}$ and $10^{2}$.
Our objective now is to select from all solutions the one that exhibits an optimal combination of high fitting accuracy and high sparsity. 
To this end, the mean squared error and the $L_1$-norm of the solution vector are computed for each solution.
The minimum and maximum mean squared errors among all solutions are denoted as $\overline{\mathrm{MSE}}_{\mathrm{min}}$ and $\overline{\mathrm{MSE}}_{\mathrm{max}}$, respectively.
Further, we define a threshold mean squared error (see \cref{fig:Pareto}) as
\be
\overline{\mathrm{MSE}}_{\mathrm{th}} = \overline{\mathrm{MSE}}_{\mathrm{min}} + r_{\lambda} (\overline{\mathrm{MSE}}_{\mathrm{max}} - \overline{\mathrm{MSE}}_{\mathrm{min}}),
\ee
where $r_{\lambda}$ is a small positive scalar, $0 < r_{\lambda} \ll 1$.
All solutions for which the mean squared errors are greater than this threshold are expected to exhibit a low fitting accuracy and are thus discarded.
From the remaining solutions, i.e., those for which the mean squared error is below $\overline{\mathrm{MSE}}_{\mathrm{th}}$, we select the sparsest one, i.e., the solution with the lowest value of the $L_1$-norm term (see \cref{fig:Pareto}).
In this way, a sparse solution with a low mean squared error is obtained, whereby the amount of sparsity is dictated by the user-defined hyperparameter $r_{\lambda}$. In this work we set $r_{\lambda}=0.02$ and keep it constant throughout the analyses.

\subsection{Thresholding}
The solution vector $\bar\bftheta$ selected in \cref{sec:Pareto} may contain values that are close to zero.
As these material parameters and their corresponding features have a vanishing effect on the model response, they are set to zero.
Specifically, we define a threshold value $\bar\theta_{\text{th}}$ (here chosen as $\bar\theta_{\text{th}}=0.01$) and we set $\bar{\theta}_i=0$ if $\bar{\theta}_i<\bar{\theta}_{\text{th}}$ for all $i$.

\subsection{Feature clustering}
\label{sec:feature_clustering}
\begin{figure}[H]
	\centering
    \includegraphics[width=0.7\textwidth]{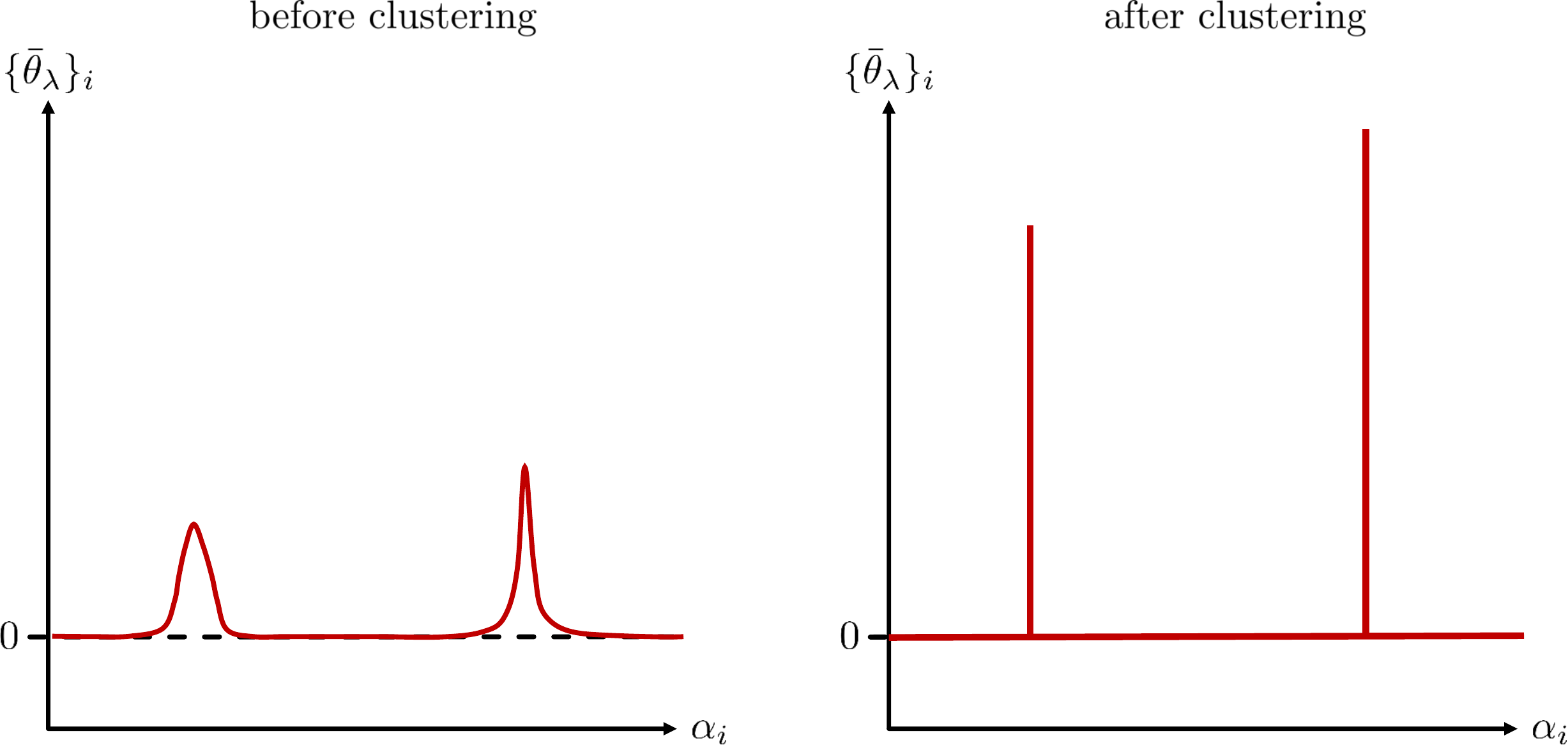}
    \caption{Qualitative illustration of the feature clustering.}
	\label{fig:clustering}
\end{figure}

Due to the fine discretization of the chosen range of $\alpha_i$ values (see \cref{sec:model_library}), the feature library contains Ogden features with very similar values of $\alpha_i$, which in turn results in a high correlation between these features.
While the solution vector obtained by solving the optimization problem with \textit{Lasso} regularization already exhibits a high degree of sparsity, the sparsity and thus interpretability of the material model can be further improved by grouping together highly correlated features in the solution.
To this end, we leverage a clustering algorithm that automatically groups together similar $\alpha_i$ values, as qualitatively illustrated in \cref{fig:clustering}.
For a similar clustering strategy that was used to group together similar Maxwell elements in viscoelastic constitutive laws discovered with EUCLID, the interested reader is referred to \cite{marino_automated_2023}.

In this paper, we adopt the clustering method \textit{DBSCAN} (Density-Based Spatial Clustering of Applications with Noise) \citep{ester_density-based_1996} as implemented in the Python library \textit{sklearn}.
\textit{DBSCAN} assigns each $\alpha_i$ whose corresponding $\{\bar\theta_\lambda\}_i$ is non-zero to a cluster.
This is done in such a way that for each $\alpha_i$ in a cluster there exists at least one value $\alpha_j$ (with $i \neq j$) in the same cluster such that their distance $|\alpha_i-\alpha_j|$ is smaller than or equal to a predefined value, here chosen as 0.01. The advantage over the k-means clustering algorithm used in \cite{marino_automated_2023} is that the needed number of clusters is obtained automatically. 
After grouping the $\alpha_i$ values into a set of clusters, an averaged value $\bar\alpha$ is computed for each cluster as
\be
\bar\alpha = \frac{\sum_{k}\alpha_{k}\{\bar\theta_\lambda\}_k}{\sum_{k}\{\bar\theta_\lambda\}_k},
\ee
where $k$ ranges over all elements in the considered cluster. 
Afterwards, $\bar\alpha$ is rounded to the closest original value of $\alpha_i$ contained in the material model library (see \cref{sec:model_library}).
Thus, we identify for each cluster one value of $\alpha_i$ for which the corresponding coefficient in $\bar\bftheta_\lambda$ is non-zero, while all other entries in $\bar\bftheta_\lambda$ are set to zero.

\subsection{Final regression without regularization}
\label{sec:regression_wo_regularization}
By solving the sparse regression problem and applying the clustering method, a small number of relevant features are automatically selected from the originally large library of candidate features.
In a final step, all features corresponding to parameters that have been identified as zero are disregarded from the library by removing the corresponding columns from the matrix $\bar\bfA$, thus obtaining a reduced system matrix $\bar\bfA_{\text{red}}$.
Afterwards, the reduced regression problem is solved for the remaining non-zero parameters $\bar\bftheta_{\text{red}}$, but this time without introducing any regularization term (i.e., we solve \cref{eq:lasso_problem} with $\lambda_p=0$).

The motivation behind this last step is twofold.
First, while the regularization term in \cref{eq:lasso_problem} forces unneeded parameters to zero, it simultaneously shrinks the absolute value of the relevant parameters.
Solving the regression problem without regularization for the small number of selected features leads to material parameters that are not affected by such shrinkage.
And second, after applying the clustering method, most entries in $\bar\bftheta_\lambda$ have been identified as zero, while the remaining non-zero parameters in $\bar\bftheta_\lambda$ are still unknown.
The final regression problem without regularization is used to calibrate those non-zero parameters.
Finally, the original parameters $\bftheta$ can be recovered from the scaled parameters $\bar\bftheta$ using \cref{eq:standardization}.

\section{Numerical verification}
\label{sec:numerical_verification}
\subsection{Synthetical data}
Before applying the proposed method for material model discovery to experimental data, we verify the method on synthetically generated data without and with artificially added noise.
Such data are obtained by assuming different hyperelastic material models and generating labeled data pairs under the assumption of uniaxial compression/tension and simple torsion.
Specifically, we consider the four material models whose strain energy density functions are provided in \cref{tab:ArtificialDiscovery}:
\begin{itemize}
\item first-order Mooney-Rivlin model (MR1)
\item one-term Ogden model (O1)
\item two-term Ogden model (O2)
\item combination of second-order Mooney-Rivlin model and one-term Ogden model (MR2O1)
\end{itemize}

The chosen material models are used to generate $n_{\text{UT}}=60$ pairs of data for uniaxial compression/tension, i.e., $(\lambda_{\text{UT}}^{(l)},P_{11}^{(l)})$, where the values of $\lambda_{\text{UT}}^{(l)}$ are equally spaced between $\lambda_{\text{UT}}^{(1)}=0.7$ and $\lambda_{\text{UT}}^{(n_{\text{UT}})}=1.3$, and $n_{\text{ST}}=60$ pairs of data for simple torsion, i.e., $(\tilde\psi^{(l)},\tau^{(l)})$, where the values of $\tilde\psi^{(l)}$ are equally spaced between $\tilde\psi^{(1)}=-1$ and $\tilde\psi^{(n_{\text{ST}})}=1$.

To emulate real experimental data, the generated data are perturbed by artificial noise by considering noisy data pairs $(\lambda_{\text{UT}}^{(l)},P_{11}^{(l)} + \eps^{(l)}_{\text{UT}})$ and $(\tilde\psi^{(l)},\tau^{(l)} + \eps^{(l)}_{\text{ST}})$, where $\eps^{(l)}_{\text{UT}} \sim \mathcal{N} \left(0, \sigma^2 \right)$ and $\eps^{(l)}_{\text{ST}} \sim \mathcal{N} \left(0, \sigma^2 \right)$ is random noise that is drawn independently from a Gaussian distribution with zero mean and standard deviation $\sigma$.
Here, we consider (beside the noiseless case) two different levels of noise, i.e. $\sigma \in \{ 0 \ \text{Pa},5 \ \text{Pa},10 \ \text{Pa} \}$.

\subsection{Results and discussion}
The synthetically generated data serve as input for the proposed algorithm for material model discovery.
Testing the method on synthetical data has the advantage that the discovered material models can be compared to the "ground truth" (in machine learning jargon), i.e., to the material models that were assumed during data generation.
\cref{tab:ArtificialDiscovery} shows the expressions of the discovered strain energy density functions in comparison with the true ones.
It is observed that the correct mathematical form of the strain energy density function is recovered for each test case.
\cref{tab:ArtificialDiscovery} further provides the mean squared errors that quantify the mismatch between the data and the predictions of the discovered models.
As expected, the $\mathrm{MSE}$ increases for increasing level of noise.

\begin{table}[H]
	\caption{Strain energy density functions of the true and discovered material models and MSE of predictions.}
	\label{tab:ArtificialDiscovery}
	\centering
	\resizebox{15cm}{!}{
        \begin{tabular}{|ll|rcl|r|l|}
        \hline
        \multicolumn{2}{|l|}{\textbf{Benchmarks\vphantom{$\frac{\int}{\int}$}{}}} & \multicolumn{3}{c|}{\textbf{Strain energy density} $\tilde{W}$ $[\text{Pa}]$}                                                                                                                                                                                                         & \multicolumn{1}{c|}{MSE $[\text{Pa}^2]$} & \multicolumn{1}{c|}{$\overline{\mathrm{MSE}}$ $[-]$} \\ \hline
        \multicolumn{1}{|l|}{MR1}   & Truth\vphantom{$\frac{\int}{\int}$}{}       & $40.00\left(I_1 - 3\right)$                                                                                                & $+$ & $20.00\left(I_2 - 3\right)$                                                                                                          & \multicolumn{1}{c|}{-}            & \multicolumn{1}{c|}{-}                \\ \cline{2-7} 
        \multicolumn{1}{|l|}{}      & $\sigma=0$\vphantom{$\frac{\int}{\int}$}{}  & $40.00\left(I_1 - 3\right)$                                                                                                & $+$ & $20.00\left(I_2 - 3\right)$                                                                                                          & 0.0000                            & 0.0000                                \\ \cline{2-7} 
        \multicolumn{1}{|l|}{}      & $\sigma=5 \ \text{Pa}$\vphantom{$\frac{\int}{\int}$}{}  & $41.25\left(I_1 - 3\right)$                                                                                                & $+$ & $17.71\left(I_2 - 3\right)$                                                                                                          & 13.4120                           & 0.0022                                \\ \cline{2-7} 
        \multicolumn{1}{|l|}{}      & $\sigma=10 \ \text{Pa}$\vphantom{$\frac{\int}{\int}$}{} & $42.51\left(I_1 - 3\right)$                                                                                                & $+$ & $15.42\left(I_2 - 3\right)$                                                                                                          & 53.6481                           & 0.0090                                \\ \hline
        \multicolumn{1}{|l|}{O1}    & Truth\vphantom{$\frac{\int}{\int}$}{}       & $2.00\left(\lambda_1^{-10.00} +   \lambda_2^{-10.00} + \lambda_3^{-10.00} - 3\right)$                                      &     &                                                                                                                                        & \multicolumn{1}{c|}{-}            & \multicolumn{1}{c|}{-}                \\ \cline{2-7} 
        \multicolumn{1}{|l|}{}      & $\sigma=0$\vphantom{$\frac{\int}{\int}$}{}  & $1.97\left(\lambda_1^{-10.03} + \lambda_2^{-10.03} +   \lambda_3^{-10.03} - 3\right)$                                      &     &                                                                                                                                        & 0.1815                            & 0.0000                                \\ \cline{2-7} 
        \multicolumn{1}{|l|}{}      & $\sigma=5 \ \text{Pa}$\vphantom{$\frac{\int}{\int}$}{}  & $1.94\left(\lambda_1^{-10.05} + \lambda_2^{-10.05} +   \lambda_3^{-10.05} - 3\right)$                                      &     &                                                                                                                                        & 13.5305                           & 0.0001                                \\ \cline{2-7} 
        \multicolumn{1}{|l|}{}      & $\sigma=10 \ \text{Pa}$\vphantom{$\frac{\int}{\int}$}{} & $1.91\left(\lambda_1^{-10.08} + \lambda_2^{-10.08} +   \lambda_3^{-10.08} - 3\right)$                                      &     &                                                                                                                                        & 53.8606                           & 0.0006                                \\ \hline
        \multicolumn{1}{|l|}{O2}    & Truth\vphantom{$\frac{\int}{\int}$}{}       & $16.00\left(\lambda_1^{-5.00\phantom{0}{}}   + \lambda_2^{-5.00\phantom{0}{}} + \lambda_3^{-5.00\phantom{0}{}} - 3\right)$ & $+$ & $\phantom{0}{}8.00\left(\lambda_1^{5.00\phantom{-0}{}}   + \lambda_2^{5.00\phantom{-0}{}} + \lambda_3^{5.00\phantom{-0}{}} - 3\right)$ & \multicolumn{1}{c|}{-}            & \multicolumn{1}{c|}{-}                \\ \cline{2-7} 
        \multicolumn{1}{|l|}{}      & $\sigma=0$\vphantom{$\frac{\int}{\int}$}{}  & $13.50\left(\lambda_1^{-5.25\phantom{0}{}} +   \lambda_2^{-5.25\phantom{0}{}} + \lambda_3^{-5.25\phantom{0}{}} - 3\right)$ & $+$ & $11.79\left(\lambda_1^{4.43\phantom{-0}{}} +   \lambda_2^{4.43\phantom{-0}{}} + \lambda_3^{4.43\phantom{-0}{}} - 3\right)$             & 0.2936                            & 0.0000                                \\ \cline{2-7} 
        \multicolumn{1}{|l|}{}      & $\sigma=5 \ \text{Pa}$\vphantom{$\frac{\int}{\int}$}{}  & $13.29\left(\lambda_1^{-5.26\phantom{0}{}} +   \lambda_2^{-5.26\phantom{0}{}} + \lambda_3^{-5.26\phantom{0}{}} - 3\right)$ & $+$ & $11.79\left(\lambda_1^{4.44\phantom{-0}{}} +   \lambda_2^{4.44\phantom{-0}{}} + \lambda_3^{4.44\phantom{-0}{}} - 3\right)$             & 13.7349                           & 0.0002                                \\ \cline{2-7} 
        \multicolumn{1}{|l|}{}      & $\sigma=10 \ \text{Pa}$\vphantom{$\frac{\int}{\int}$}{} & $13.16\left(\lambda_1^{-5.26\phantom{0}{}} +   \lambda_2^{-5.26\phantom{0}{}} + \lambda_3^{-5.26\phantom{0}{}} - 3\right)$ & $+$ & $11.75\left(\lambda_1^{4.45\phantom{-0}{}} +   \lambda_2^{4.45\phantom{-0}{}} + \lambda_3^{4.45\phantom{-0}{}} - 3\right)$             & 54.0784                           & 0.0007                                \\ \hline
        \multicolumn{1}{|l|}{MR2O1} & Truth\vphantom{$\frac{\int}{\int}$}{}       & $30.00\left(I_2 - 3\right)^2$                                                                                            & $+$ & $\phantom{0}{}2.00\left(\lambda_1^{-10.00} +   \lambda_2^{-10.00} + \lambda_3^{-10.00} - 3\right)$                                     & \multicolumn{1}{c|}{-}            & \multicolumn{1}{c|}{-}                \\ \cline{2-7} 
        \multicolumn{1}{|l|}{}      & $\sigma=0$\vphantom{$\frac{\int}{\int}$}{}  & $29.25\left(I_2 - 3\right)^2$                                                                                            & $+$ & $\phantom{0}{}1.97\left(\lambda_1^{-10.04} +   \lambda_2^{-10.04} + \lambda_3^{-10.04} - 3\right)$                                     & 0.3179                            & 0.0000                                \\ \cline{2-7} 
        \multicolumn{1}{|l|}{}      & $\sigma=5 \ \text{Pa}$\vphantom{$\frac{\int}{\int}$}{}  & $31.69\left(I_2 - 3\right)^2$                                                                                            & $+$ & $\phantom{0}{}1.91\left(\lambda_1^{-10.07} +   \lambda_2^{-10.07} + \lambda_3^{-10.07} - 3\right)$                                     & 13.7564                           & 0.0002                                \\ \cline{2-7} 
        \multicolumn{1}{|l|}{}      & $\sigma=10 \ \text{Pa}$\vphantom{$\frac{\int}{\int}$}{} & $34.16\left(I_2 - 3\right)^2$                                                                                            & $+$ & $\phantom{0}{}1.86\left(\lambda_1^{-10.10} +   \lambda_2^{-10.10} + \lambda_3^{-10.10} - 3\right)$                                     & 53.9747                           & 0.0007                                \\ \hline
        \end{tabular}
    }
\end{table}

For the benchmark case MR1, the material coefficients in the strain energy density function are exactly recovered for the noiseless case.
The deviation between the true and identified parameters increases for increasing noise, as expected.
For the benchmark cases O1 and MR2O1, the true and discovered parameters are in excellent agreement.
However, the material parameters are not exactly recovered even in the noiseless case.
The reason for this is that computing the averaged values $\bar\alpha$ after the feature clustering (\cref{sec:feature_clustering}) cannot exactly recover the ground truth exponents in the Ogden terms, which leads to small deviations in the identified parameters.
For the benchmark case O2, the true and identified parameters exhibit slightly larger deviations.
One explanation for these deviations could be that the model library comprises many material models and many combinations of material parameters that are equally suited to describe the material response.
The proposed algorithm does not guarantee to find exactly the same model and parameters that were assumed during data generation, but may instead provide an adequate surrogate model that is equally suited to predict the data.
We emphasise that, although the true and identified material parameters show some deviations, the $\mathrm{MSE}$ for the benchmark case O2 is similar to the $\mathrm{MSE}$ for the other benchmarks.
Thus, the deviation in the model parameters does not seem to seriously affect the fitting accuracy of the model.

\begin{figure}[H]
	\centering
    \begin{subfigure}[b]{0.35\textwidth}
        \centering
        \includegraphics[width=\textwidth]{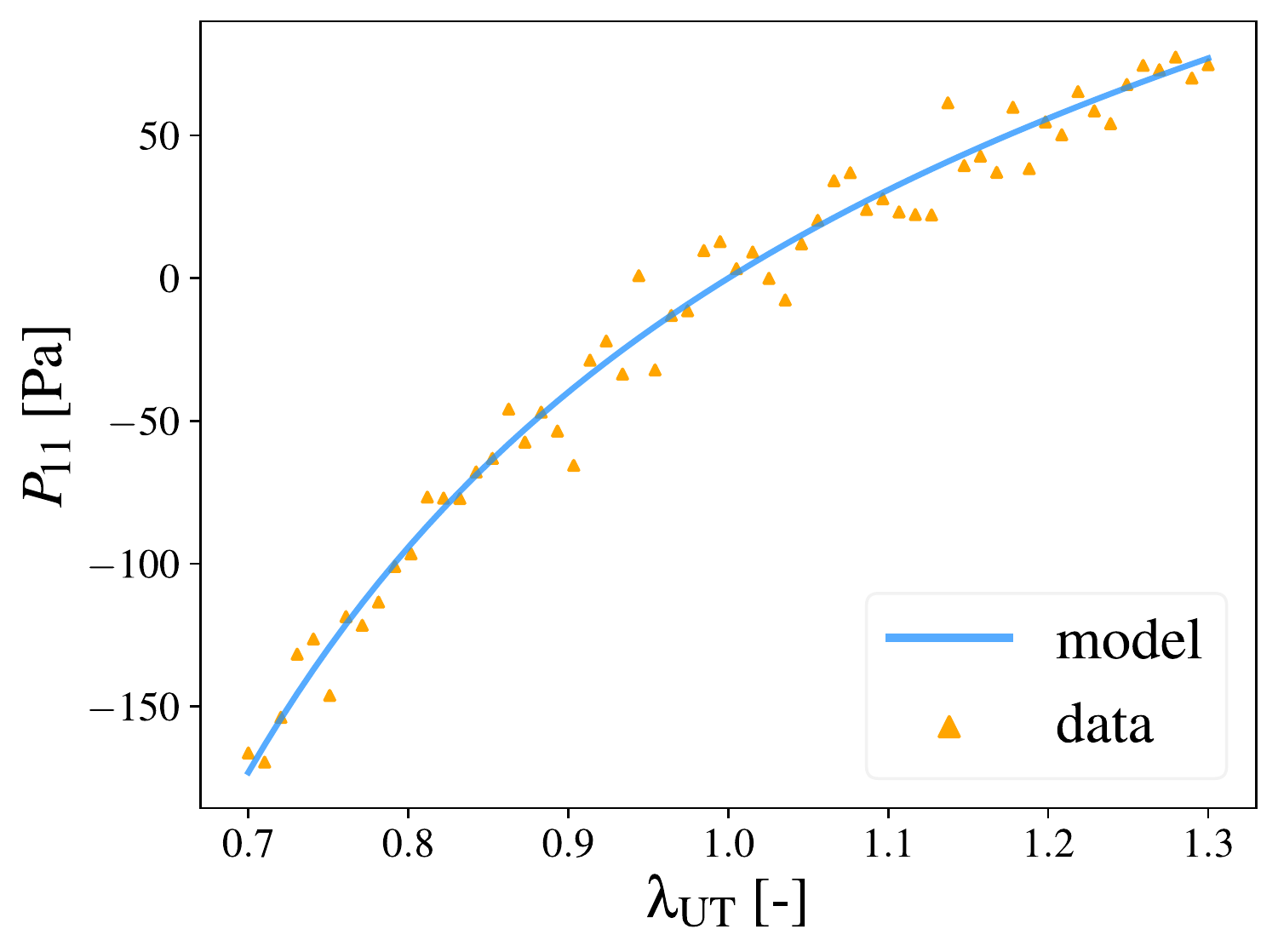}
        \caption{MR1 under UT}
    \end{subfigure}
    \begin{subfigure}[b]{0.35\textwidth}
        \centering
        \includegraphics[width=\textwidth]{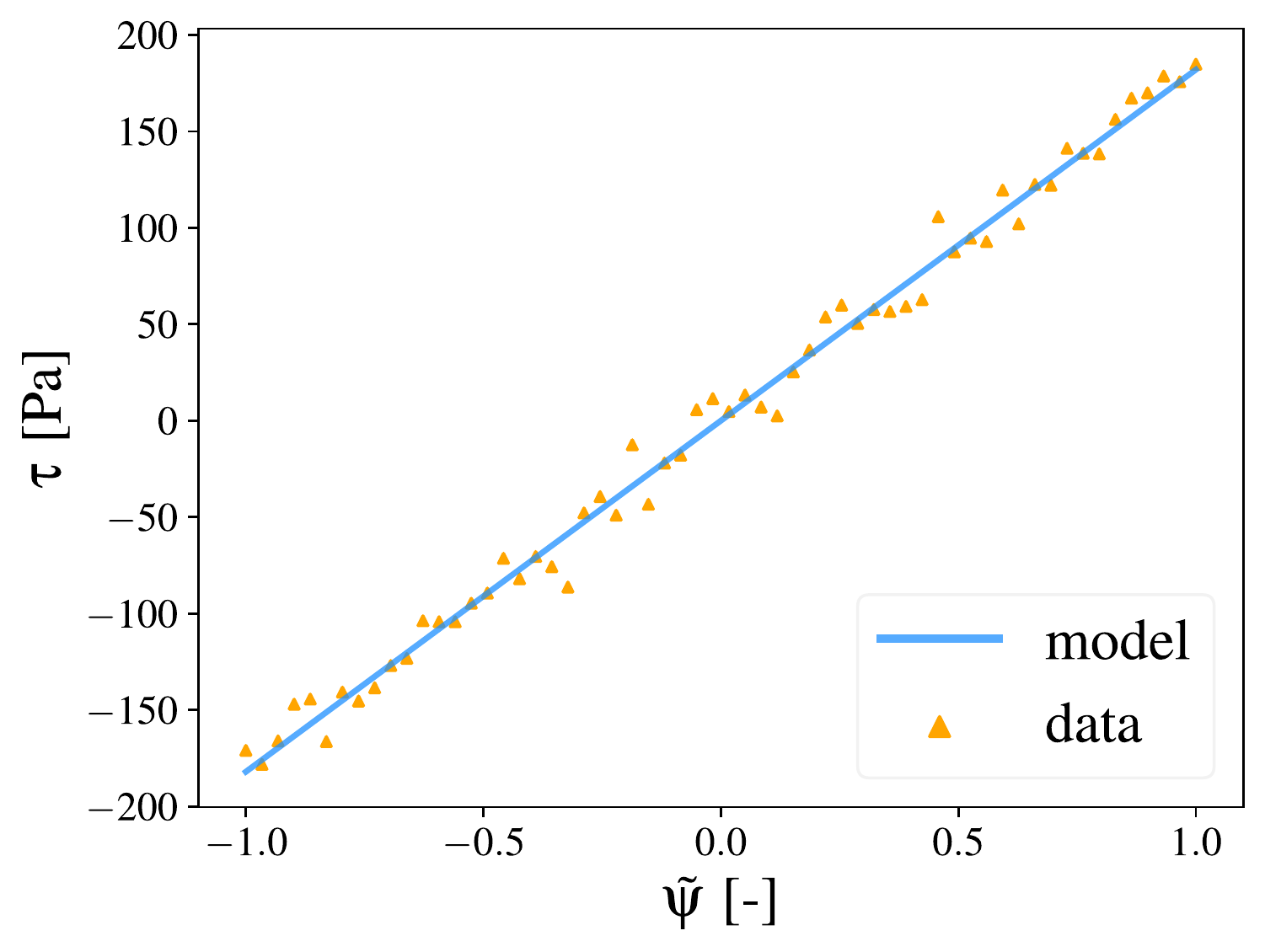}
        \caption{MR1 under ST}
    \end{subfigure}
    \begin{subfigure}[b]{0.35\textwidth}
        \centering
        \includegraphics[width=\textwidth]{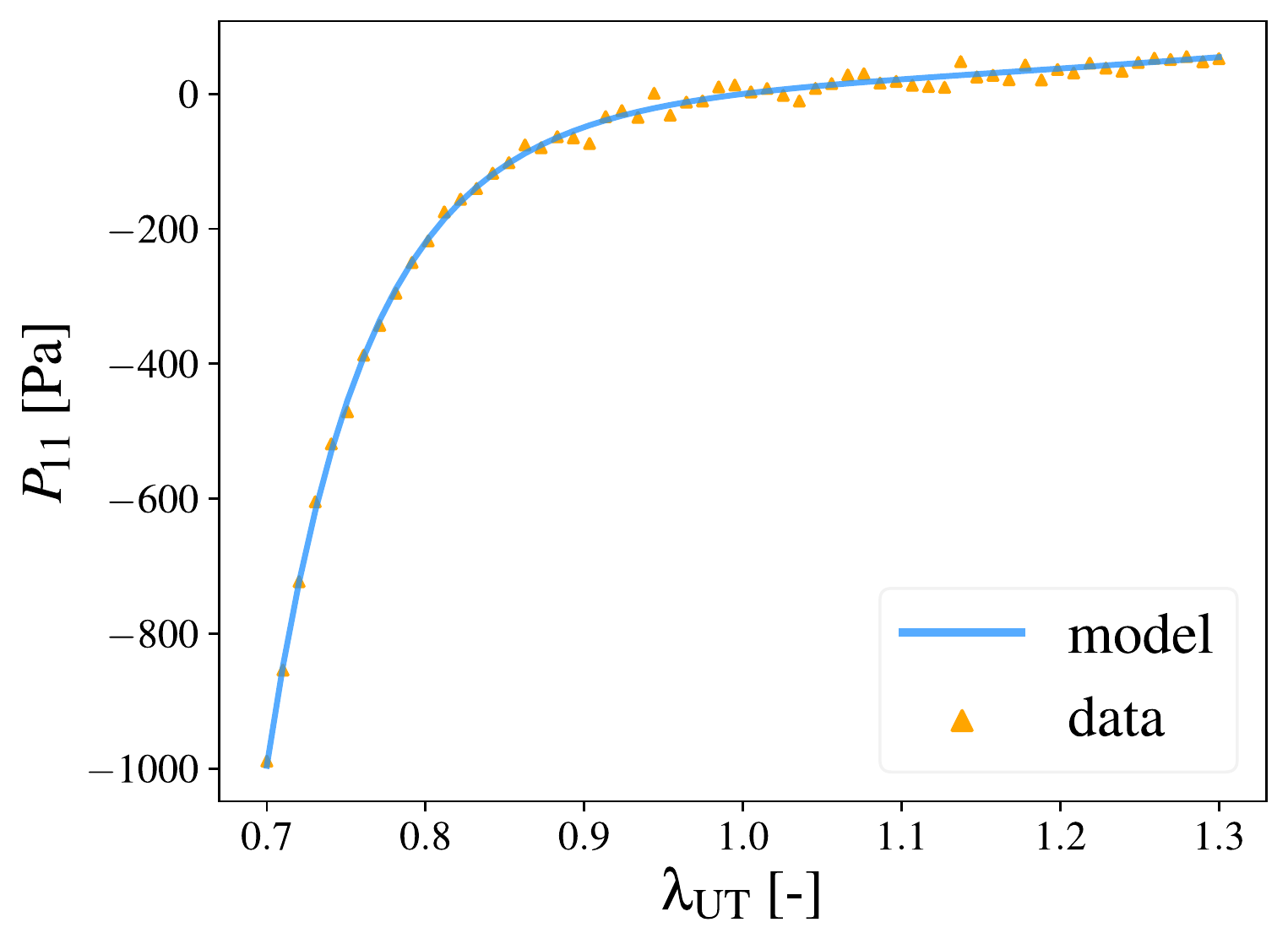}
        \caption{O1 under UT}
    \end{subfigure}
    \begin{subfigure}[b]{0.35\textwidth}
        \centering
        \includegraphics[width=\textwidth]{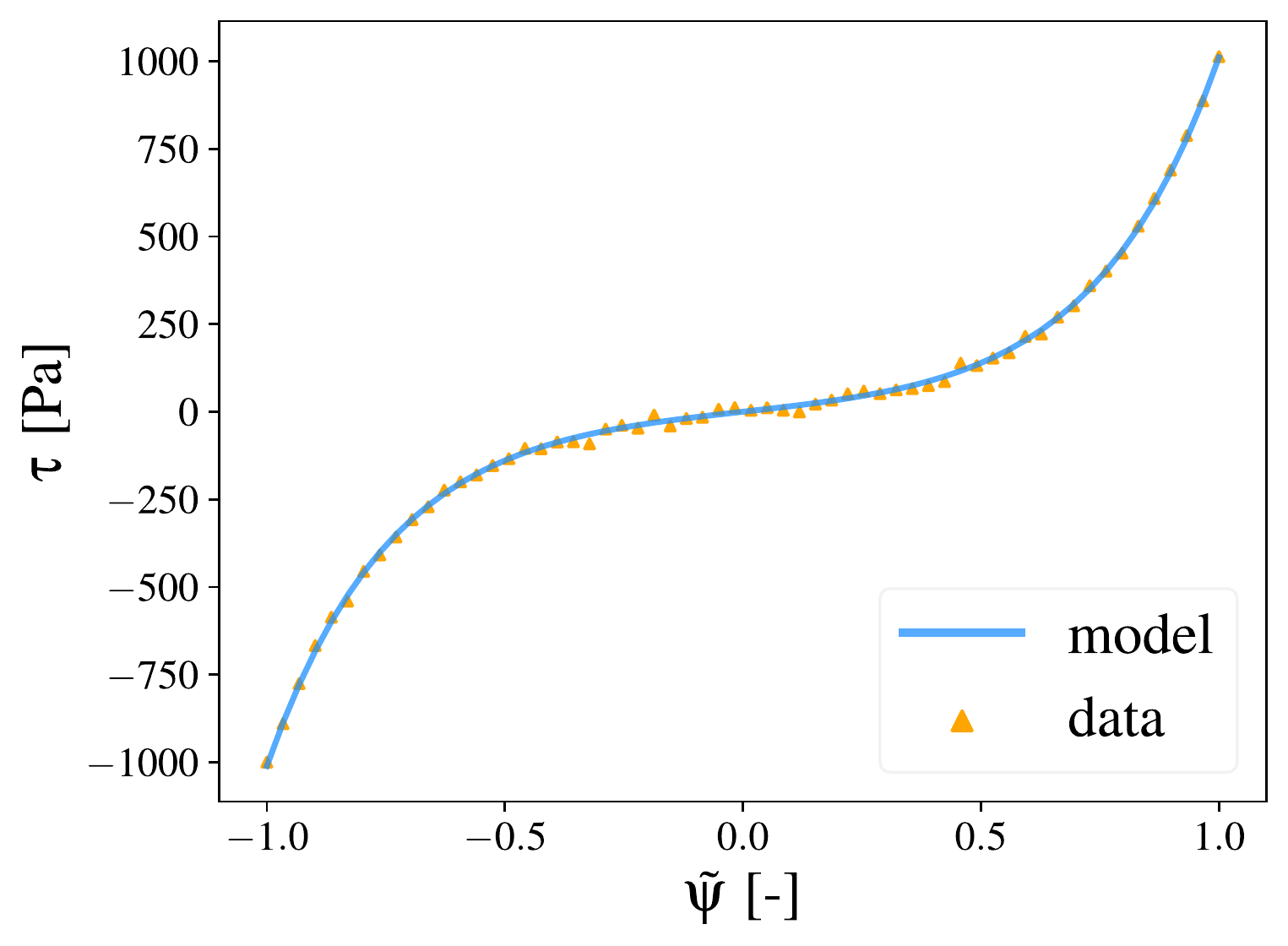}
        \caption{O1 under ST}
    \end{subfigure}
    \begin{subfigure}[b]{0.35\textwidth}
        \centering
        \includegraphics[width=\textwidth]{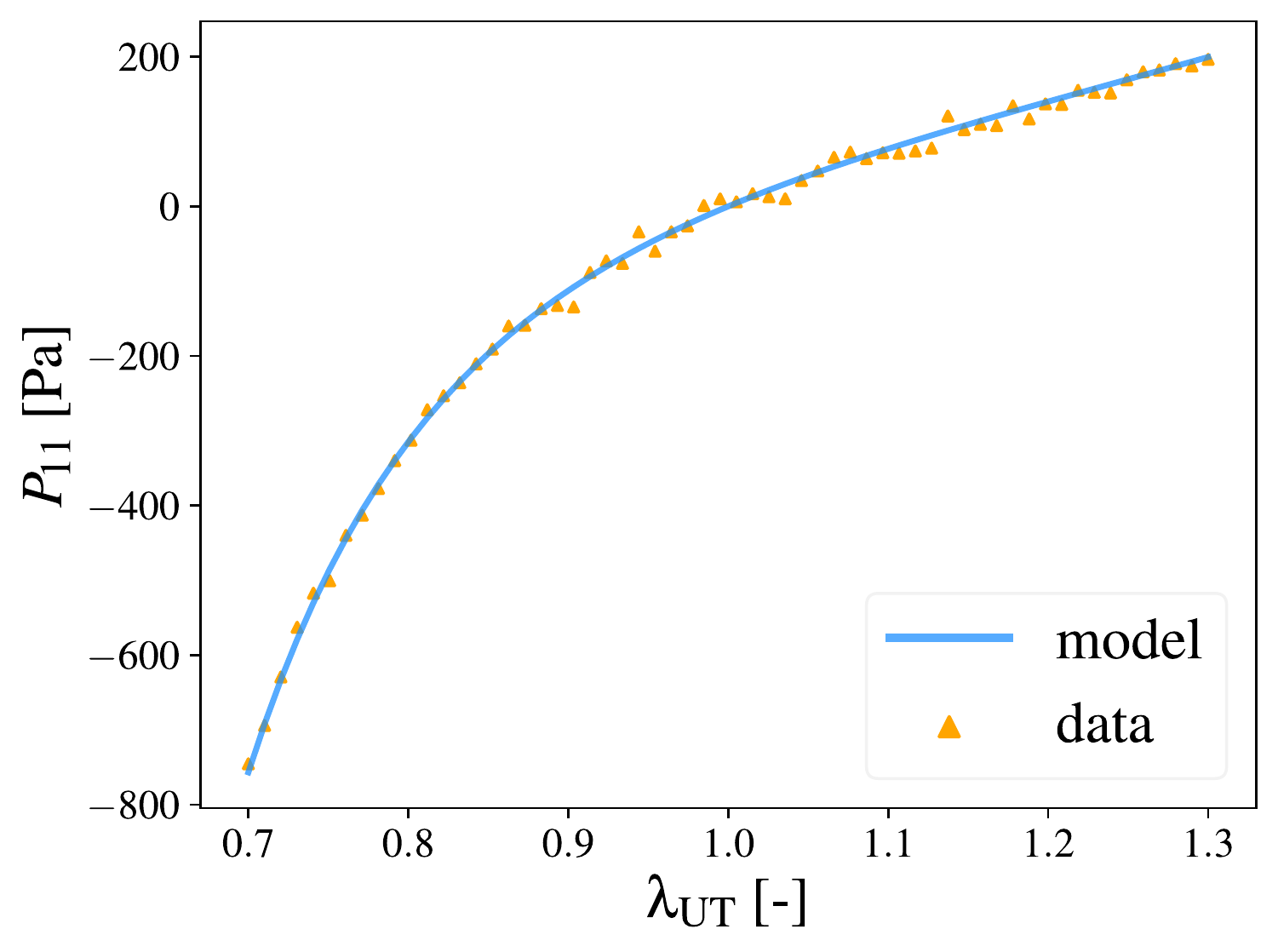}
        \caption{O2 under UT}
    \end{subfigure}
    \begin{subfigure}[b]{0.35\textwidth}
        \centering
        \includegraphics[width=\textwidth]{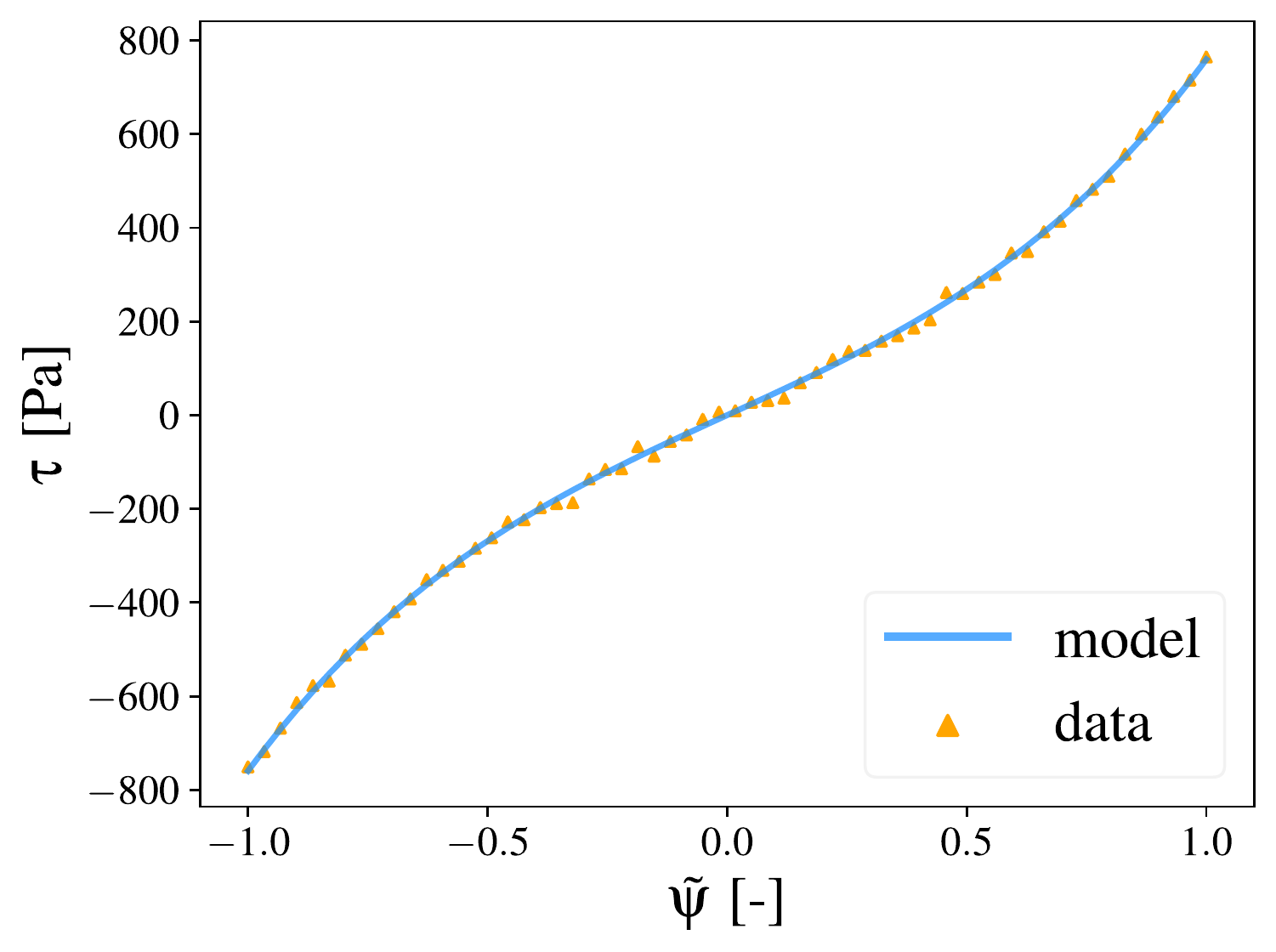}
        \caption{O2 under ST}
    \end{subfigure}
    \begin{subfigure}[b]{0.35\textwidth}
        \centering
        \includegraphics[width=\textwidth]{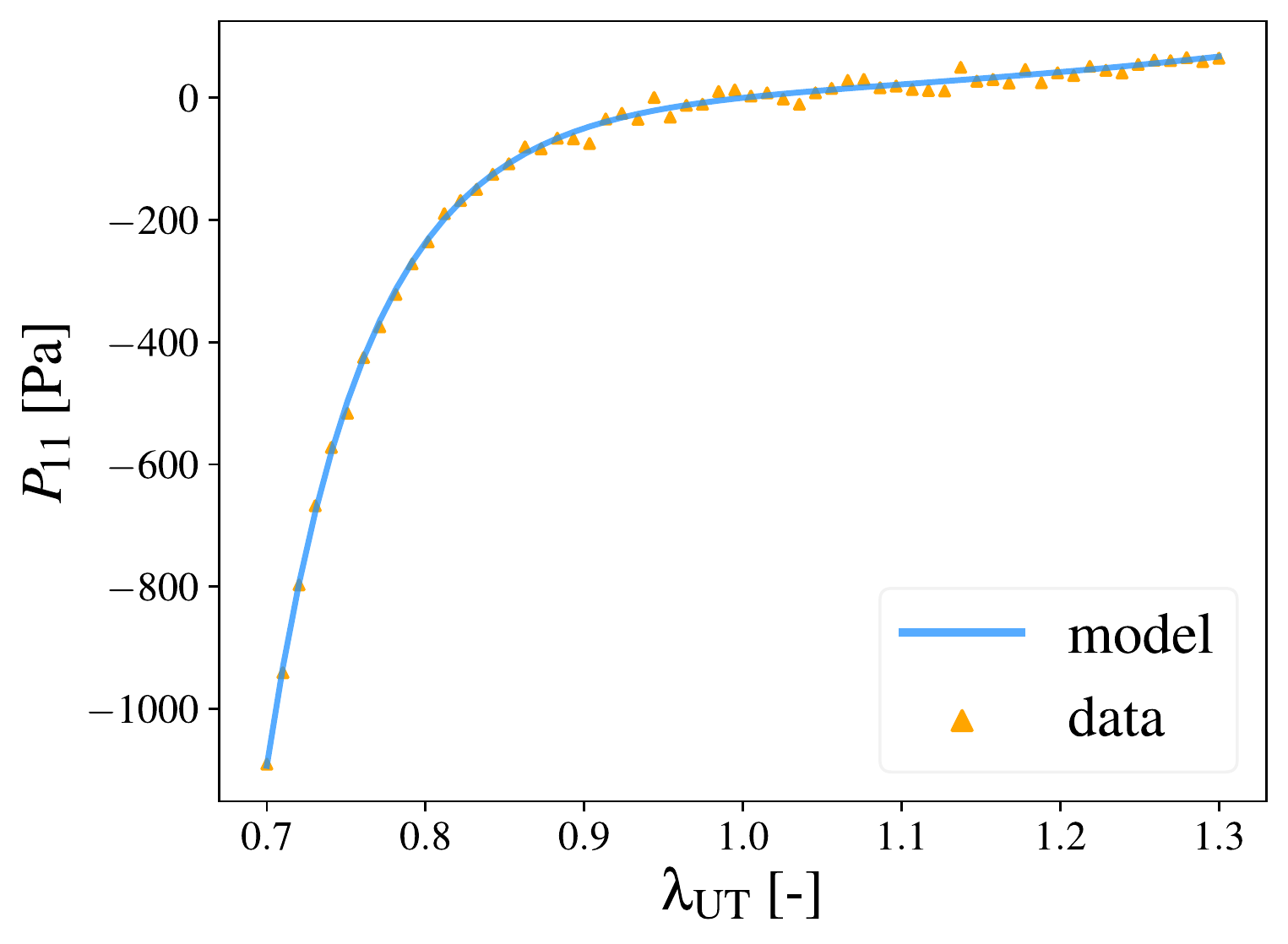}
        \caption{MR2O1 under UT}
    \end{subfigure}
 	\begin{subfigure}[b]{0.35\textwidth}
        \centering
        \includegraphics[width=\textwidth]{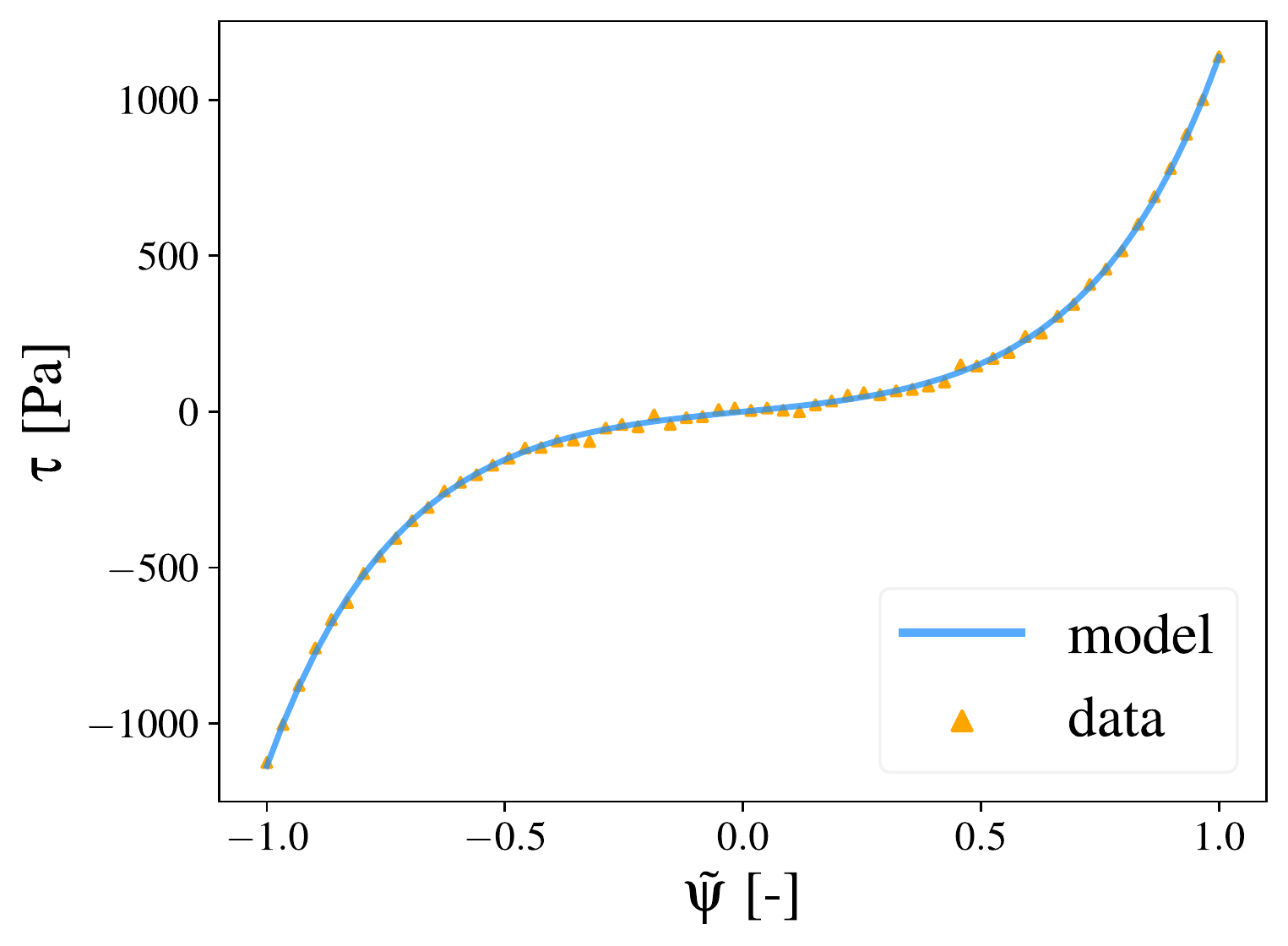}
        \caption{MR2O1 under ST}
    \end{subfigure}
    \caption{Stretch versus stress response under uniaxial compression/tension and (normalized) twist versus torque response under simple torsion of the discovered material models in comparison with the synthetical data (noise level $\sigma = 10 \ \text{Pa}$).}
	\label{fig:PlotsArtificial}
\end{figure}

\cref{fig:PlotsArtificial} illustrates the material response of the discovered material models in direct comparison with the synthetically generated data for the case with the highest level of noise, revealing an excellent agreement.

\begin{figure}[H]
	\centering
    \begin{subfigure}[b]{0.5\textwidth}
        \centering
        \includegraphics[width=\textwidth]{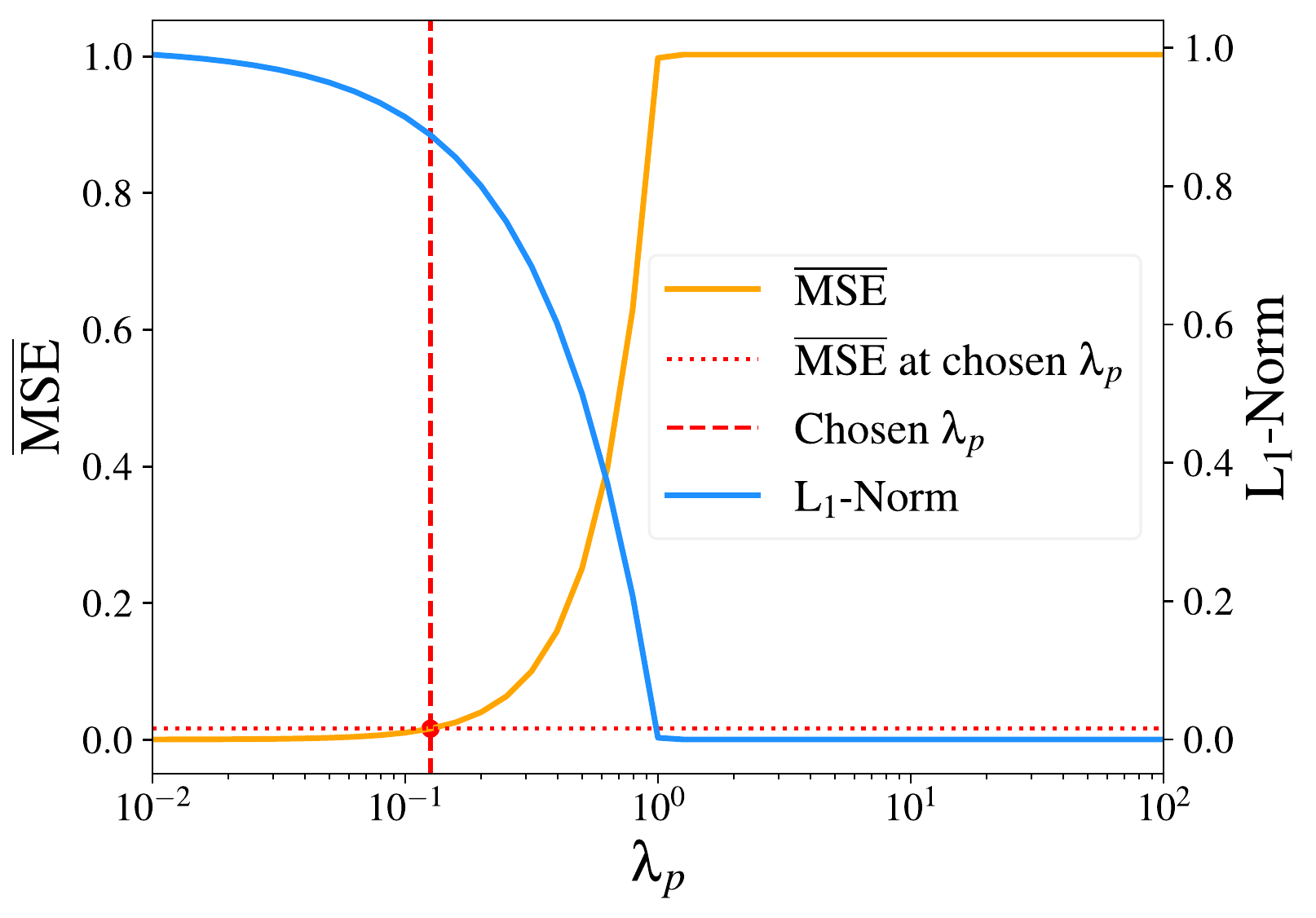}
    \end{subfigure}
    \caption{Illustration of the Pareto analysis for benchmark case MR2O1 (noise level $\sigma = 10 \ \text{Pa}$).}
	\label{fig:ParetoMR2O1}
\end{figure}

In the following, we discuss in more detail some of the intermediate steps in the sparsity promoting algorithm for material model discovery (see Steps 3-8 in \cref{fig:flowchart_algorithm}) in the context of benchmark case MR2O1 with noise level $\sigma = 10 \ \text{Pa}$.
\cref{fig:ParetoMR2O1} illustrates the Pareto analysis, which is leveraged for the choice of the hyperparameter $\lambda_p$ in the sparse regression problem.
It can be seen that the mean squared error $\overline{\mathrm{MSE}}$ increases and the $L_1$-norm of the material parameter vector $\|\bar\bftheta\|_1$ decreases for increasing values of $\lambda_p$. For values of $\lambda_p$ exceeding approximately $1$, the material parameter vector obtained from the solution of the sparse regression problem contains all zeros, so that its $L_1$-norm remains constant at zero and the corresponding $\overline{\mathrm{MSE}}$ saturates.
Using the method described in \cref{sec:Pareto}, the appropriate value of $\lambda_p$ is chosen such that the selected model exhibits both a high fitting accuracy and a low model complexity. Note also that for values of $\lambda_p$ lower than the chosen one, the $\overline{\mathrm{MSE}}$ is very weakly sensitive to the specific choice of $\lambda_p$.

\begin{figure}[H]
	\centering
    \begin{subfigure}[b]{0.8\textwidth}
        \centering
        \includegraphics[width=\textwidth]{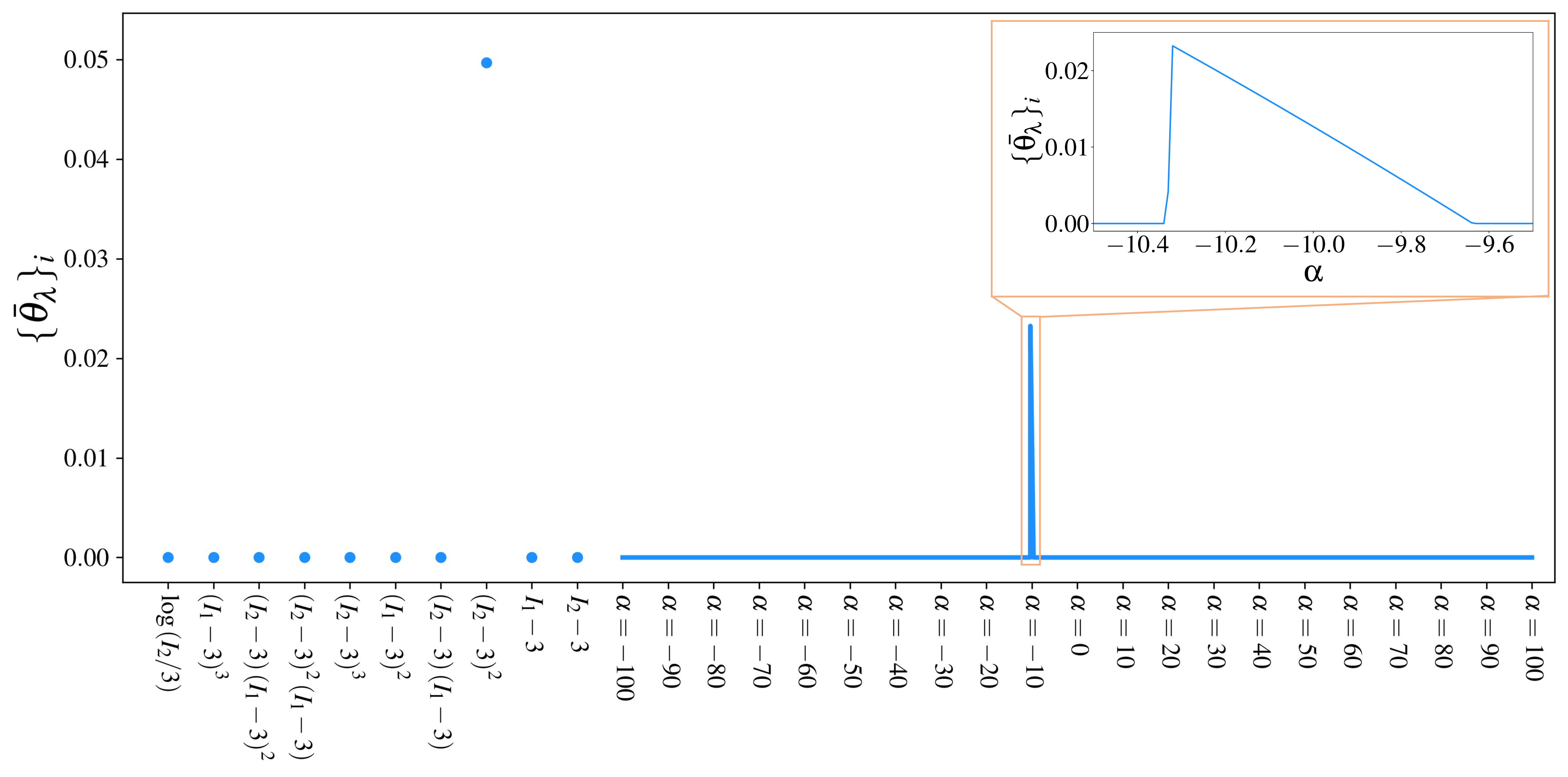}
        \caption{Before clustering}
    \end{subfigure}
    \begin{subfigure}[b]{0.8\textwidth}
        \centering
        \includegraphics[width=\textwidth]{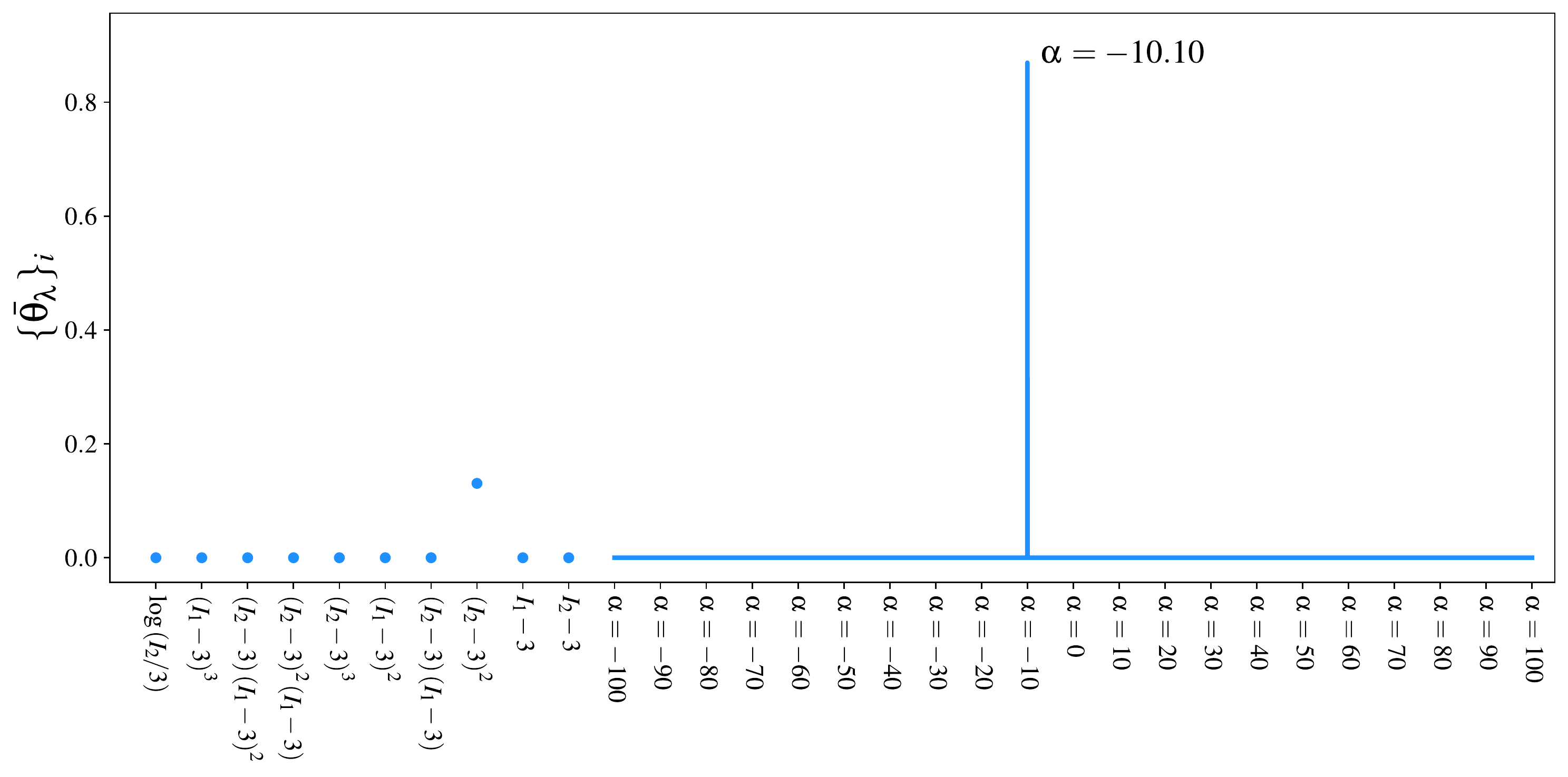}
        \caption{After clustering and final regression without regularization}
    \end{subfigure}
    \caption{Illustration of the feature clustering for benchmark case MR2O1 (noise level $\sigma = 10 \ \text{Pa}$).}
	\label{fig:clusteringMR2O1}
\end{figure}

In \cref{fig:clusteringMR2O1}, the effect of the clustering algorithm (see \cref{sec:feature_clustering}) is illustrated.
Before applying the clustering method, many values of $\alpha_i$ are zero while some non-zero values agglomerate in a small region near the ground truth value (equal to 10, see Table \ref{tab:ArtificialDiscovery}). 
The clustering algorithm identifies these values as one cluster. 
Afterwards, the values of $\alpha_i$ belonging to the cluster are averaged, such that the discovered material model exhibits only one Ogden term as the ground truth model, and the feature coefficients are calibrated, as illustrated in \cref{sec:regression_wo_regularization}.

\section{Experimental validation}
\label{sec:experimental_validation}
\subsection{Experimental data}
In the context of this work, we use multi-modal large-strain mechanical testing data from 81 cylindrical specimens extracted from three different human brains.
The data acquisition and preprocessing are briefly described in the following.
For details, the reader is referred to \cite{hinrichsen_inverse_2022}.

81 cylindrical specimens from different regions of the brain are prepared.
Each specimen is labeled by a brain number and a specimen number as $\text{HBE}\_\langle\text{brain~no.}\rangle\_\langle\text{specimen~no.}\rangle$.
The specimens have a radius of approximately $r_{\text{out}} \approx 4 \ \text{mm}$. The heights of the specimens and the region that they stem from are recorded in \cref{tab:HHBE}.

\begin{figure}
\centering
\includegraphics[width=0.8\linewidth]{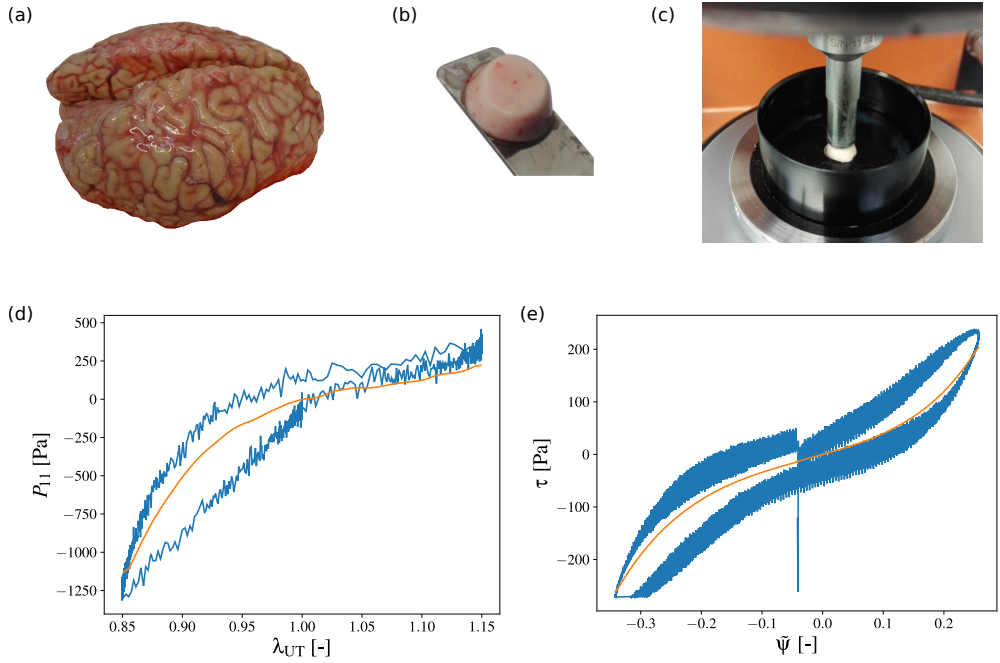}
\caption{Experimental setup and data preprocessing. Cylindrical samples (b) are extracted from multiple locations within the human brains (a) and then mounted to the rheometer (c). The raw data from the rheometer (blue) is preprocessed to obtain the hyperelastic response (orange) (see (d) and (e)). This is exemplarily shown for a specimen from the motor cortex (HBE\_1\_8). First, the noise is reduced using a moving average as well as a lowpass filter. Subsequently, the hyperelastic response is obtained by averaging the loading and unloading curve.}\label{fig:preProc}
\end{figure}

\cref{fig:preProc} shows one of the human brains and an exemplary cylindrical sample before and after mounting it to the rheometer for the mechanical testing. Each loading mode (compression/tension and torsional shear) consists of three cycles. Here, we use the data from the third cycle representing the preconditioned material response as brain tissue was found to show substantial pre-conditioning behavior \citep{budday_mechanical_2017}.
We limit ourselves to hyperelastic material models (neglecting poro- and viscoelastic effects).
The strain rates needed to obtain a purely hyperelastic response are (in theory infinitely) high or low and therefore not feasible in actual experiments.
Thus, the experimental data obtained from the rheometer shows a considerable hysteresis (see \cref{fig:preProc}) and we extract the hyperelastic response through a preprocessing procedure.
First, a moving average as well as a low pass filter is applied to reduce high frequency noises. Afterwards, the hyperelastic response is approximated by the averaged loading and unloading curves (see also \cite{budday_mechanical_2017}). The raw data and the extracted hyperelastic response are illustrated in \cref{fig:preProc}.
To reduce the dimensionality of the data while still preserving the characteristic shape of the curve, we reduce the points per deformation mode to 60.

This finally leads to a number of $n_{\text{UT}} = 60$ data pairs $(\lambda_{\text{UT}}^{(l)},P_{11}^{(l)})$ and $n_{\text{ST}} = 60$ data pairs $(\tilde\psi^{(l)},\tau^{(l)})$ acquired from the mechanical tests under uniaxial compression/tension and simple torsion, respectively.
The values of $\lambda_{\text{UT}}^{(l)}$ are uniformly distributed between $\lambda_{\text{UT}}^{(1)}=0.85$ and $\lambda_{\text{UT}}^{(n_{\text{UT}})}=1.15$ and the values of $\tilde\psi^{(l)}$ are uniformly distributed between $\tilde\psi^{(1)} = -0.3$ and $\tilde\psi^{(n_{\text{ST}})} = 0.3$.

\subsection{Results and discussion}
The proposed algorithm for material model discovery is applied to all experimentally acquired data sets, keeping all hyperparameters and algorithmic settings constant.
The resulting material parameters are listed in \cref{tab:BrainDiscovery} and \cref{tab:BrainDiscovery2}.
Many material parameters are identified as zero, meaning that the discovered expressions of the strain energy density functions are concise and interpretable.
The material model discovered for most of the specimens (i.e., about $60\%$ of the specimens) is the one-term Ogden model, which has also previously been used to fit human brain tissue data \citep{budday_mechanical_2017,budday_fifty_2020,hinrichsen_inverse_2022}. Further models that were identified are the two-term Ogden model (about $32\%$ of the specimens), and combinations of the one-term Ogden model with a Mooney-Rivlin term (about $7\%$ of the specimens).


To measure the ability of the discovered material models to fit the given experimental data, the mean squared errors $\mathrm{MSE}$ and $\overline{\mathrm{MSE}}$ are provided in \cref{tab:BrainDiscovery} and \cref{tab:BrainDiscovery2}.
The scaled mean squared error $\overline{\mathrm{MSE}}$ ranges from a minimum value of $0.01$ to a maximum value of $0.71$, and
the average $\overline{\mathrm{MSE}}$ over all experimental data sets is $0.17$.

\begin{figure}[H]
	\centering
 	\begin{subfigure}[b]{0.35\textwidth}
         \centering
         \includegraphics[width=\textwidth]{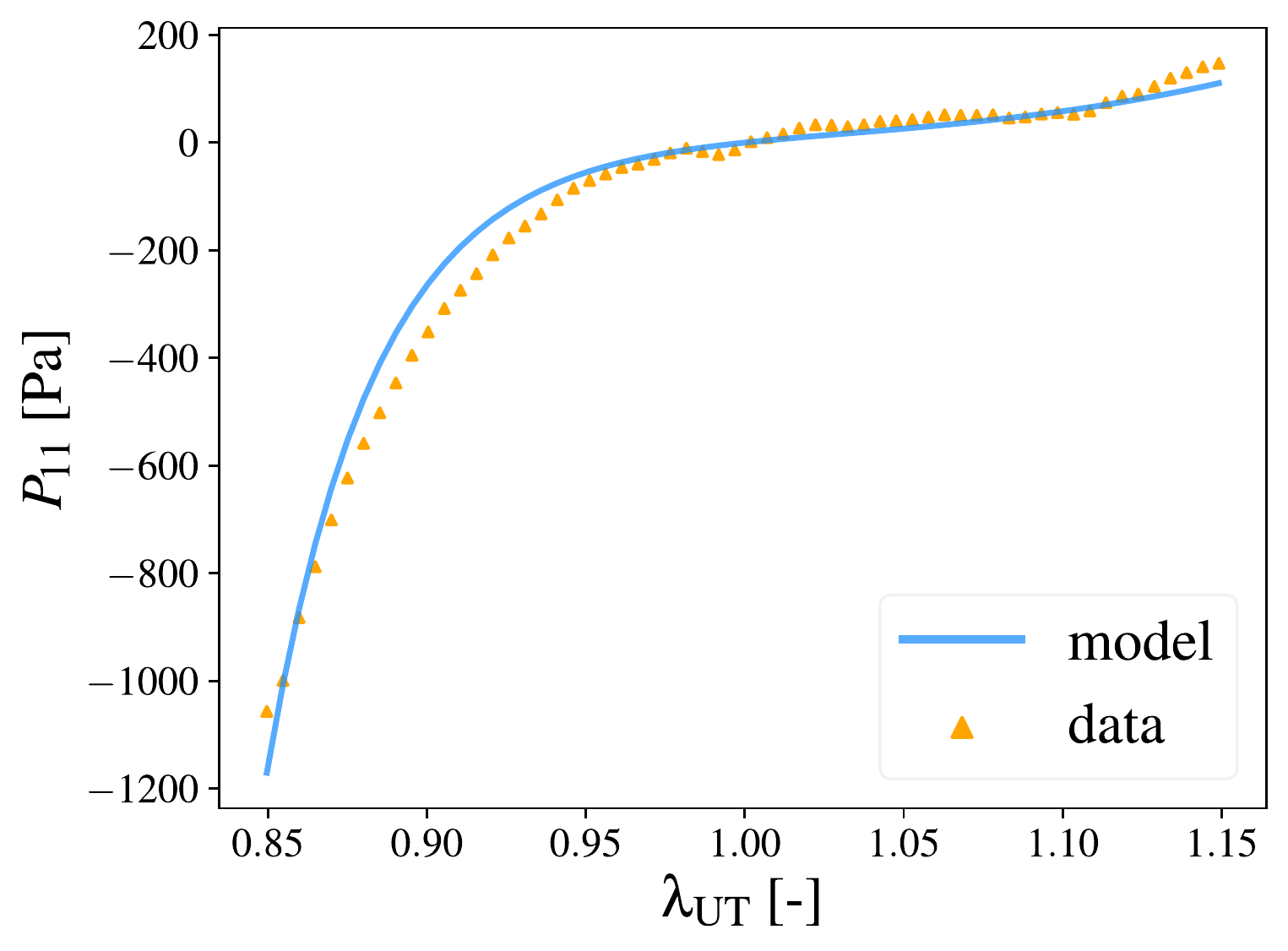}
         \caption{HBE\_01\_26 under UT}
     \end{subfigure}
 	\begin{subfigure}[b]{0.35\textwidth}
         \centering
         \includegraphics[width=\textwidth]{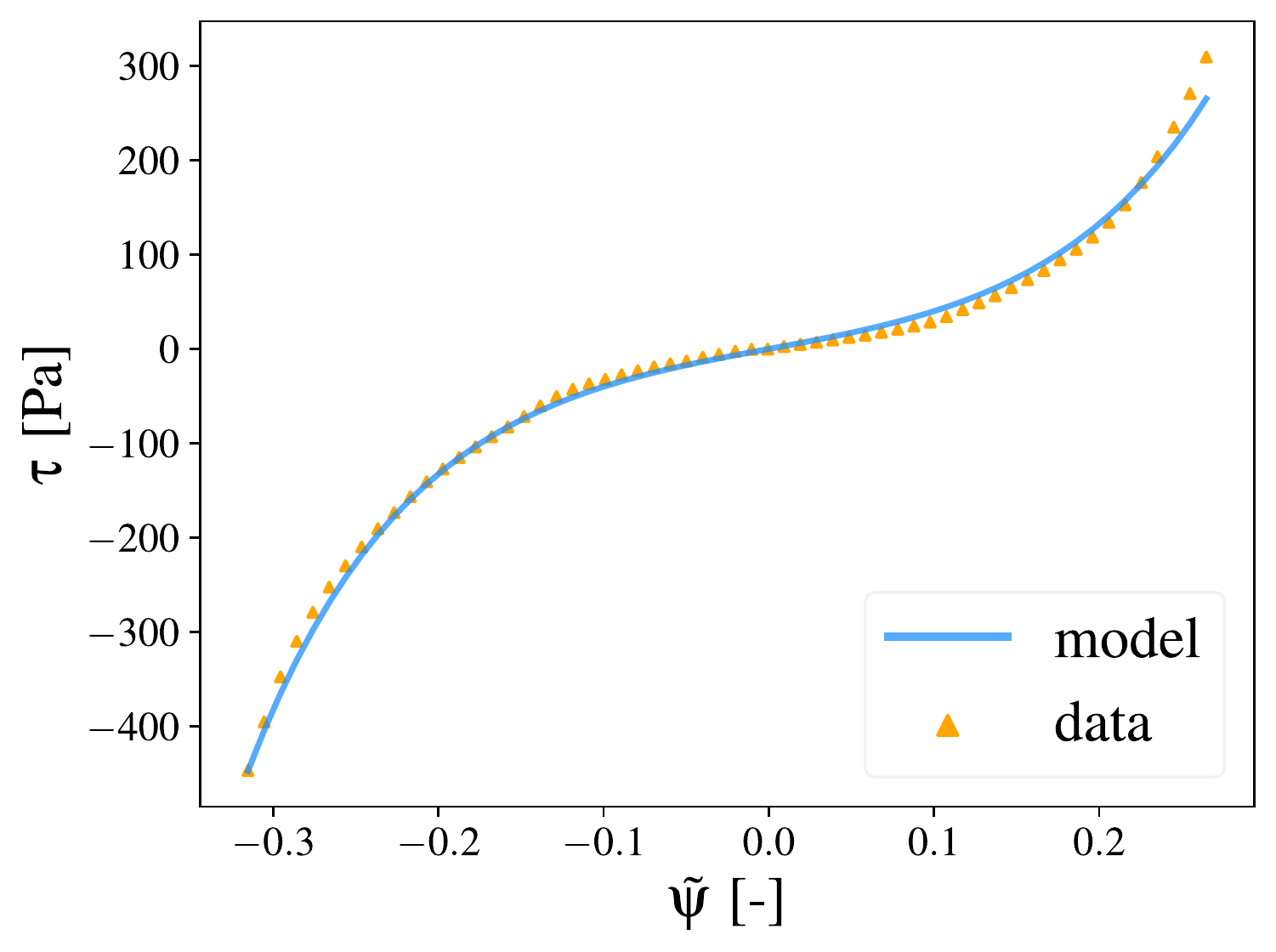}
         \caption{HBE\_01\_26 under ST}
     \end{subfigure}
    \caption{Stress versus stretch response under uniaxial compression/tension and torque versus (normalized) twist response under simple torsion of a discovered material model in comparison with the experimental data. The mean squared error $\overline{\mathrm{MSE}} = 0.0091$ is the smallest in the data set.}
	\label{fig:PlotsBrain_smallMSE}
\end{figure}

\begin{figure}[H]
	\centering
 	\begin{subfigure}[b]{0.35\textwidth}
         \centering
         \includegraphics[width=\textwidth]{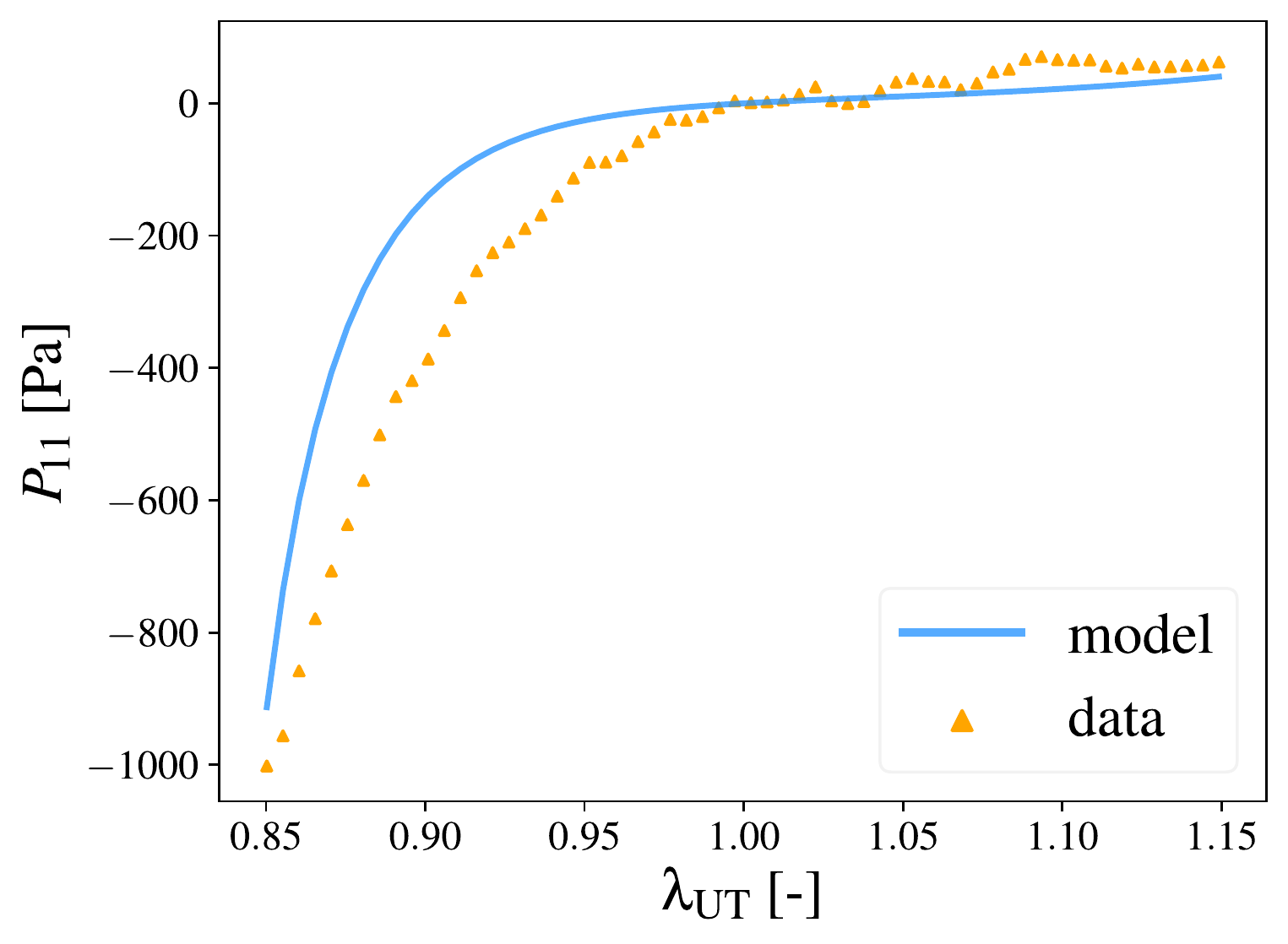}
         \caption{HBE\_01\_29 under UT}
     \end{subfigure}
 	\begin{subfigure}[b]{0.35\textwidth}
         \centering
         \includegraphics[width=\textwidth]{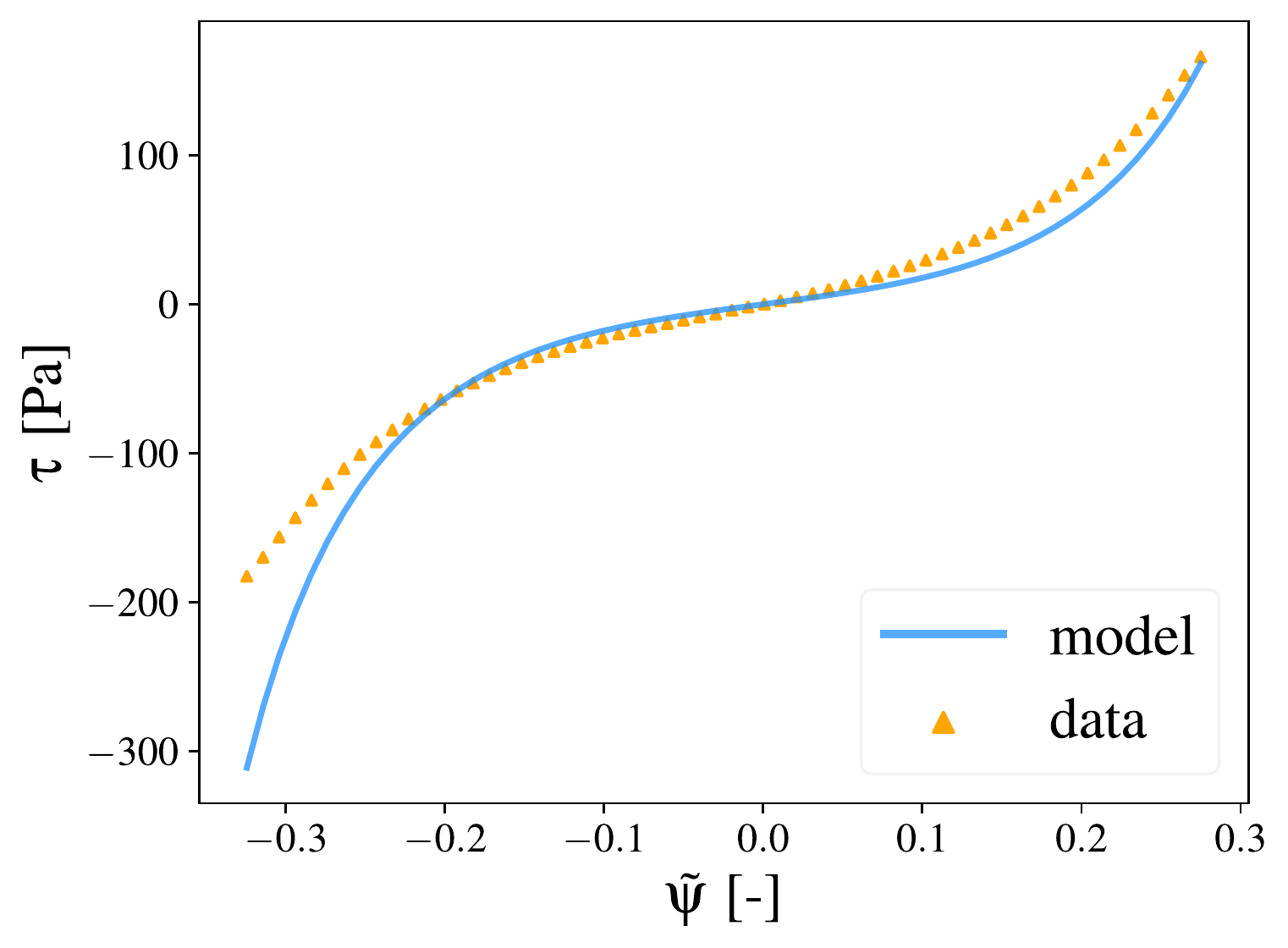}
         \caption{HBE\_01\_29 under ST}
     \end{subfigure}
 	\begin{subfigure}[b]{0.35\textwidth}
         \centering
         \includegraphics[width=\textwidth]{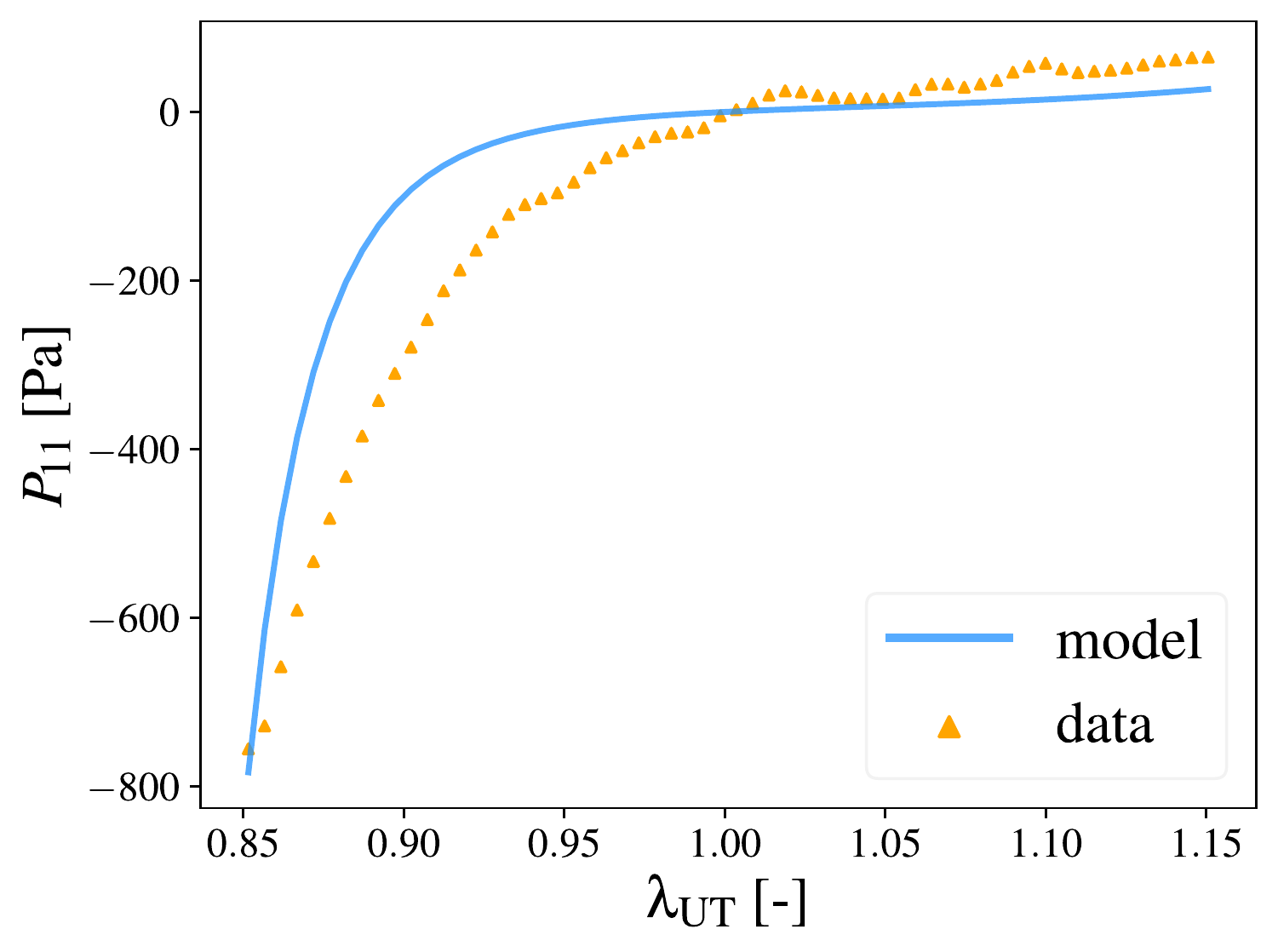}
         \caption{HBE\_02\_15 under UT}
     \end{subfigure}
 	\begin{subfigure}[b]{0.35\textwidth}
         \centering
         \includegraphics[width=\textwidth]{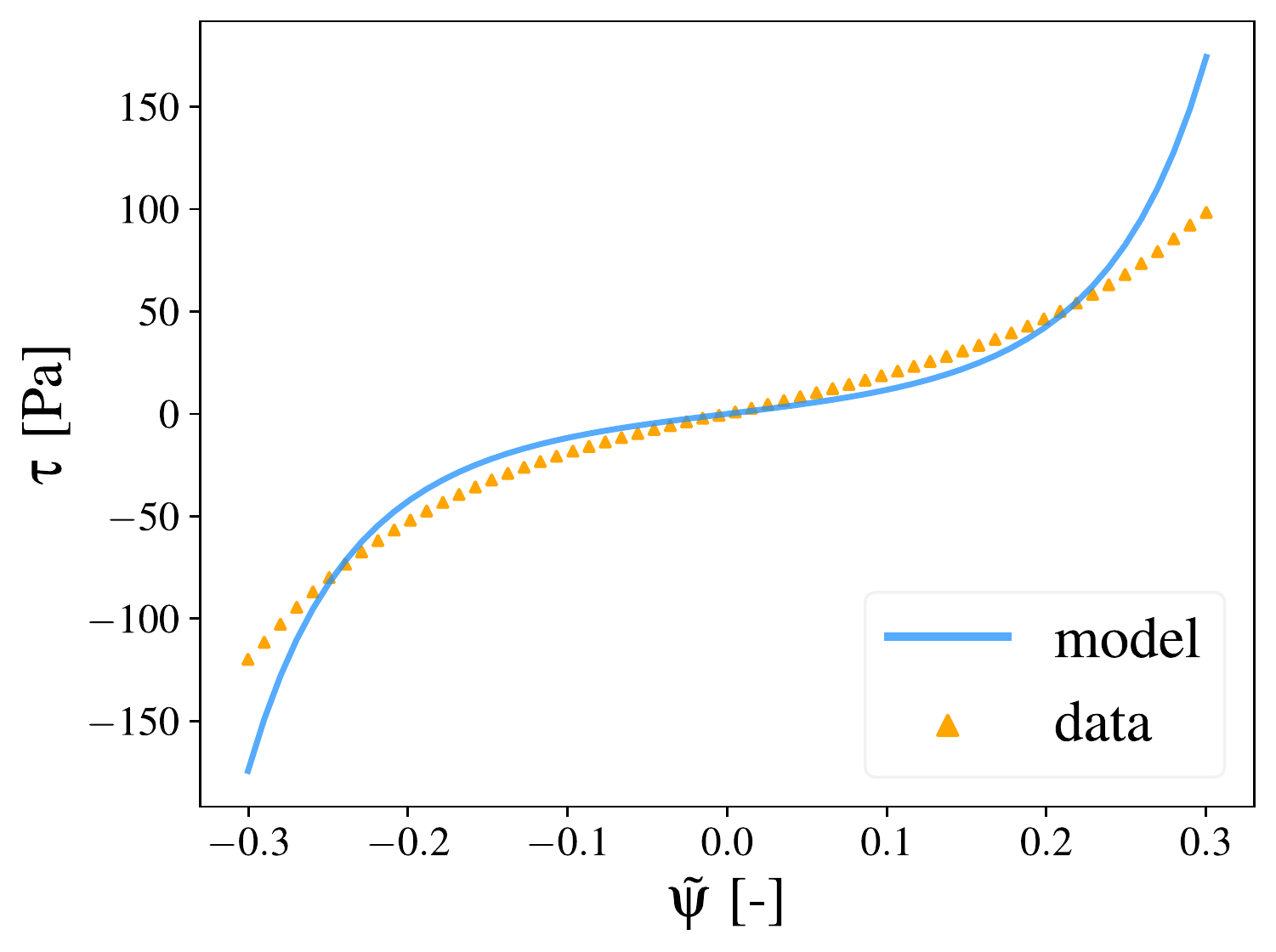}
         \caption{HBE\_02\_15 under ST}
     \end{subfigure}
 	\begin{subfigure}[b]{0.35\textwidth}
         \centering
         \includegraphics[width=\textwidth]{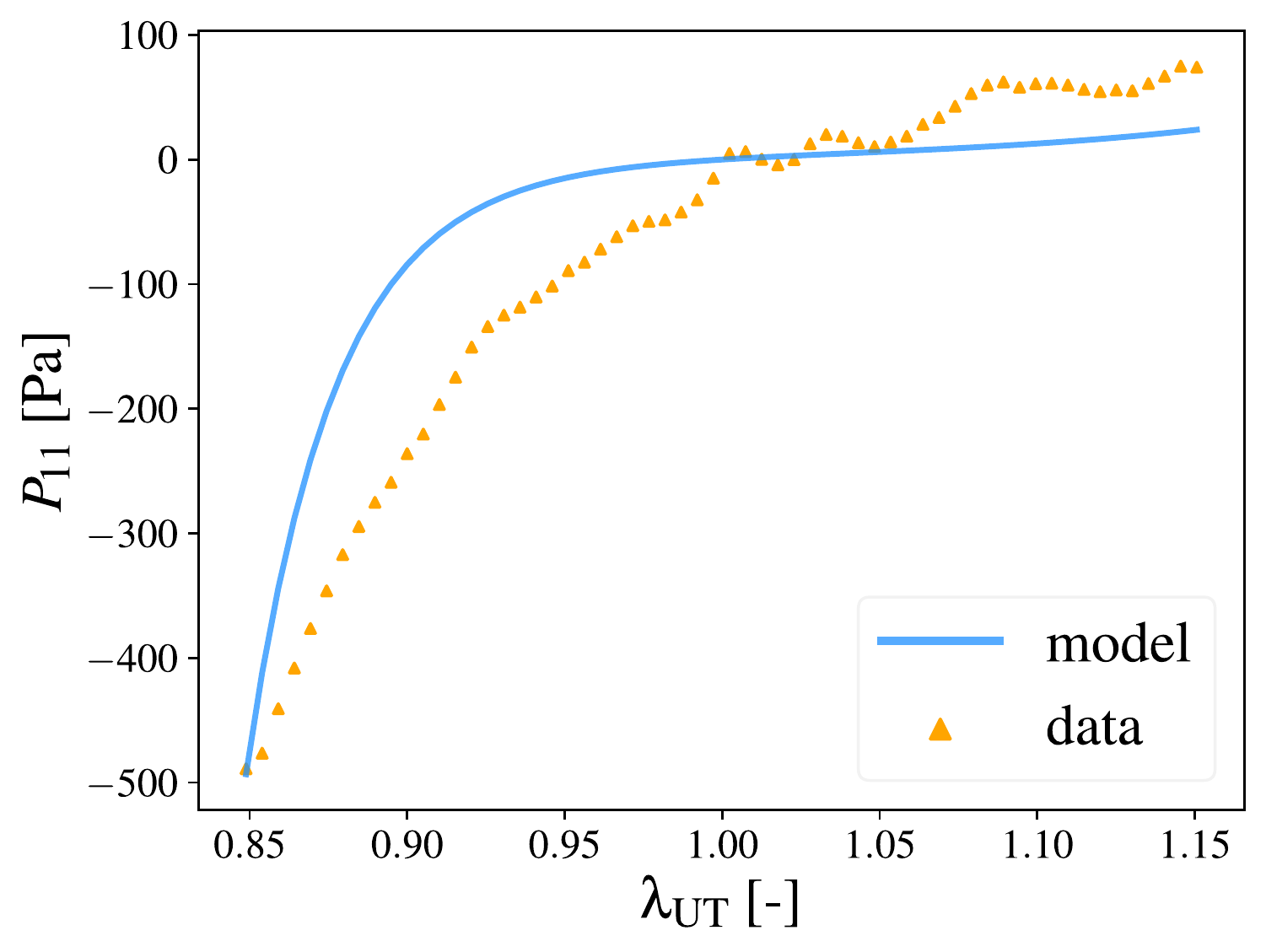}
         \caption{HBE\_03\_15 under UT}
     \end{subfigure}
 	\begin{subfigure}[b]{0.35\textwidth}
         \centering
         \includegraphics[width=\textwidth]{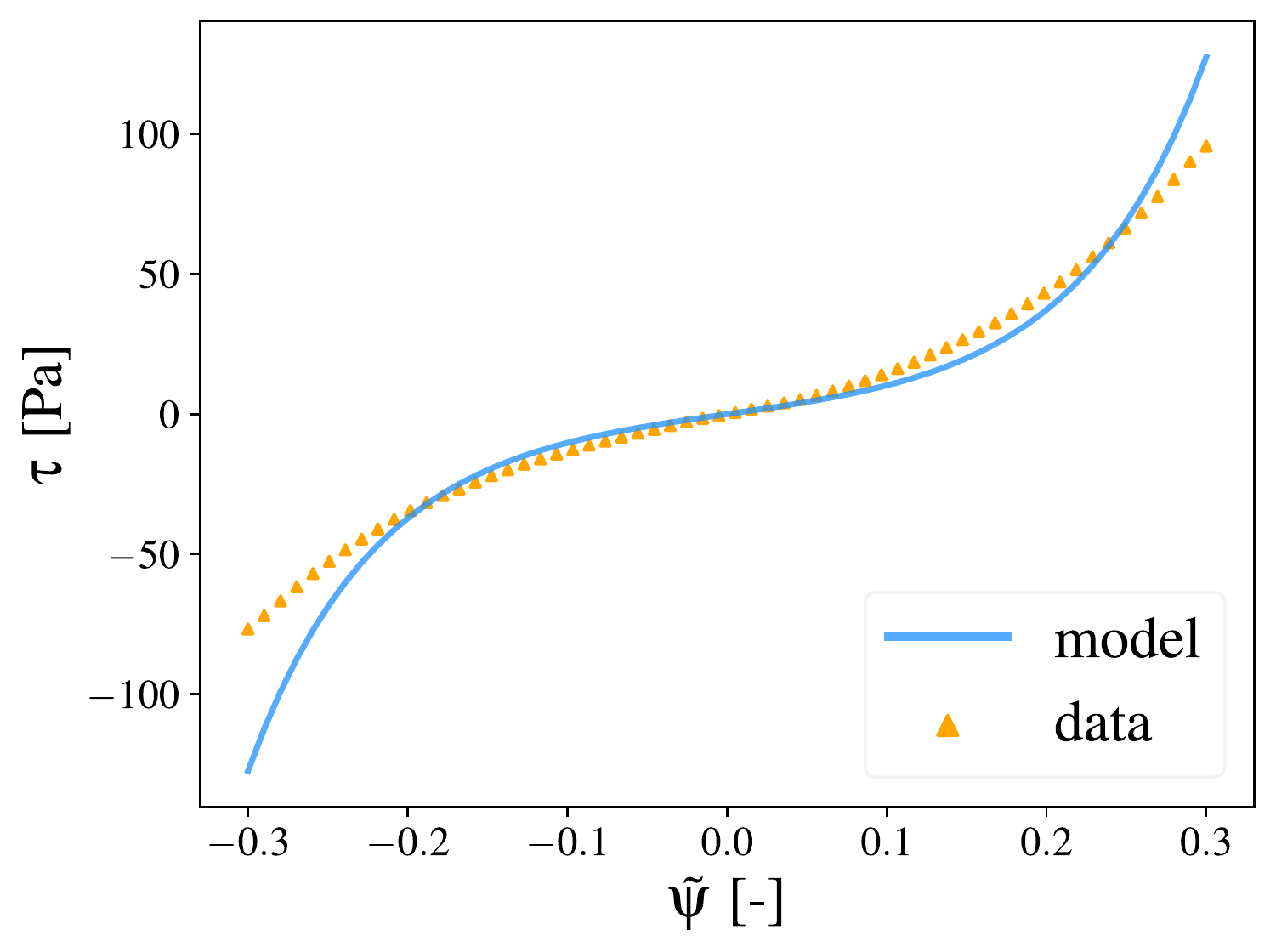}
         \caption{HBE\_03\_15 under ST}
     \end{subfigure}
    \caption{Stress versus stretch response under uniaxial compression/tension and  torque versus (normalized) twist response under simple torsion of a discovered material model in comparison with the experimental data. The mean squared errors ($\overline{\mathrm{MSE}} = 0.1657$,  $\overline{\mathrm{MSE}} = 0.1671$,  $\overline{\mathrm{MSE}} = 0.1664$ for HBE\_01\_29, HBE\_02\_15, HBE\_03\_15, respectively) are close to the average $\overline{\mathrm{MSE}}$ in the data set.}
	\label{fig:PlotsBrain_averageMSE}
\end{figure}

\cref{fig:PlotsBrain_smallMSE} shows the stress versus stretch response under uniaxial compression/tension and torque versus normalized twist response under simple torsion of a discovered material model for which the value of $\overline{\mathrm{MSE}}$ is small in comparison to the other experiments.
It is evident that the discovered material model is very well suited to describe the given data. \cref{fig:PlotsBrain_averageMSE} shows the same plots for discovered material models for which the values of $\overline{\mathrm{MSE}}$ are close to the average value of $\overline{\mathrm{MSE}}$ over all experiments.
The models are in qualitative agreement with the data.
However, especially for the response under uniaxial compression/tension, the models are not capable to precisely match the given data.
These deviations could possibly be attributed to modeling assumptions that do not hold true during the mechanical experiments.
For example, the geometries of some of the specimens were not ideally cylindrical, because the softer specimens deformed under their own weight. 
Further, it was assumed throughout the derivations in this paper that the strain state in the specimens is homogeneous under uniaxial compression/tension,
and it was assumed that the specimens are fixed in longitudinal direction at the boundaries, but free to move in the transversal directions.
During the experiments, however, both ends of the specimens were glued, meaning that the cross-sectional area of the specimens remained unchanged at the glued boundaries.
Therefore, the sides of the cylindrical specimens bulged into a barrel shape, resulting in non-homogeneous strain states in the specimens.
The constraint that the specimens are fixed in both longitudinal and transversal directions may add additional stiffness to the experimentally measured stress versus stretch response, which could explain the underestimated stress by the discovered material models during uniaxial compression/tension in \cref{fig:PlotsBrain_averageMSE}.
Another source for the disagreement between the discovered models and the data could be that only isotropic material behavior has been included in the material model library.
Thus, possible effects from material anisotropy in the experimental data cannot be predicted by the discovered models.

\begin{figure}[H]
	\centering
 	\begin{subfigure}[b]{0.35\textwidth}
         \centering
         \includegraphics[width=\textwidth]{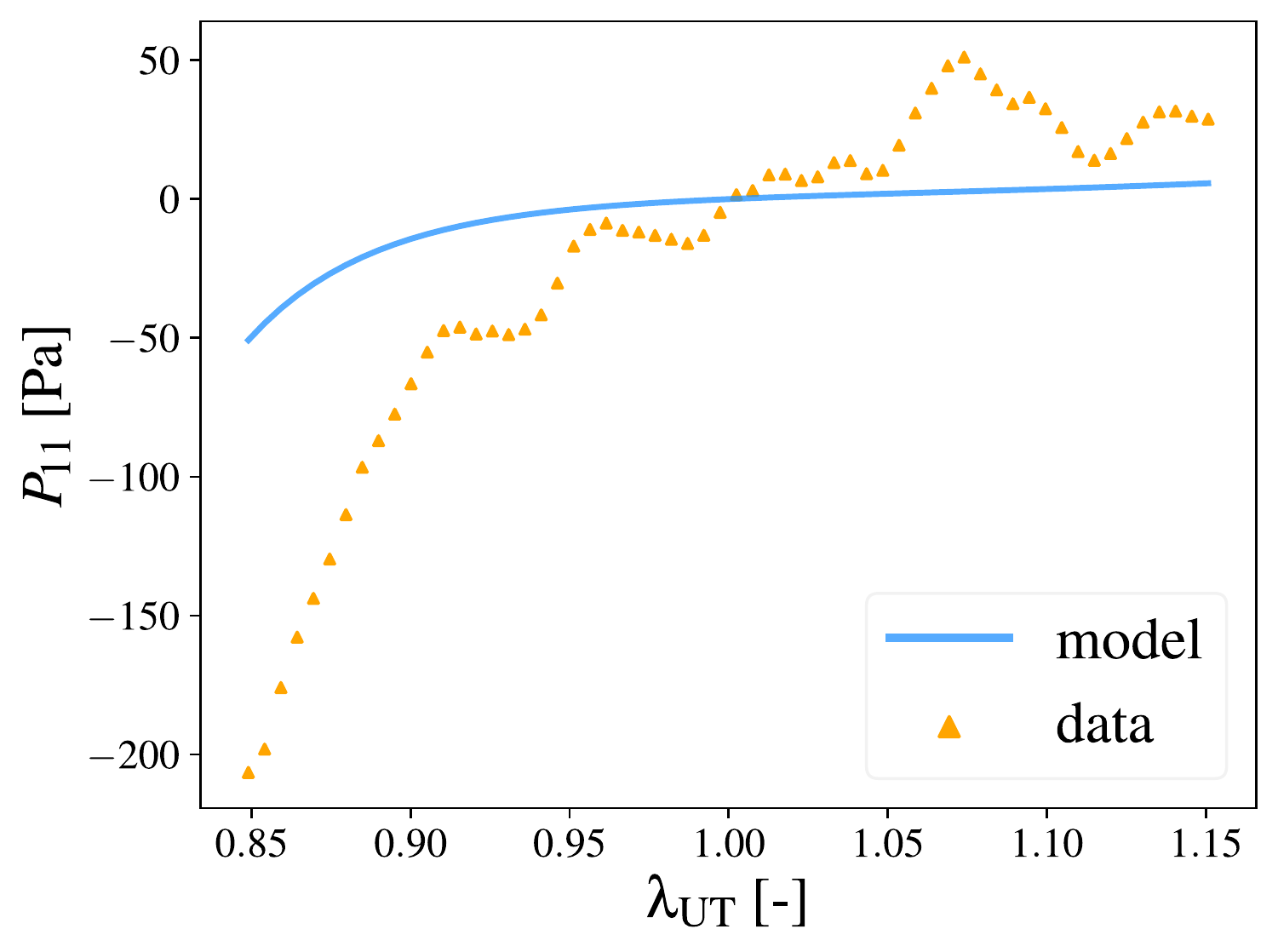}
         \caption{HBE\_01\_24 under UT}
         \label{fig:PlotsBrain_highMSE_1}
     \end{subfigure}
 	\begin{subfigure}[b]{0.35\textwidth}
         \centering
         \includegraphics[width=\textwidth]{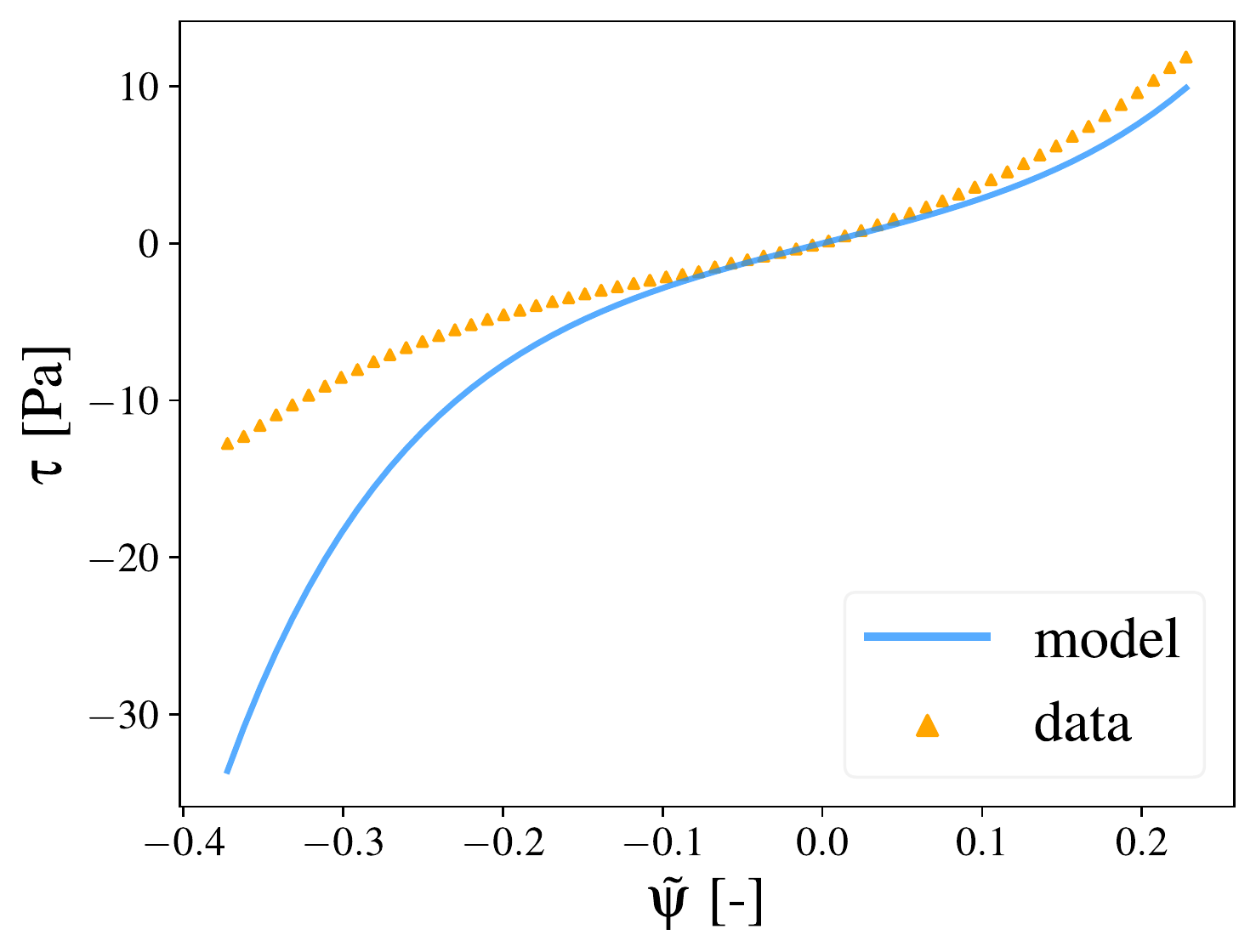}
         \caption{HBE\_01\_24 under ST}
     \end{subfigure}
 	\begin{subfigure}[b]{0.35\textwidth}
         \centering
         \includegraphics[width=\textwidth]{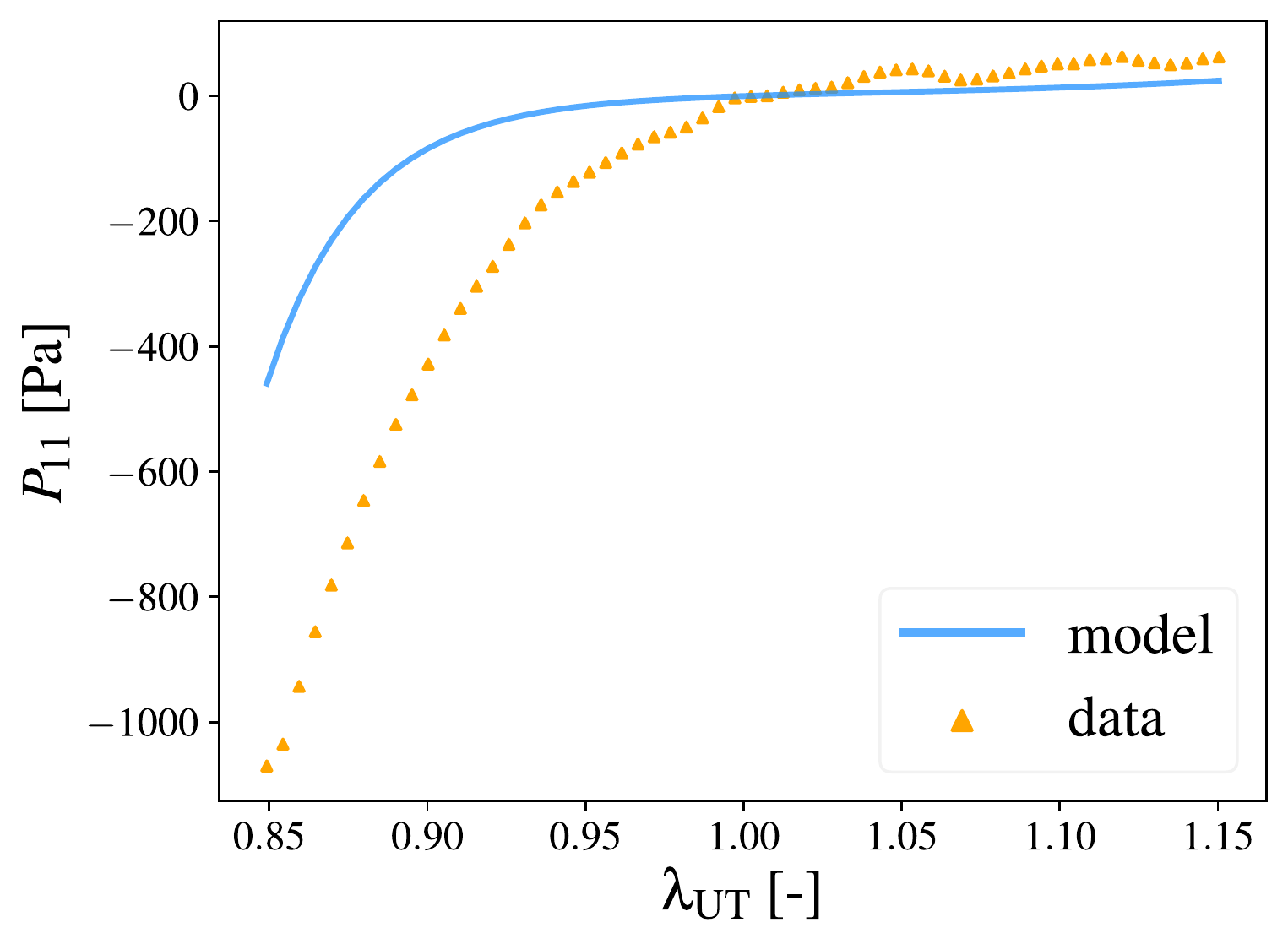}
         \caption{HBE\_01\_30 under UT}
         \label{fig:PlotsBrain_highMSE_2}
     \end{subfigure}
 	\begin{subfigure}[b]{0.35\textwidth}
         \centering
         \includegraphics[width=\textwidth]{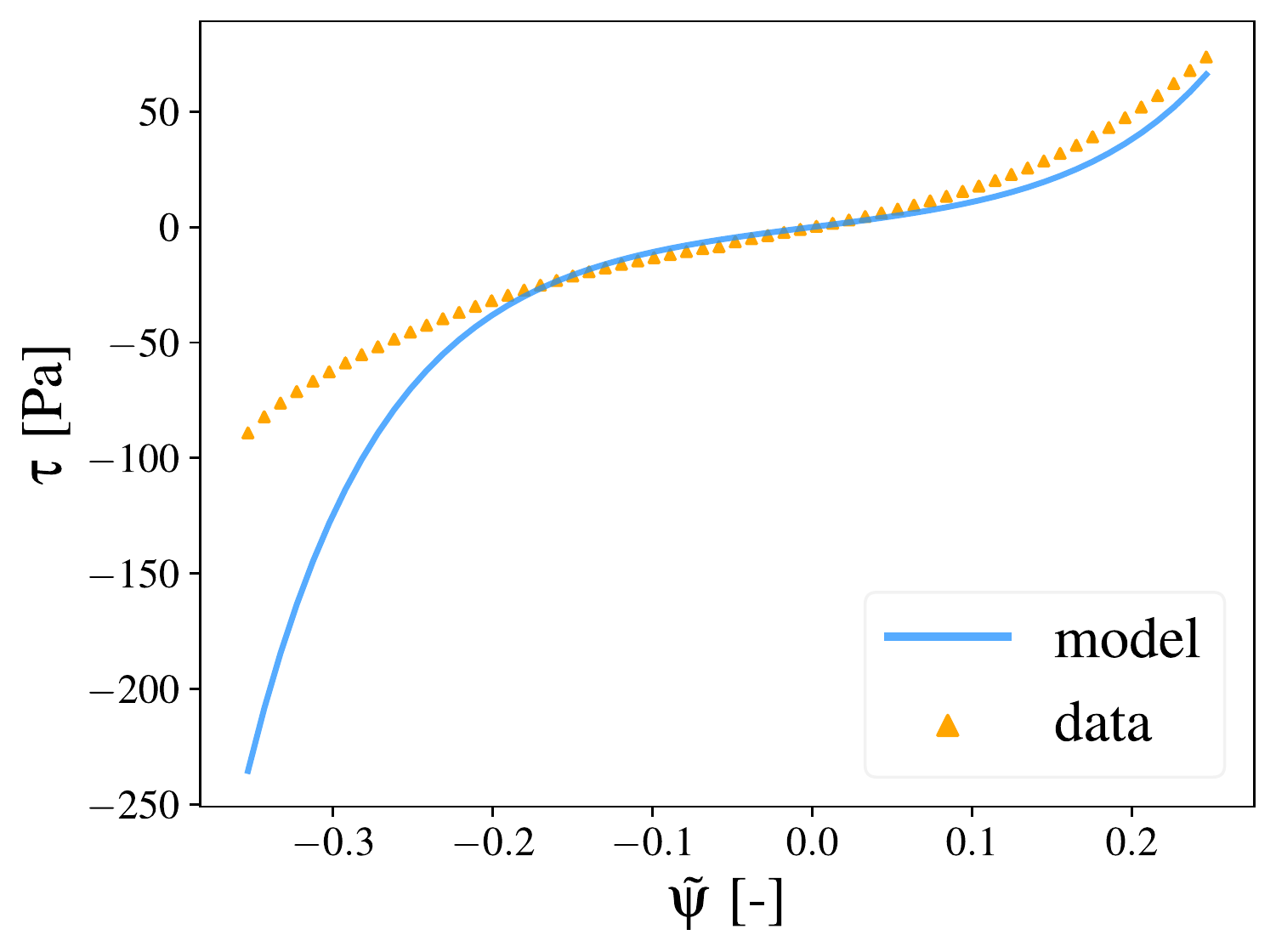}
         \caption{HBE\_01\_30 under ST}
     \end{subfigure}
    \caption{Stress versus stretch response under uniaxial compression/tension and torque versus (normalized) twist response under simple torsion of discovered material models in comparison with the experimental data. The mean squared errors ($\overline{\mathrm{MSE}} = 0.7064$, $\overline{\mathrm{MSE}} = 0.6107$ for HBE\_01\_24, HBE\_01\_30, respectively) are the largest in the data set.}
	\label{fig:PlotsBrain_highMSE}
\end{figure}

Finally, \cref{fig:PlotsBrain_highMSE} shows the stretch versus stress and normalized twist versus torque plots of discovered material models for which the value of $\overline{\mathrm{MSE}}$ is large in comparison to the other experiments.
As the large mean squared error indicates, the mismatch between the discovered model and the data is significant.
Such large discrepancies are exceptional for the considered experimental data sets (see \cref{tab:BrainDiscovery} and \cref{tab:BrainDiscovery2}).
For cyclic shear tests, the reason for the large deviation between the model and the data could be erroneous experimental measurements: the two shown samples were not loaded symmetrically during cyclic shear tests but were already twisted before the shear test started. This can happen when a torque is induced during axial loading (for example due to inhomogeneous microstructure of the sample). In this case, the rheometer applies a deformation until the torque goes back to zero.
The non-monotonic stress versus stretch response under uniaxial compression/tension in \cref{fig:PlotsBrain_highMSE_1} could be a consequence of a low signal-to-noise ratio since this sample is very soft. The non-increasing constant stress plateau under uniaxial compression/tension in \cref{fig:PlotsBrain_highMSE_2} could indicate a failure of the glue that connects the specimen with the testing machine. 


\section{Conclusion}
\label{sec:conclusion}

EUCLID, a recently proposed computational framework for automatically discovering constitutive models as interpretable symbolic mathematical expressions, was formulated in this work in a supervised setting and validated on the mechanical testing data of human brain specimens.
It was shown that the proposed model selection strategy, which comprises sparse regression and feature clustering, is able to automatically discover interpretable constitutive models with satisfactory fitting accuracy to the given uniaxial compression/tension and simple torsion data. The most commonly discovered model was the one-term Ogden model, which has also previously been widely used to fit human brain tissue data.
The presented results constitute the first experimental validation of EUCLID.
However, several aspects of the proposed method may be improved or extended in future studies.
For example, the method proposed in this work is informed by labeled data pairs due to the limited access to full-field displacement data for human brain tissues.
The experimental validation of EUCLID driven by unlabeled data, i.e., full-field displacement and net reaction force data (see \cite{flaschel_automated_2023}), is the subject of ongoing research.
Moreover, we restricted the model library in which EUCLID searches for a model to the class of isotropic hyperelastic models.
As the tissue response is in fact not just hyperelastic but affected by dissipative effects as well as possibly not perfectly isotropic, an important future study could be to further generalize the model library, such that EUCLID can not only select a constitutive model for the human brain tissue within a predefined class of models, but further distinguish between different modeling classes (such as hyperelasticity, viscoelasticity, elastoplasticity), as shown by \cite{flaschel_automated_2023-1} for synthetical data, as well as automatically detect possible anisotropies and identify the corresponding parameters.


\appendix

\section{Derivatives}
\label{sec:derivatives}

\subsection{Derivatives of the strain invariants}
\label{sec:derivatives_strain_invariants}
The derivatives of the strain invariants with respect to the deformation gradient are
\be
\frac{\partial I_1}{\partial \bfF} = 2\bfF, \quad
\frac{\partial I_2}{\partial \bfF} = 2I_1\bfF - 2\bfF\bfC.
\ee

\subsection{Derivatives of the principal stretches}
\label{sec:derivatives_principal_stretches}
The derivatives of the principal directions depend on the eigenvectors of $\bfC=\bfF^T\bfF$, denoted by $\bfN_i$, and the eigenvectors of $\bfb=\bfF\bfF^T$, denoted by $\bfn_i$.
Under the assumption $\lambda_1\neq\lambda_2\neq\lambda_3\neq\lambda_1$, it is (see \cite{holzapfel_nonlinear_2000})
\be
\frac{\partial \lambda_1}{\partial \bfF} = \bfn_1\otimes\bfN_1, \quad
\frac{\partial \lambda_2}{\partial \bfF} = \bfn_2\otimes\bfN_2, \quad
\frac{\partial \lambda_3}{\partial \bfF} = \bfn_3\otimes\bfN_3.
\ee

\subsection{Derivatives of the feature vectors}
\label{sec:derivatives_feature_vectors}
The derivatives of the feature vector $\bfQ_I$ with respect to the strain invariants are
\be
\begin{aligned}
\frac{\partial \bfQ_I}{\partial I_1} &= 
\left[ m(I_1-3)^{m-1}(I_2-3)^{n-m} : n\in \{1,\dots,N_{\text{Mooney}}\}, m\in\{0,\dots,n\}\right]^T
\oplus 
\left[ ~0~ \right],\\
\frac{\partial \bfQ_I}{\partial I_2} &= 
\left[ (n-m)(I_1-3)^m(I_2-3)^{n-m-1} : n\in \{1,\dots,N_{\text{Mooney}}\}, m\in\{0,\dots,n\}\right]^T
\oplus 
\left[ ~1/I_2~ \right],
\end{aligned}
\ee
which can be used to calculate the derivatives of the strain energy density contribution $\tilde{W}_I$ with respect to the strain invariants, i.e.,
\be
\frac{\partial \tilde{W}_I}{\partial I_a} = \bftheta_I \cdot \frac{\partial \bfQ_I}{\partial I_a}.
\ee

The components of the feature vector $\bfQ_{\lambda}$ are defined as
\be
\left\{Q_{\lambda}(\lambda_1,\lambda_2,\lambda_3)\right\}_i = \lambda_1^{\alpha_i} + \lambda_2^{\alpha_i} + \lambda_3^{\alpha_i} - 3.
\ee
Assuming incompressibility $\lambda_1\lambda_2\lambda_3 \overset{!}{=} 1$, the components of the feature vector $\bfQ_{\lambda}$ can be written as
\be
\left\{Q_{\lambda}(\lambda_1,\lambda_2)\right\}_i = \lambda_1^{\alpha_i} + \lambda_2^{\alpha_i} + \lambda_1^{-\alpha_i}\lambda_2^{-\alpha_i} - 3.
\ee
Thus, the derivatives of the components of the feature vector $\bfQ_{\lambda}$ with respect to the principal stretches are
\be
\begin{aligned}
\frac{\partial \left\{Q_{\lambda}(\lambda_1,\lambda_2)\right\}_i}{\partial \lambda_1}
&= \alpha_i \left(\lambda_1^{\alpha_i-1} - \lambda_1^{-\alpha_i-1}\lambda_2^{-\alpha_i}\right),\\
\frac{\partial \left\{Q_{\lambda}(\lambda_1,\lambda_2)\right\}_i}{\partial \lambda_2}
&= \alpha_i \left(\lambda_2^{\alpha_i-1} - \lambda_1^{-\alpha_i}\lambda_2^{-\alpha_i-1}\right),
\end{aligned}
\ee
which can be used to calculate the derivatives of the strain energy density contribution $\tilde{W}_{\lambda}$ with respect to the principal stretches, i.e.,
\be
\frac{\partial \tilde{W}_{\lambda}}{\partial \lambda_i} = \bftheta_{\lambda} \cdot \frac{\partial \bfQ_{\lambda}}{\partial \lambda_i}.
\ee

\subsection{Derivatives for uniaxial compression and tension}
\label{sec:derivatives_uniaxial_tension}
In the following, we seek to simplify the expression for $P_{11}(\lambda_{\text{UT}};\bftheta)$ (see \cref{eq:constitutive_map_uniaxial_tension}) by leveraging the characteristics of the kinematic state under uniaxial compression and tension.
Considering such kinematic state, the deformation gradient and the resulting right Cauchy-Green strain tensor read
\be
\begin{aligned}
\bfF =
	\begin{bmatrix}
		\lambda_{\text{UT}} & 0 & 0\\
		0 & \frac{1}{\sqrt{\lambda_{\text{UT}}}} & 0\\
		0 & 0 & \frac{1}{\sqrt{\lambda_{\text{UT}}}}\\
	\end{bmatrix},
\qquad
\bfC =
	\begin{bmatrix}
		\lambda_{\text{UT}}^2 & 0 & 0\\
		0 & \frac{1}{\lambda_{\text{UT}}} & 0\\
		0 & 0 & \frac{1}{\lambda_{\text{UT}}}\\
	\end{bmatrix}.
\end{aligned}
\ee
Thus, the strain invariants simplify to 
\be
I_1 = \lambda_{\text{UT}}^2 + \frac{2}{\lambda_{\text{UT}}}, \quad
I_2 = 2 \lambda_{\text{UT}} + \frac{1}{\lambda_{\text{UT}}^2},
\ee
and the principal stretches simplify to 
\be
\lambda_1 = \lambda_{\text{UT}}, \quad
\lambda_2 = \lambda_3 = \frac{1}{\sqrt{\lambda_{\text{UT}}}}.
\ee
The simple structure of the kinematic state is used to simplify the feature vector $\bfQ$. The feature vectors $\bfQ_I$ and $\bfQ_{\lambda}$ are
\be
\bfQ_I(\lambda_{\text{UT}}) = 
\left[ \left(\lambda_{\text{UT}}-3\right)^m\left(2 \lambda_{\text{UT}} + \frac{1}{\lambda_{\text{UT}}^2}-3\right)^{n-m} : n\in \{1,\dots,N_{\text{Mooney}}\}, m\in\{0,\dots,n\}\right]^T
\oplus 
\left[ \log \left(\frac{2}{3} \lambda_{\text{UT}} + \frac{1}{3\lambda_{\text{UT}}^2}\right)\right],
\ee
and
\be
\left\{Q_{\lambda}(\lambda_{\text{UT}})\right\}_i = \lambda_{\text{UT}}^{\alpha_i} +  2\left(\frac{1}{\sqrt{\lambda_{\text{UT}}}}\right)^{\alpha_i} - 3,
\ee
whose derivatives with respect to $\lambda_{\text{UT}}$
are trivial.
We finally obtain the following expression for the feature vector derivative in \cref{eq:constitutive_map_uniaxial_tension}
\be
\bfQ'_{\text{UT}}(\lambda_{\text{UT}})
= \frac{\partial\bfQ(\lambda_{\text{UT}})}{\partial \lambda_{\text{UT}}},
\ee
which after substitution in \cref{eq:constitutive_map_uniaxial_tension} results in a simplified expression for $P_{11}(\lambda_{\text{UT}};\bftheta)$.


\subsection{Derivatives for simple torsion}
\label{sec:derivatives_simple_torsion}
In the following, we seek to simplify the expression for $\tau(\tilde\psi;\bftheta)$ (see \cref{eq:constitutive_map_simple_torsion}) by leveraging the characteristics of the kinematic state under simple torsion.
Considering simple torsion, the deformation gradient and the resulting right Cauchy-Green strain tensor read
\be
\begin{aligned}
\bfF =
	\begin{bmatrix}
		1 & 0 & 0\\
		0 & 1 & F_{\vartheta z}\\
		0 & 0 & 1\\
	\end{bmatrix},
\qquad
\bfC =
	\begin{bmatrix}
		1 & 0 & 0\\
		0 & 1 & F_{\vartheta z}\\
		0 & F_{\vartheta z} & 1 + F_{\vartheta z}^2\\
	\end{bmatrix},
\end{aligned}
\ee
where $F_{\vartheta z} = \rho\tilde{\psi}$.
Thus, the strain invariants simplify to 
\be
I_1 = 
I_2 = F_{\vartheta z}^2 + 3.
\ee
Defining the auxiliary variable $\bar{C} = 1 + \frac{1}{2}F_{\vartheta z}^2$, the eigenvalues of $\bfC$ (which are the squares of the principal stretches) read
\be
\lambda_1^2 = \bar{C} - \sqrt{\bar{C}^2 - 1},
\quad
\lambda_2^2 = 1,
\quad
\lambda_3^2 = \bar{C} + \sqrt{\bar{C}^2 - 1},
\ee
which results in the principal stretches
\be
\lambda_1 = \sqrt{\bar{C} - \sqrt{\bar{C}^2 - 1}},
\quad
\lambda_2 = 1,
\quad
\lambda_3 = \sqrt{\bar{C} + \sqrt{\bar{C}^2 - 1}}.
\ee
The simple structure of the kinematic state is used to simplify the feature vector $\bfQ$. The feature vector $\bfQ_I$ is
\be
\bfQ_I(F_{\vartheta z}) = 
\left[ F_{\vartheta z}^{2n} : n\in \{1,\dots,N_{\text{Mooney}}\}, m\in\{0,\dots,n\}\right]^T
\oplus 
\left[ \log \left(F_{\vartheta z}^2/3 + 1\right)\right],
\ee
whose derivative with respect to $F_{\vartheta z}$ is trivial, and the feature vector $\bfQ_{\lambda}$ is
\be
\left\{Q_{\lambda}(\bar{C})\right\}_i = \left(\sqrt{\bar{C} - \sqrt{\bar{C}^2 - 1}}\right)^{\alpha_i} + \left(\sqrt{\bar{C} + \sqrt{\bar{C}^2 - 1}}\right)^{\alpha_i} - 2,
\ee
whose derivative with respect to $F_{\vartheta z}$ can be computed using the chain rule
\be
\frac{\partial \bfQ_{\lambda}}{\partial F_{\vartheta z}}
= \frac{\partial \bfQ_{\lambda}}{\partial \bar{C}}
\frac{\partial \bar{C}}{\partial F_{\vartheta z}}
= \frac{\partial \bfQ_{\lambda}}{\partial \bar{C}}
F_{\vartheta z}.
\ee
Thus, the derivative $\frac{\partial \bfQ}{\partial F_{\vartheta z}}$ is known and it depends on $\rho$ and $\tilde{\psi}$ through $\bar{C} = 1 + \frac{1}{2}F_{\vartheta z}^2$ and $F_{\vartheta z}=\rho\tilde{\psi}$.
With reference to \cref{eq:constitutive_map_simple_torsion}, it is
\be
\bfQ'_{\text{ST}}\left(\tilde\psi\right) = \int_0^1 \ 2\pi\rho^2 \frac{\partial \bfQ}{\partial F_{\vartheta z}} \ \mathrm{d}\rho,
\ee
where the integral over $\rho$ can be computed through numerical quadrature. 

\section{Algorithm}
\label{algorithm}
\begin{figure}[H]
\centering
\begin{tikzpicture}[node distance=1.5cm]
\node (pro1) [process] {Step 1: Compute $\bfA_{\mathrm{UT}}$, $\bfb_{\mathrm{UT}}$ and $\bfA_{\mathrm{ST}}$, $\bfb_{\mathrm{ST}}$, and assemble $\bfA$ and $\bfb$};
\node (pro2) [process, below of=pro1] {Step 2: Apply feature scaling to obtain $\bar\bfA$ and $\bar\bfb$};
\node (pro3) [process, below of=pro2] {Step 3: Solve the Lasso problem for a set of distinct values of $\lambda_p$};
\node (pro4) [process, below of=pro3] {Step 4: Discard all solutions from Step 3 with $\overline{\mathrm{MSE}}(\bar\bftheta) > \overline{\mathrm{MSE}}_{\mathrm{th}}$};
\node (pro5) [process, below of=pro4] {Step 5: Select the solution remaining from Step 4 with the smallest value of $\| \bar\bftheta \|_1$};
\node (pro6) [process, below of=pro5] {Step 6: For all $i$, if $\bar{\theta}_i<\bar{\theta}_{\text{th}}$, set $\bar{\theta}_i=0$};
\node (pro7) [process, below of=pro6] {Step 7: Use the clustering method to group together similar values of $\alpha_i$};
\node (pro8) [process, below of=pro7] {Step 8: Apply regression without regularization to calibrate the non-zero entries in $\bar\bftheta$};
\node (pro9) [process, below of=pro8] {Step 9: Compute $\bftheta$ from $\bar\bftheta$};
\draw [arrow] (pro1) -- (pro2);
\draw [arrow] (pro2) -- (pro3);
\draw [arrow] (pro3) -- (pro4);
\draw [arrow] (pro4) -- (pro5);
\draw [arrow] (pro5) -- (pro6);
\draw [arrow] (pro6) -- (pro7);
\draw [arrow] (pro7) -- (pro8);
\draw [arrow] (pro8) -- (pro9);
\end{tikzpicture}
\caption{Step-by-step description of the sparsity promoting algorithm for material model discovery.}
\label{fig:flowchart_algorithm}
\end{figure}
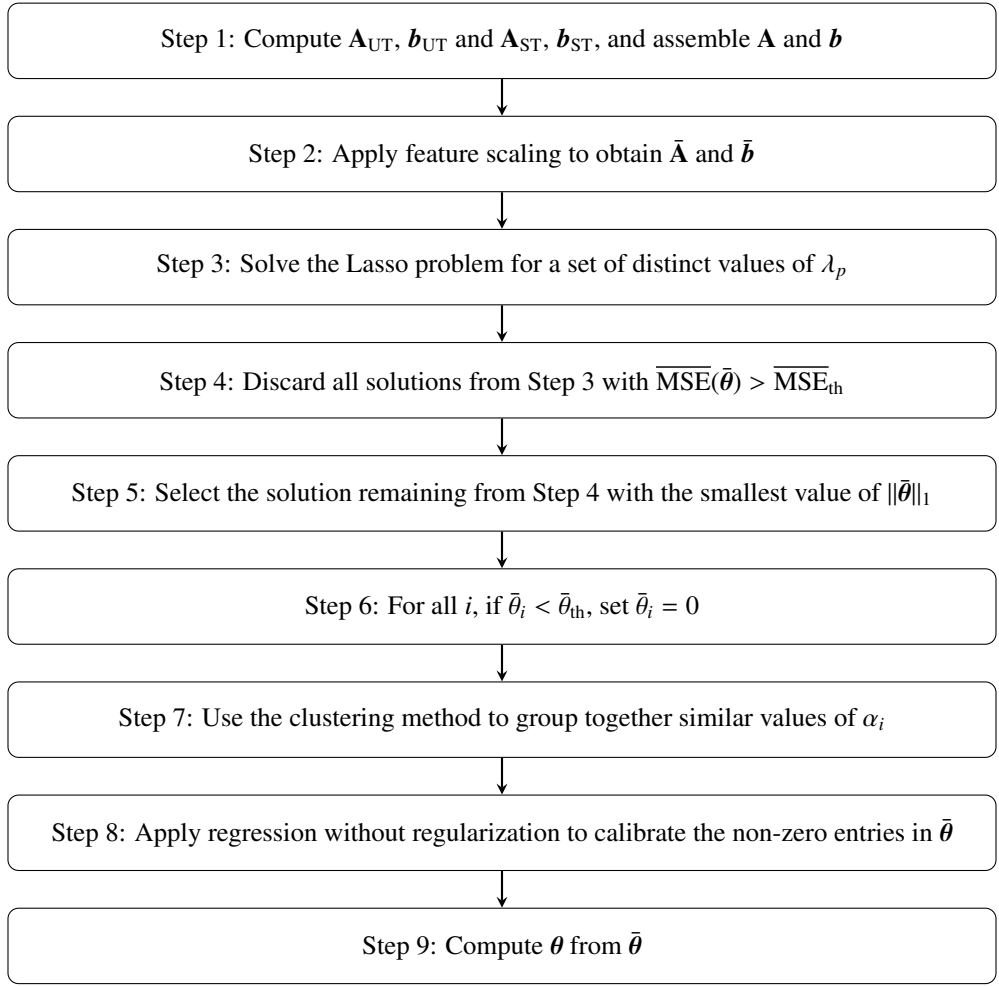

\cref{fig:flowchart_algorithm} provides a step-by-step description of the proposed algorithm for material model discovery.

\section{Specimen information}
\label{sec:specimen_dimensions}
The corresponding heights and regions of the different specimens are recorded in \cref{tab:HHBE}, while the radius of each specimen is $r_{\text{out}} \approx 4 \text{mm}$. \cref{tab:abbreviations} lists the abbreviations of the brain regions and their meaning (see \cite{hinrichsen_inverse_2022}).

\begin{table}[H]
	\caption{Heights and regions of the brain tissue specimens.}
	\label{tab:HHBE}
	\centering
	\begin{tabular}{|lll|lll|lll|lll|}
	\hline
	\textbf{HBE} & Region & $h$ [mm] & \textbf{HBE} & Region  & $h$ [mm] & \textbf{HBE} & Region  & $h$ [mm] & \textbf{HBE} & Region  & $h$ [mm] \\ \hline
	01\_03 & C         & 4.89159    & 02\_01 & CR        & 2.9083     & 03\_01 & BS         & 6.1803     & 03\_21 & CC       & 4.32991    \\
	01\_04 & BG       & 4.44005    & 02\_02 & C          & 4.36397    & 03\_02 & BS       & 5.17463    & 03\_22 & CC      & 3.91166    \\
	01\_05 & CR       & 4.64004    & 02\_03 & C           & 3.56882    & 03\_03 & BS       & 6.52776    & 03\_23 & CR      & 5.91647    \\
	01\_06 & CR       & 4.19198    & 02\_04 & CR        & 3.70545    & 03\_04 & Am       & 4.04604    & 03\_25 & C       & 4.7642     \\
	01\_07 & BG       & 3.86355    & 02\_05 & CC        & 3.71248    & 03\_05 & M        & 5.39193    & 03\_26 & Am     & 4.79861    \\
	01\_08 & C          & 5.1278     & 02\_06 & CC         & 4.27716    & 03\_06 & M        & 5.20739    & 03\_27 & BG      & 5.46595    \\
	01\_09 & BG       & 3.6299     & 02\_07 & CC         & 3.8597     & 03\_07 & CC         & 3.7975     & 03\_28 & BG       & 5.80823    \\
	01\_10 & CR       & 3.79468    & 02\_08 & CC         & 4.0629     & 03\_08 & CC        & 4.29526    & 03\_29 & CR      & 5.61761    \\
	01\_12 & CC       & 4.29779    & 02\_09 & M          & 5.16654   & 03\_09 & BG       & 5.64718    & 03\_30 & CR      & 4.97806    \\
	01\_13 & CR       & 3.89796    & 02\_10 & BG         & 4.00143    & 03\_10 & BG      & 4.96097    & 03\_31 & C        & 4.68306    \\
	01\_14 & C         & 4.59097    & 02\_13 & M          & 5.28094    & 03\_11 & CC       & 5.15737    & 03\_32 & CR      & 4.01945    \\
	01\_21 & C         & 4.42721    & 02\_14 & BG         & 4.84645    & 03\_12 & BG      & 5.96805    & 03\_33 & CR      & 6.07937    \\
	01\_22 & CR       & 3.7167      & 02\_15 & BG         & 4.69894    & 03\_13 & CC       & 3.27287    & 03\_34 & CR      & 4.66727    \\
	01\_23 & C         & 3.78048    & 02\_16 & CR          & 4.96181    & 03\_14 & CB      & 6.77129    & 03\_35 & CR      & 5.81572    \\
	01\_24 & CR       & 2.79804    & 02\_17 & C            & 4.63            & 03\_15 & CB      & 6.74769    & 03\_36 & C       & 3.8775     \\
	01\_25 & BS       & 3.61422    & 02\_18 & BG          & 3.91647     & 03\_16 & CB      & 6.40546    & 03\_37 & M      & 5.80837    \\
	01\_26 & BS       & 4.11383    & 02\_19 & C	         & 4.79345    & 03\_17 & CB       & 5.18112    & 03\_38 & M      & 6.11145    \\
	01\_27 & CB       & 4.9045      & 02\_20 & BS          & 5.41687    & 03\_18 & CR      & 6.19653    & 03\_39 & C        & 4.75102    \\
	01\_28 & CB       & 4.6799      & 02\_21 & BS          & 4.36067    & 03\_19 & C         & 4.86192    & 03\_40 & CR      & 5.26147    \\
	01\_29 & M        & 3.121        & 02\_22 & CB          & 5.37975    & 03\_20 & BG      & 6.97399    & 03\_41 & C        & 4.96854    \\
	01\_30 & M       & 4.74689    &	             &       &                    &               &       &                    &                &        &            \\ \hline
	\end{tabular}
\end{table}

\begin{table}[H]
	\caption{Abbreviations and full names of the brain regions.}
	\label{tab:abbreviations}
	\centering
    \begin{tabular}{|p{0.62 cm}|p{2.7 cm}||p{0.62 cm}|p{2.7 cm}||p{0.62 cm}|p{2.7 cm}||p{0.62 cm}|p{2.7 cm}|}
    \hline
              Am &        Amygdala  & BG &   Basal ganglia & BS &      Brain stem & C &          Cortex \\
    \hline
              CB &      Cerebellum & CC & Corpus callosum & CR &  Corona radiata & M &        Midbrain \\
    \hline
    	\end{tabular}
\end{table}

\section{Material parameters}
The discovered material models and the corresponding parameters are reported in \cref{tab:BrainDiscovery} and \cref{tab:BrainDiscovery2}. The tables further record different measures for the fitting accuracy of the discovered models as well as the computational time needed for the discovery process.

\begin{table}[H]
	\caption{Material parameters, mean squared errors and time to solution for the material model discovery from experimental data - brain~no.~01 and~02. Note that model parameters that are zero throughout all experiments are not shown in the table. All material parameters are in $\text{Pa}$, except for the dimensionless parameters $\alpha_i$.}
	\label{tab:BrainDiscovery}
	\centering
	\resizebox{16cm}{!}{
    \begin{tabular}{c|rrrrrrrr|r|r|r|}
    \cline{2-12}
    \textbf{}                         & \multicolumn{8}{c|}{\textbf{Material parameters\vphantom{$\frac{\int}{\int}$}{}}}                                                                                                                                                                                                                                                                           & \multicolumn{1}{c|}{MSE}   & \multicolumn{1}{c|}{$\overline{\mathrm{MSE}}$} & \multicolumn{1}{c|}{\textbf{Time}} \\ \hline
    \multicolumn{1}{|c|}{\textbf{HBE}} & \multicolumn{1}{c}{$\mu_1$} & \multicolumn{1}{c}{$\alpha_1$} & \multicolumn{1}{c}{$\mu_2$} & \multicolumn{1}{c}{$\alpha_2$} & \multicolumn{1}{c}{$I_2 - 3$} & \multicolumn{1}{c}{$\left(I_2 - 3\right)^2$} & \multicolumn{1}{c}{$\left(I_2 - 3\right)\left(I_1 - 3\right)$} & \multicolumn{1}{c|}{$\left(I_1 - 3\right)^2$} & \multicolumn{1}{c|}{$[\text{Pa}^2]$} & \multicolumn{1}{c|}{$[-]$}                     & \multicolumn{1}{c|}{$[s]$}         \\ \hline
    \multicolumn{1}{|c|}{01\_03}      & 29.85                       & -38.84                         & -                           & -                              & -                             & -                                            & -                                                              & -                                             & 1364.68                    & 0.26                                           & 688                                \\ \hline
    \multicolumn{1}{|c|}{01\_04}      & 125.13                      & -21.14                         & -                           & -                              & -                             & -                                            & -                                                              & -                                             & 260.16                     & 0.08                                           & 713                                \\ \hline
    \multicolumn{1}{|c|}{01\_05}      & 0.13                        & -68.28                         & 6.64                        & -33.86                         & -                             & -                                            & -                                                              & -                                             & 373.01                     & 0.51                                           & 718                                \\ \hline
    \multicolumn{1}{|c|}{01\_06}      & -                           & -                              & -                           & -                              & 33.50                         & -                                            & -                                                              & -                                             & 189.60                     & 0.50                                           & 638                                \\ \hline
    \multicolumn{1}{|c|}{01\_07}      & 7.38                        & -49.08                         & 69.29                       & -25.57                         & -                             & -                                            & -                                                              & -                                             & 2861.05                    & 0.27                                           & 719                                \\ \hline
    \multicolumn{1}{|c|}{01\_08}      & 350.45                      & -18.80                         & -                           & -                              & -                             & -                                            & -                                                              & -                                             & 1327.16                    & 0.07                                           & 741                                \\ \hline
    \multicolumn{1}{|c|}{01\_09}      & 122.55                      & -25.75                         & -                           & -                              & -                             & -                                            & -                                                              & -                                             & 878.70                     & 0.14                                           & 715                                \\ \hline
    \multicolumn{1}{|c|}{01\_10}      & 1.80                        & -57.89                         & -                           & -                              & -                             & -                                            & -                                                              & -                                             & 1027.67                    & 0.43                                           & 666                                \\ \hline
    \multicolumn{1}{|c|}{01\_12}      & 33.75                       & -20.06                         & -                           & -                              & -                             & -                                            & -                                                              & -                                             & 16.97                      & 0.09                                           & 680                                \\ \hline
    \multicolumn{1}{|c|}{01\_13}      & 0.85                        & -54.19                         & 49.02                       & -23.15                         & -                             & -                                            & -                                                              & -                                             & 251.27                     & 0.18                                           & 670                                \\ \hline
    \multicolumn{1}{|c|}{01\_14}      & 274.80                      & -21.10                         & -                           & -                              & -                             & -                                            & -                                                              & -                                             & 606.65                     & 0.04                                           & 680                                \\ \hline
    \multicolumn{1}{|c|}{01\_21}      & 2.09                        & -54.38                         & 10.20                       & -24.62                         & -                             & -                                            & -                                                              & -                                             & 872.84                     & 0.49                                           & 675                                \\ \hline
    \multicolumn{1}{|c|}{01\_22}      & 2.46                        & -45.70                         & 45.62                       & -25.99                         & -                             & -                                            & -                                                              & -                                             & 291.84                     & 0.16                                           & 674                                \\ \hline
    \multicolumn{1}{|c|}{01\_23}      & 35.76                       & -36.88                         & -                           & -                              & -                             & -                                            & -                                                              & -                                             & 983.44                     & 0.23                                           & 790                                \\ \hline
    \multicolumn{1}{|c|}{01\_24}      & 16.35                       & -20.05                         & -                           & -                              & -                             & -                                            & -                                                              & -                                             & 140.42                     & 0.71                                           & 684                                \\ \hline
    \multicolumn{1}{|c|}{01\_25}      & 0.10                        & -70.62                         & 62.52                       & -25.82                         & -                             & -                                            & -                                                              & -                                             & 540.38                     & 0.18                                           & 717                                \\ \hline
    \multicolumn{1}{|c|}{01\_26}      & 207.64                      & -24.58                         & -                           & -                              & -                             & 392.93                                       & -                                                              & -                                             & 136.29                     & 0.01                                           & 758                                \\ \hline
    \multicolumn{1}{|c|}{01\_27}      & 2.37                        & -52.66                         & 64.85                       & -24.86                         & -                             & -                                            & -                                                              & -                                             & 970.58                     & 0.23                                           & 708                                \\ \hline
    \multicolumn{1}{|c|}{01\_28}      & 119.72                      & -18.48                         & -                           & -                              & -                             & 160.40                                       & -                                                              & -                                             & 46.12                      & 0.02                                           & 727                                \\ \hline
    \multicolumn{1}{|c|}{01\_29}      & 0.01                        & -78.58                         & 90.36                       & -28.71                         & -                             & -                                            & -                                                              & -                                             & 1138.84                    & 0.17                                           & 781                                \\ \hline
    \multicolumn{1}{|c|}{01\_30}      & 55.97                       & -28.04                         & -                           & -                              & -                             & -                                            & -                                                              & -                                             & 3227.89                    & 0.61                                           & 743                                \\ \hline
    \multicolumn{1}{|c|}{02\_01}      & 6.03                        & -51.39                         & -                           & -                              & -                             & -                                            & -                                                              & -                                             & 1499.17                    & 0.35                                           & 816                                \\ \hline
    \multicolumn{1}{|c|}{02\_02}      & 1.37                        & -51.25                         & 76.88                       & -22.85                         & -                             & -                                            & -                                                              & -                                             & 275.26                     & 0.11                                           & 816                                \\ \hline
    \multicolumn{1}{|c|}{02\_03}      & 56.66                       & -34.54                         & -                           & -                              & -                             & -                                            & -                                                              & -                                             & 1369.33                    & 0.20                                           & 750                                \\ \hline
    \multicolumn{1}{|c|}{02\_04}      & 9.11                        & -46.83                         & -                           & -                              & -                             & -                                            & -                                                              & -                                             & 1014.31                    & 0.32                                           & 767                                \\ \hline
    \multicolumn{1}{|c|}{02\_05}      & 1.06                        & -51.48                         & 16.30                       & -21.06                         & -                             & -                                            & -                                                              & -                                             & 64.85                      & 0.20                                           & 700                                \\ \hline
    \multicolumn{1}{|c|}{02\_06}      & 0.37                        & -46.55                         & 16.32                       & -24.76                         & -                             & -                                            & -                                                              & -                                             & 26.25                      & 0.18                                           & 686                                \\ \hline
    \multicolumn{1}{|c|}{02\_07}      & 53.93                       & -18.72                         & -                           & -                              & -                             & -                                            & -                                                              & -                                             & 20.69                      & 0.05                                           & 710                                \\ \hline
    \multicolumn{1}{|c|}{02\_08}      & 0.49                        & -50.67                         & 42.52                       & -10.72                         & -                             & -                                            & -                                                              & -                                             & 42.58                      & 0.18                                           & 730                                \\ \hline
    \multicolumn{1}{|c|}{02\_09}      & 134.96                      & -25.83                         & -                           & -                              & -                             & 280.13                                       & -                                                              & -                                             & 211.21                     & 0.03                                           & 763                                \\ \hline
    \multicolumn{1}{|c|}{02\_10}      & 7.25                        & -47.26                         & -                           & -                              & -                             & -                                            & -                                                              & -                                             & 808.74                     & 0.34                                           & 692                                \\ \hline
    \multicolumn{1}{|c|}{02\_13}      & 0.94                        & -51.11                         & 72.32                       & -25.43                         & -                             & -                                            & -                                                              & -                                             & 272.57                     & 0.10                                           & 746                                \\ \hline
    \multicolumn{1}{|c|}{02\_14}      & 92.21                       & -24.97                         & -                           & -                              & -                             & -                                            & -                                                              & -                                             & 166.90                     & 0.06                                           & 766                                \\ \hline
    \multicolumn{1}{|c|}{02\_15}      & 2.21                        & -52.14                         & 58.47                       & -26.20                         & -                             & -                                            & -                                                              & -                                             & 591.73                     & 0.17                                           & 837                                \\ \hline
    \multicolumn{1}{|c|}{02\_16}      & 3.88                        & -46.16                         & 19.90                       & -28.55                         & -                             & -                                            & -                                                              & -                                             & 334.23                     & 0.24                                           & 723                                \\ \hline
    \multicolumn{1}{|c|}{02\_17}      & 7.69                        & -44.42                         & 19.19                       & -29.34                         & -                             & -                                            & -                                                              & -                                             & 1013.33                    & 0.34                                           & 692                                \\ \hline
    \multicolumn{1}{|c|}{02\_18}      & 88.59                       & -22.14                         & -                           & -                              & -                             & -                                            & -                                                              & -                                             & 92.16                      & 0.06                                           & 732                                \\ \hline
    \multicolumn{1}{|c|}{02\_19}      & 28.46                       & -37.27                         & -                           & -                              & -                             & -                                            & -                                                              & -                                             & 746.94                     & 0.23                                           & 754                                \\ \hline
    \multicolumn{1}{|c|}{02\_20}      & 4.99                        & -46.94                         & 18.22                       & -25.82                         & -                             & -                                            & -                                                              & -                                             & 363.63                     & 0.22                                           & 784                                \\ \hline
    \multicolumn{1}{|c|}{02\_21}      & 73.98                       & -24.99                         & -                           & -                              & -                             & -                                            & -                                                              & -                                             & 117.98                     & 0.07                                           & 751                                \\ \hline
    \multicolumn{1}{|c|}{02\_22}      & 2.96                        & -48.92                         & 25.56                       & -24.22                         & -                             & -                                            & -                                                              & \textbf{-}                                    & 298.52                     & 0.24                                           & 731                                \\ \hline
    \end{tabular}
    }
\end{table}

\begin{table}[H]
	\caption{Material parameters, mean squared errors and time to solution for the material model discovery from experimental data - brain~no.~03. Note that model parameters that are zero throughout all experiments are not shown in the table. All material parameters are in $\text{Pa}$, except for the dimensionless parameters $\alpha_i$.}
	\label{tab:BrainDiscovery2}
	\centering
	\resizebox{16cm}{!}{
    \begin{tabular}{crrrrrrrr|r|r|r|}
    \cline{2-12}
    \multicolumn{1}{r|}{\textbf{}}    & \multicolumn{8}{c|}{\textbf{Material parameters\vphantom{$\frac{\int}{\int}$}{}}}                                                                                                                                                                                                                                                                           & \multicolumn{1}{c|}{MSE}   & \multicolumn{1}{c|}{$\overline{\mathrm{MSE}}$} & \multicolumn{1}{c|}{\textbf{Time}} \\ \hline
    \multicolumn{1}{|c|}{\textbf{HBE}} & \multicolumn{1}{c}{$\mu_1$} & \multicolumn{1}{c}{$\alpha_1$} & \multicolumn{1}{c}{$\mu_2$} & \multicolumn{1}{c}{$\alpha_2$} & \multicolumn{1}{c}{$I_2 - 3$} & \multicolumn{1}{c}{$\left(I_2 - 3\right)^2$} & \multicolumn{1}{c}{$\left(I_2 - 3\right)\left(I_1 - 3\right)$} & \multicolumn{1}{c|}{$\left(I_1 - 3\right)^2$} & \multicolumn{1}{c|}{$[\text{Pa}^2]$} & \multicolumn{1}{c|}{$[-]$}                     & \multicolumn{1}{c|}{$[s]$}         \\ \hline
    \multicolumn{1}{|c|}{03\_01}      & 4.13                        & -46.12                         & 43.08                       & -26.43                         & -                             & -                                            & -                                                              & -                                             & 548.09                     & 0.20                                           & 703                                \\ \hline
    \multicolumn{1}{|c|}{03\_02}      & 38.05                       & -19.71                         & -                           & -                              & -                             & -                                            & -                                                              & -                                             & 9.05                       & 0.04                                           & 694                                \\ \hline
    \multicolumn{1}{|c|}{03\_03}      & 33.32                       & -30.51                         & -                           & -                              & -                             & -                                            & -                                                              & -                                             & 105.64                     & 0.11                                           & 793                                \\ \hline
    \multicolumn{1}{|c|}{03\_04}      & 4.24                        & -43.24                         & 63.20                       & -23.46                         & -                             & -                                            & -                                                              & -                                             & 241.05                     & 0.10                                           & 753                                \\ \hline
    \multicolumn{1}{|c|}{03\_05}      & 85.82                       & -22.43                         & -                           & -                              & -                             & -                                            & -                                                              & -                                             & 77.84                      & 0.05                                           & 710                                \\ \hline
    \multicolumn{1}{|c|}{03\_06}      & 71.81                       & -28.86                         & -                           & -                              & -                             & -                                            & -                                                              & -                                             & 123.95                     & 0.04                                           & 717                                \\ \hline
    \multicolumn{1}{|c|}{03\_07}      & 0.53                        & -50.57                         & 12.87                       & -22.03                         & -                             & -                                            & -                                                              & -                                             & 19.29                      & 0.16                                           & 709                                \\ \hline
    \multicolumn{1}{|c|}{03\_08}      & 0.97                        & -48.88                         & 12.33                       & -17.28                         & -                             & -                                            & -                                                              & -                                             & 30.73                      & 0.22                                           & 711                                \\ \hline
    \multicolumn{1}{|c|}{03\_09}      & 95.08                       & -20.61                         & -                           & -                              & -                             & -                                            & -                                                              & -                                             & 33.18                      & 0.02                                           & 726                                \\ \hline
    \multicolumn{1}{|c|}{03\_10}      & 22.46                       & -30.21                         & -                           & -                              & -                             & -                                            & -                                                              & -                                             & 58.02                      & 0.13                                           & 776                                \\ \hline
    \multicolumn{1}{|c|}{03\_11}      & 52.98                       & -16.10                         & -                           & -                              & -                             & -                                            & -                                                              & -                                             & 4.30                       & 0.02                                           & 708                                \\ \hline
    \multicolumn{1}{|c|}{03\_12}      & 55.48                       & -29.03                         & -                           & -                              & -                             & -                                            & -                                                              & -                                             & 183.62                     & 0.09                                           & 744                                \\ \hline
    \multicolumn{1}{|c|}{03\_13}      & 8.98                        & -11.23                         & -                           & -                              & -                             & -                                            & 41.30                                                          & -                                             & 3.30                       & 0.04                                           & 685                                \\ \hline
    \multicolumn{1}{|c|}{03\_14}      & 5.19                        & -41.87                         & -                           & -                              & -                             & -                                            & -                                                              & -                                             & 86.43                      & 0.30                                           & 767                                \\ \hline
    \multicolumn{1}{|c|}{03\_15}      & 52.15                       & -29.04                         & -                           & -                              & -                             & -                                            & -                                                              & -                                             & 344.86                     & 0.17                                           & 702                                \\ \hline
    \multicolumn{1}{|c|}{03\_16}      & 4.14                        & -42.41                         & -                           & -                              & -                             & -                                            & -                                                              & -                                             & 77.03                      & 0.33                                           & 697                                \\ \hline
    \multicolumn{1}{|c|}{03\_17}      & 22.37                       & -34.83                         & -                           & -                              & -                             & -                                            & -                                                              & -                                             & 212.28                     & 0.18                                           & 700                                \\ \hline
    \multicolumn{1}{|c|}{03\_18}      & 52.44                       & -18.07                         & -                           & -                              & -                             & -                                            & -                                                              & -                                             & 7.86                       & 0.02                                           & 725                                \\ \hline
    \multicolumn{1}{|c|}{03\_19}      & 35.71                       & -27.60                         & -                           & -                              & -                             & -                                            & -                                                              & -                                             & 76.98                      & 0.11                                           & 708                                \\ \hline
    \multicolumn{1}{|c|}{03\_20}      & 69.96                       & -24.83                         & -                           & -                              & -                             & -                                            & -                                                              & -                                             & 181.74                     & 0.11                                           & 719                                \\ \hline
    \multicolumn{1}{|c|}{03\_21}      & 6.67                        & -33.26                         & -                           & -                              & -                             & -                                            & -                                                              & -                                             & 8.02                       & 0.12                                           & 748                                \\ \hline
    \multicolumn{1}{|c|}{03\_22}      & 23.35                       & -19.87                         & -                           & -                              & -                             & -                                            & -                                                              & 102.69                                        & 7.63                       & 0.05                                           & 781                                \\ \hline
    \multicolumn{1}{|c|}{03\_23}      & 4.50                        & -46.22                         & 11.66                       & -26.94                         & -                             & -                                            & -                                                              & -                                             & 295.75                     & 0.26                                           & 737                                \\ \hline
    \multicolumn{1}{|c|}{03\_25}      & 63.51                       & -29.29                         & -                           & -                              & -                             & -                                            & -                                                              & -                                             & 366.86                     & 0.13                                           & 795                                \\ \hline
    \multicolumn{1}{|c|}{03\_26}      & 62.21                       & -20.92                         & -                           & -                              & -                             & -                                            & -                                                              & -                                             & 45.17                      & 0.06                                           & 770                                \\ \hline
    \multicolumn{1}{|c|}{03\_27}      & 68.54                       & -23.96                         & -                           & -                              & -                             & -                                            & -                                                              & -                                             & 70.19                      & 0.05                                           & 764                                \\ \hline
    \multicolumn{1}{|c|}{03\_28}      & 191.05                      & -18.54                         & -                           & -                              & -                             & -                                            & -                                                              & -                                             & 50.77                      & 0.01                                           & 740                                \\ \hline
    \multicolumn{1}{|c|}{03\_29}      & 1.84                        & -51.56                         & 28.07                       & -19.34                         & -                             & -                                            & -                                                              & -                                             & 388.02                     & 0.35                                           & 693                                \\ \hline
    \multicolumn{1}{|c|}{03\_30}      & 58.05                       & -21.27                         & -                           & -                              & -                             & -                                            & -                                                              & -                                             & 28.16                      & 0.04                                           & 688                                \\ \hline
    \multicolumn{1}{|c|}{03\_31}      & 15.53                       & -40.39                         & -                           & -                              & -                             & -                                            & -                                                              & -                                             & 478.73                     & 0.25                                           & 728                                \\ \hline
    \multicolumn{1}{|c|}{03\_32}      & 32.07                       & -25.14                         & -                           & -                              & -                             & -                                            & -                                                              & -                                             & 23.06                      & 0.06                                           & 730                                \\ \hline
    \multicolumn{1}{|c|}{03\_33}      & 54.99                       & -23.71                         & -                           & -                              & -                             & -                                            & -                                                              & -                                             & 45.31                      & 0.05                                           & 745                                \\ \hline
    \multicolumn{1}{|c|}{03\_34}      & 1.21                        & -48.04                         & 42.59                       & -24.79                         & -                             & -                                            & -                                                              & -                                             & 115.99                     & 0.12                                           & 742                                \\ \hline
    \multicolumn{1}{|c|}{03\_35}      & 29.28                       & -25.46                         & -                           & -                              & -                             & -                                            & -                                                              & -                                             & 28.11                      & 0.09                                           & 775                                \\ \hline
    \multicolumn{1}{|c|}{03\_36}      & 34.02                       & -32.20                         & -                           & -                              & -                             & -                                            & -                                                              & -                                             & 345.65                     & 0.21                                           & 760                                \\ \hline
    \multicolumn{1}{|c|}{03\_37}      & 105.48                      & -23.43                         & -                           & -                              & -                             & -                                            & -                                                              & -                                             & 131.43                     & 0.05                                           & 766                                \\ \hline
    \multicolumn{1}{|c|}{03\_38}      & 81.85                       & -22.01                         & -                           & -                              & -                             & -                                            & -                                                              & -                                             & 49.05                      & 0.03                                           & 725                                \\ \hline
    \multicolumn{1}{|c|}{03\_39}      & 83.47                       & -26.77                         & -                           & -                              & -                             & -                                            & -                                                              & -                                             & 237.35                     & 0.08                                           & 693                                \\ \hline
    \multicolumn{1}{|c|}{03\_40}      & 42.47                       & -23.87                         & -                           & -                              & -                             & -                                            & -                                                              & -                                             & 22.89                      & 0.05                                           & 692                                \\ \hline
    \multicolumn{1}{|c|}{03\_41}      & 2.90                        & -41.58                         & 46.88                       & -29.99                         & -                             & -                                            & -                                                              & -                                             & 365.95                     & 0.13                                           & 677                                \\ \hline
                                      &                             &                                &                             &                                &                               &                                              & \multicolumn{1}{r|}{}                                          & \textbf{Average}                              & \textbf{399.57}            & \textbf{0.17}                                & \textbf{729}                       \\ \cline{9-12} 
    \end{tabular}
    }
\end{table}

\section*{Acknowledgments}

MF and LDL acknowledge funding by the Swiss National Science Foundation (SNF) through grant N. $200021\_204316$ ``Unsupervised data-driven discovery of material laws''.
SB and PS acknowledge funding by the German Research Foundation (DFG) through grants N. $460333672/\text{CRC}1540/1-2023$ ``Exploring Brain Mechanics (EBM)'' and BU 3728/1-1.
\RR{PS acknowledges support from the European Research Council (ERC) through grant N. $101052785$, ``Configurational
Mechanics of Soft Materials (SoftFrac)''.}

\section*{Code and data availability}
Codes and data related to the EUCLID project are available at \url{https://euclid-code.github.io/}.
The codes corresponding to this work are publically available in the ETH library \citep{yu_supplementary_2023} and the mechanical testing data of the human brain tissue are available upon request.

\RR{\section*{CRediT author statement}
MF:
Conceptualization,
Methodology,
Software,
Investigation,
Writing - Original Draft,
Supervision.
HY:
Methodology,
Software,
Formal analysis,
Investigation,
Writing - Original Draft.
NR:
Resources,
Data Curation,
Investigation,
Writing - Original Draft.
JH:
Resources,
Data Curation,
Investigation,
Writing - Original Draft.
SB:
Conceptualization,
Resources,
Writing - Review \& Editing,
Funding acquisition.
PS:
Conceptualization,
Resources,
Writing - Review \& Editing,
Funding acquisition.
SK:
Conceptualization,
Methodology,
Writing - Review \& Editing.
LDL:
Conceptualization,
Methodology,
Writing - Review \& Editing,
Supervision,
Funding acquisition.}

\bibliographystyle{elsarticle-harv}
\bibliography{BibMF}

\end{document}